\documentclass[12pt]{article}
\usepackage[utf8]{inputenc}
\usepackage[a4paper, total={16.5cm, 22.5cm}]{geometry}
\usepackage{amsmath}
\usepackage{amssymb}
\usepackage{cite}
\usepackage{relsize}
\usepackage{subcaption}
\usepackage[linktocpage=true]{hyperref}

\usepackage{tikz}
\usetikzlibrary{positioning,arrows}
\usetikzlibrary{decorations.pathmorphing}
\usetikzlibrary{decorations.markings}
\usepackage{pgfplots}
\usepackage{xparse}
\usetikzlibrary{positioning,arrows,patterns}
\usetikzlibrary{decorations.markings}
\usetikzlibrary{calc}

\def\bc{\begin{center}}
\def\ec{\end{center}}
\def\be{\begin{equation}}
\def\ee{\end{equation}}
\def\beqn{\begin{eqnarray}}
\def\eeqn{\end{eqnarray}}
\def\no{\nonumber}
\def\nn{\no\\}
\def\eqn#1{(\ref{#1})}
\def\ba{\begin{array}{c}}
\def\ea{\end{array}}
\def\bat{\begin{array}{cc}}
\def\bi{\begin{itemize}}
\def\ei{\end{itemize}}
\def\cA{{\cal A}}
\def\cL{{\cal L}}
\def\cQ{{\cal Q}}

\def\cM{{\cal M}}

\def\cO{{\cal O}}

\newcommand{\lsim}{~{}_{\textstyle\sim}^{\textstyle <}~}

\newcommand{\e}{\mathrm{e}}

\newcommand{\DS}{\Delta S=1}  

\newcommand{\gf}{\gamma_5}

\begin{document}

\thispagestyle{empty}
\hfill\mbox{IFIC/17-56, FTUV/17-1218}
\vspace{2cm}

\begin{center}

{
\fontsize{22}{26}\selectfont\sc \bf\boldmath
%
Direct CP violation in $K^0\to\pi\pi$:\\[14pt] Standard Model Status
}

\vspace{1.4cm}

{\sc
Hector Gisbert and Antonio Pich 
}

\vspace*{.7cm}

{\sl
Departament de F\'\i sica Te\`orica, IFIC, Universitat de Val\`encia -- CSIC\\
Apt. Correus 22085, E-46071 Val\`encia, Spain
}

\end{center}

\vspace*{0.1cm}

\begin{abstract}
\noindent
In 1988 the NA31 experiment presented the first evidence of direct CP violation in the $K^0\to\pi\pi$ decay amplitudes. A clear signal with a $7.2\,\sigma$ statistical significance was later established with the full data samples from the NA31, E731, NA48 and KTeV experiments, confirming that CP violation is associated with a $\Delta S=1$ quark transition, as predicted by the Standard Model. However, the theoretical  prediction for the measured ratio $\varepsilon'/\varepsilon$ has been a subject of strong controversy along the years.
Although the underlying physics was already clarified in 2001, the recent release of improved lattice data has revived again the theoretical debate.
We review the current status, discussing in detail the different ingredients that enter into the calculation of this observable and the reasons why seemingly contradictory predictions were obtained in the past by several groups. 
An update of the Standard Model prediction is presented and the prospects for future improvements are analysed. Taking into account all known short-distance and long-distance contributions, one obtains  
$\mbox{Re}\left(\varepsilon'/\varepsilon\right) = (15 \pm 7)\cdot 10^{-4}$, in good agreement with the experimental measurement.

\end{abstract}

\vfill\mbox{}

\setcounter{page}{0}

\newpage

\tableofcontents
\addtocontents{toc}{~\hfill\textbf{Page}\par}
\newpage

\section{Historical prelude}
\label{sec:intro}

The investigation of kaon decays \cite{Cirigliano:2011ny} has a fruitful record of very important scientific achievements, being at the origin of many of the fundamental ingredients that have given rise to the current structure of the electroweak Standard Model (SM) \cite{Glashow:1961tr,Weinberg:1967tq,Salam:1968rm}: the flavour concept of strangeness \cite{GellMann:1953zza,Pais:1952zz}, parity violation \cite{Dalitz:1954cq,Lee:1956qn}, meson-antimeson oscillations \cite{GellMann:1955jx}, quark mixing \cite{Cabibbo:1963yz}, the GIM mechanism \cite{Glashow:1970gm} and CP violation \cite{Christenson:1964fg,Kobayashi:1973fv}. Since their discovery in 1947 \cite{Rochester:1947mi}, kaons have played a very significant role in our understanding of fundamental physics, providing accurate tests of quantum mechanics and uncovering the existence of higher-mass scales such as the charm \cite{Glashow:1970gm,Gaillard:1974hs} and the top quarks \cite{Kobayashi:1973fv,Buras:1992uf}. Nowadays, the kaon decay data continue having a major impact on flavour phenomenology and impose very stringent constraints on plausible scenarios of new physics. 

The measured ratios of $K_L\to 2\pi$ and $K_S\to 2\pi$ decay amplitudes,
\be\label{eq:eta_def}
\eta_{_{00}}\,\equiv\, \frac{A(K_L\to\pi^0\pi^0)}{A(K_S\to\pi^0\pi^0)}\,\equiv\, \varepsilon - 2\,\varepsilon'\, ,
\qquad\qquad
\eta_{_{+-}}\,\equiv\, \frac{A(K_L\to\pi^+\pi^-)}{A(K_S\to\pi^+\pi^-)}\,\equiv\, \varepsilon +\varepsilon'\, ,
\ee
exhibit a clear violation of the CP symmetry at the per-mill level \cite{Olive:2016xmw},
%
\be\label{eq:epsilon_exp}
|\varepsilon|\,  =\, \frac{1}{3} \left| \eta_{_{00}} + 2\, \eta_{_{+-}} \right|\, =\,
(2.228\pm 0.011)\cdot 10^{-3}\, ,
\ee
which originates in a $\Delta S=2$ weak transition between the $K^0$ and the $\bar K^0$ states \cite{Christenson:1964fg,Cirigliano:2011ny}. A more subtle effect is the existence of a tiny difference between $\eta_{_{00}}$ and $\eta_{_{+-}}$ that has been experimentally established through the ratio \cite{Batley:2002gn,Lai:2001ki,Fanti:1999nm,Barr:1993rx,Burkhardt:1988yh,Abouzaid:2010ny,AlaviHarati:2002ye,AlaviHarati:1999xp,Gibbons:1993zq}
\be\label{eq:exp}
\mathrm{Re} \left(\varepsilon'/\varepsilon\right)\; =\;
\frac{1}{3} \left( 1   -\left|\frac{\eta_{_{00}}}{\eta_{_{+-}}}\right|\right) \; =\;
(16.6 \pm 2.3) \cdot  10^{-4}\, ,
\ee
demonstrating the existence of direct CP violation in the $K^0\to 2\pi$ decay amplitudes. This measurement plays a crucial role in our understanding of the dynamical origin of the CP violation, since it confirms that it is associated with a $\Delta S=1$  transition, as predicted by the 
SM Cabibbo-Kobayashi-Maskawa (CKM) mechanism \cite{Cabibbo:1963yz,Kobayashi:1973fv}.

The theoretical prediction of $\varepsilon'/\varepsilon$ has a quite controversial history \cite{Gilman:1978wm,Buchalla:1989we,Buras:1993dy,Buras:1996dq,Bosch:1999wr,Buras:2000qz,Ciuchini:1995cd,Ciuchini:1992tj,Bertolini:1995tp,Bertolini:1997nf,Bertolini:1998vd,Bertolini:2000dy,Hambye:1999yy,Pallante:1999qf,Pallante:2000hk,Pallante:2001he,Bijnens:2000im,Hambye:2003cy,Buras:2003zz,Pich:2004ee}
because the first next-to-leading order (NLO) calculations \cite{Buras:1993dy,Buras:1996dq,Bosch:1999wr,Buras:2000qz,Ciuchini:1995cd,Ciuchini:1992tj} claimed SM values one order of magnitude smaller than \eqn{eq:exp}, contradicting the clear signal observed in 1988 by the CERN NA31 collaboration \cite{Barr:1993rx,Burkhardt:1988yh} and giving support to the null result obtained by the E731 experiment at Fermilab \cite{Gibbons:1993zq}. The final confirmation that $\mathrm{Re} \left(\varepsilon'/\varepsilon\right)\sim 10^{-3}$, with the NA48 \cite{Batley:2002gn,Lai:2001ki,Fanti:1999nm} and KTeV \cite{Abouzaid:2010ny,AlaviHarati:2002ye,AlaviHarati:1999xp} data,  triggered then a large number of new-physics explanations for that exciting ``flavour anomaly''.\footnote{Many papers addressing the claimed discrepancy can be found at the Inspire data basis. We refrain from quoting them here.} However, it was soon realized that the former SM predictions had missed completely the important role of the final pion dynamics \cite{Pallante:1999qf,Pallante:2000hk}. Once long-distance contributions were properly taken into account, the theoretical prediction was found to be in good agreement with the experimental value, albeit with unfortunately large uncertainties of non-perturbative origin \cite{Pallante:2001he}.

Numerical QCD simulations on a discretised space-time volume are an appropriate tool to address non-perturbative problems. However, lattice calculations of the $K\to 2\pi$ amplitudes face many technical challenges, associated with their Minkowskian nature (the physical amplitudes cannot be extracted from standard Euclidean lattice simulations \cite{Maiani:1990ca}), the presence of several competing operators with a very involved dynamical interplay, and the vacuum quantum numbers of the isoscalar $\pi\pi$ final state (a large vacuum-state contribution must be subtracted, which deteriorates the signal to noise ratio). For many years, a quantitative lattice corroboration of the known enhancement of the $\Delta I =1/2$ amplitude remained unsuccessful, while attempts to estimate $\varepsilon'/\varepsilon$ were unreliable, often obtaining negative values due to an insuficient signal in the isoscalar decay amplitude \cite{Pekurovsky:1998jd,Noaki:2001un,Blum:2001xb,Bhattacharya:2004qu}. 
The situation has changed in recent years, thanks to the development of more sophisticated techniques and the increasing power of modern computers. A quite successful calculation of the $\Delta I =3/2$ $K^+\to \pi^+\pi^0$ amplitude has been achieved by the RBC-UKQCD collaboration \cite{Blum:2011ng,Blum:2012uk,Blum:2015ywa}, and the first statistically-significant signal of the $\Delta I =1/2$ enhancement has recently been reported \cite{Boyle:2012ys}, confirming the qualitative dynamical understanding achieved long time ago with analytical methods \cite{Pich:1995qp,Pich:1990mw,Jamin:1994sv,Bertolini:1997ir,Antonelli:1995nv,Antonelli:1995gw,Hambye:1998sma,Bardeen:1986vz,Buras:2014maa,Bijnens:1998ee}.\footnote{A large enhancement of the isoscalar $K\to 2\pi$ amplitude has also been found at unphysical quark masses ($M_K \sim 2 M_\pi$), using PACS-CS gauge configurations generated with the Iwasaki action and the $\cO(a)$-improved Wilson fermion action \cite{Ishizuka:2015oja}.}

From its most recent lattice data, the RBC-UKQCD collaboration has also extracted a first estimate for the direct CP-violation ratio: $\mathrm{Re}(\varepsilon'/\varepsilon) = (1.38 \pm 5.15 \pm 4.59) \cdot 10^{-4}$ \cite{Bai:2015nea,Blum:2015ywa}. Although the quoted errors are still large, the low central value would imply a $2.1\,\sigma$ deviation from the experimental value in Eq.~\eqn{eq:exp}. This discrepancy has revived some of the old SM calculations predicting low values of $\varepsilon'/\varepsilon$ \cite{Buras:2015xba,Buras:2016fys,Buras:2015yba} (missing again the crucial pion dynamics), and has triggered new studies of possible contributions  from physics beyond the SM \cite{Buras:2014sba,Buras:2015yca,Blanke:2015wba,Buras:2015kwd,Buras:2016dxz,Buras:2015jaq,Kitahara:2016otd,Kitahara:2016nld,Endo:2016aws,Endo:2016tnu,Cirigliano:2016yhc,Alioli:2017ces,Bobeth:2016llm,Bobeth:2017xry,Crivellin:2017gks,Chobanova:2017rkj,Bobeth:2017ecx,Endo:2017ums}. 

Before claiming any evidence for new physics, one should realize the technical limitations of the current lattice result. In order to access the Minkowskian kinematics with physical interacting pions, evading the Maini-Testa no-go theorem \cite{Maiani:1990ca}, the RBC-UKQCD simulation follows the elegant method developed by Lellouch and Lüscher \cite{Lellouch:2000pv} to relate the infinite-volume and finite-volume results. At finite volumes there is a discrete spectrum and the box size can be tuned to get pions with the desired momentum; moreover, the Lüscher quantization condition \cite{Luscher:1986pf,Luscher:1990ux} allows one to compute the needed S-wave phase shift of the final $\pi\pi$ state. The $(\pi\pi)_I$ phase shifts, $\delta_I$ ($I=0,2$), play a crucial role in the calculation and provide a quantitative test of the lattice result.
While the extracted $I=2$ phase shift is only $1\,\sigma$ away from its physical value, the lattice analysis of Ref.~\cite{Bai:2015nea} finds a result for
$\delta_{0}$ which disagrees with the experimental value by $2.9\,\sigma$, a much larger discrepancy than the one quoted for $\varepsilon'/\varepsilon$. Obviously, nobody is looking for any new-physics contribution to the $\pi\pi$ elastic scattering  phase shifts.
Nevertheless, although it is still premature to derive strong physics implications from these lattice results, they look already quite impressive and show that substantial improvements could be achieved in the near future \cite{Feng:2017voh}. 

Meanwhile, it seems worth to revise and update the analytical SM calculation of $\varepsilon'/\varepsilon$ \cite{Pallante:2001he}, which is already 16 years old. A very detailed study of electromagnetic and isospin-violating corrections, which play a very important role in $\varepsilon'/\varepsilon$, was accomplished later \cite{Cirigliano:2003nn,Cirigliano:2003gt,Cirigliano:2009rr}. Although the main numerical implications for the $\varepsilon'/\varepsilon$ prediction were reported in some unpublished conference proceedings \cite{Pich:2004ee} and have been quoted in more recent reviews~\cite{Cirigliano:2011ny}, a complete phenomenological analysis including properly these corrections has never been presented. The penguin matrix elements that dominate the CP-violating $K\to 2\pi$ amplitudes are also quite sensitive to the numerical inputs adopted for the light quark masses. Thanks to the impressive lattice progress achieved in recent years \cite{Aoki:2016frl}, the quark masses are nowadays determined with a much better precision and their impact on $\varepsilon'/\varepsilon$ must be investigated. Moreover, we have now a better understanding of several non-perturbative ingredients entering the calculation such as chiral low-energy constants and large-$N_C$ relations \cite{Ecker:1988te,Ecker:1989yg,Pich:2002xy,Cirigliano:2006hb,Kaiser:2007zz,Cirigliano:2004ue,Cirigliano:2005xn,RuizFemenia:2003hm,Jamin:2004re,Rosell:2004mn,Rosell:2006dt,Pich:2008jm,GonzalezAlonso:2008rf,Pich:2010sm,Bijnens:2014lea,Rodriguez-Sanchez:2016jvw,Ananthanarayan:2017qmx}. 

A convenient decomposition of the $K\to 2\pi$ amplitudes \cite{Cirigliano:2003gt} that allows us to incorporate electromagnetic and isospin-violating corrections, closely following the familiar isospin notation, is presented in section~\ref{sec:K2pi}. We also review there the phenomenological expressions needed to compute $\varepsilon'/\varepsilon$ and the isospin-breaking corrections computed in Refs.~\cite{Cirigliano:2003nn,Cirigliano:2003gt,Cirigliano:2009rr}. The short-distance contributions to $\varepsilon'/\varepsilon$ are detailed in section~\ref{sec:SD}, where large logarithmic corrections $\sim\alpha_s^k(\mu) \log^n{(M_W/\mu)}$ are summed up with the renormalization group, at the NLO ($k=n,n+1$).
The hadronic matrix elements of the relevant four-quark operators are discussed in section~\ref{sec:hadronic}. Their chiral $SU(3)_L\otimes SU(3)_R$ symmetry properties are analysed, emphasizing the reasons why the strong and electroweak penguin operators $Q_6$ and $Q_8$ dominate the CP-odd kaon decay amplitudes into final pions with $I=0$ and 2, respectively. Using the large-$N_C$ limit, with $N_C$ the number of QCD colours, we also provide there a first simplified estimate of $\varepsilon'/\varepsilon$ that exhibits the presence of a subtle numerical cancellation. This estimate allows us to easily understand the numerical values quoted in Ref.~\cite{Buras:2015yba}.

Sections~\ref{sec:eft}, \ref{sec:matching} and \ref{sec:K2pi-chpt} present a much more powerful effective field theory (EFT) approach to the problem. The low-energy realization of the short-distance $\Delta S=1$ effective Lagrangian is analysed in section~\ref{sec:eft}, using the well-known techniques of Chiral Perturbation Theory ($\chi$PT) \cite{Weinberg:1978kz,Gasser:1983yg,Gasser:1984gg,Ecker:1994gg,Pich:1995bw,Bijnens:1999sh,Bijnens:1999hw} that make possible to pin down the long-distance contributions to the $K\to\pi\pi$ amplitudes and unambiguously determine the logarithmic chiral corrections. Section~\ref{sec:matching} discusses the matching between the short-distance Lagrangian and $\chi$PT and shows how the chiral couplings can be determined at $N_C\to\infty$. The kaon decay amplitudes are worked out in section~\ref{sec:K2pi-chpt}, at the NLO in the chiral expansion. The one-loop chiral corrections are rather large and have a very important impact on $\varepsilon'/\varepsilon$ because they destroy the numerical cancellation present in simplified analyses. This is explained in section~\ref{sec:result}, which contains the updated determination of the CP-violating ratio. A detailed discussion of the current result and the prospects for future improvements are finally given in section~\ref{sec:outlook}. The input values adopted for the different parameters entering the analysis are collected in the appendix.

\section{Isospin decomposition of the $\mathbf{K\boldsymbol{\to} \boldsymbol{\pi\pi}}$ amplitudes}
\label{sec:K2pi}

The $K\to 2\pi$ decay amplitudes can be parametrized in the general form\footnote{
Including electromagnetic corrections, this parametrization holds for the infrared-finite amplitudes after the Coulomb and infrared parts are removed and treated in combination with real photon emission
\cite{Cirigliano:2003gt}.
} \cite{Cirigliano:2003gt}
%
\begin{eqnarray}  
A(K^0 \to \pi^+ \pi^-) &=&  
\cA_{1/2} + {1 \over \sqrt{2}} \left( \cA_{3/2} + \cA_{5/2} \right) \; =\; 
 A_{0}\,  e^{i \chi_0}  + { 1 \over \sqrt{2}}\,    A_{2}\,  e^{i\chi_2 } \, ,
\nn
A(K^0 \to \pi^0 \pi^0) &=& 
\cA_{1/2} - \sqrt{2} \left( \cA_{3/2} + \cA_{5/2}  \right) \; =\;
A_{0}\,  e^{i \chi_0}  - \sqrt{2}\,    A_{2}\,  e^{i\chi_2 }\, ,
\nn
A(K^+ \to \pi^+ \pi^0) &=&
{3 \over 2}  \left( \cA_{3/2} - {2 \over 3} \cA_{5/2} \right) \; =\;
{3 \over 2}\, A_{2}^{+}\,   e^{i\chi_2^{+}},
\label{eq:2pipar}
\end{eqnarray}
which expresses the three physical amplitudes in terms of three complex quantities $\cA_{\Delta I}$ that are generated by the $\Delta I = \frac{1}{2},
\frac{3}{2},\frac{5}{2}$ components of the electroweak effective Hamiltonian, in the limit of isospin conservation. Writing $\cA_{1/2} \equiv A_{0}\,  e^{i \chi_0}$, $\cA_{3/2} + \cA_{5/2} \equiv A_{2}\,  e^{i\chi_2 }$ and $\cA_{3/2} - {2 \over 3} \cA_{5/2} \equiv A_{2}^{+}\,   e^{i\chi_2^{+}}$, our notation closely follows the usual isospin decomposition. In the CP-conserving limit the amplitudes, $A_0$, $A_2$ and $A_2^+$ are real and positive by definition.

In the SM, $\cA_{5/2} = 0$ in the absence of electromagnetic interactions. If isospin symmetry is assumed, $A_0$ and $A_2 = A_2^+$ correspond to the decay amplitudes into the $I=0$ and 2 $(\pi\pi)_I$ final states. The phases $\chi_0$ and $\chi_2 = \chi_2^{+}$ can then be identified with the S-wave $\pi\pi$ scattering phase shifts $\delta_I$ at $\sqrt{s}=M_K$, up to isospin-breaking effects \cite{Cirigliano:2009rr,Cirigliano:2003gt}. 

In the isospin limit (keeping the physical meson masses in the phase space), $A_0$, $A_2$ and the phase difference $\chi_0-\chi_2$ can be directly extracted from the measured $K\to\pi\pi$ branching ratios \cite{Antonelli:2010yf}:
\begin{eqnarray}
A_0 &=& (2.704 \pm 0.001) \cdot 10^{-7} \mbox{ GeV}, \nn
A_2 &=& (1.210 \pm 0.002) \cdot 10^{-8} \mbox{ GeV}, \nn
\chi_0 - \chi_2 &=& (47.5 \pm 0.9)^{\circ}.
\label{eq:isoamps}
\end{eqnarray}
They exhibit a strong enhancement of the isoscalar amplitude with respect to the isotensor one  (the so-called ``$\Delta I =\frac{1}{2}$ rule''),
\be\label{eq:omega}
\omega \,\equiv\, \frac{\mathrm{Re} A_2}{\mathrm{Re} A_0}\,\approx\,\frac{1}{22}\, ,
\ee
and a large phase-shift difference between the two isospin amplitudes, which mostly originates in the strong S-wave rescattering of the two final pions with $I=0$.

When CP violation is turned on, the amplitudes $A_0$, $A_2$ and $A_2^+$ acquire imaginary parts. The direct CP-violating signal is generated by the interference of the two possible $K^0\to\pi\pi$ decay amplitudes, with different weak and strong phases. To first order in CP violation,
\begin{equation}
\varepsilon'\,  =\, \frac{1}{3} \left( \eta_{+-} - \eta_{00}\right)\, =\,
- \frac{i}{\sqrt{2}} \: e^{i ( \chi_2 - \chi_0 )} \:\hat\omega\;
\left[
\frac{\mathrm{Im} A_{0}}{ \mathrm{Re} A_{0}} \, - \,
\frac{\mathrm{Im} A_{2}}{ \mathrm{Re} A_{2}} \right] ,
\label{eq:cp1}
\end{equation}
with 
\be
\hat\omega\, =\, \frac{\omega}{1 - \frac{\omega}{\sqrt{2}}\,  e^{i ( \chi_2 - \chi_0 )} - \omega^2\,  e^{2 i ( \chi_2 - \chi_0 )}}\,\approx\, \omega\, .
\ee
Thus, $\varepsilon'$ is suppressed by the small dynamical ratio $\omega$. 
The global phase $\phi_{\varepsilon'} =\chi_2 - \chi_0 + \pi/2 = (42.5\pm 0.9)^\circ$
is very close to 
$\phi_\varepsilon \approx \tan^{-1}{
\left[
2 (m_{K_L}-m_{K_S})/(\Gamma_{K_S}-\Gamma_{K_L})
\right]}
= (43.52\pm 0.05)^\circ$
\cite{Olive:2016xmw}, the so-called superweak phase.
This implies that $\cos{(\phi_{\varepsilon'} -\phi_\varepsilon)}\approx 1$ and $\varepsilon'/\varepsilon$ is approximately real.

Eq.~\eqn{eq:cp1} involves a very delicate numerical balance between the two isospin contributions. In order to minimize hadronic uncertainties, the CP-conserving amplitudes $\mathrm{Re} A_{I}$ are usually set to their experimentally determined values. A first-principle calculation is only needed for the CP-odd amplitudes $\mathrm{Im} A_{0}$ and $\mathrm{Im} A_{2}$, which are dominated by the strong and electromagnetic penguin contributions, respectively, to be discussed in the next section. However, naive estimates of $\mathrm{Im} A_{I}$ result in a large numerical cancellation between the two terms, leading to unrealistically low values of $\varepsilon'/\varepsilon$ \cite{Buras:1993dy,Buras:1996dq,Bosch:1999wr,Buras:2000qz,Ciuchini:1995cd,Ciuchini:1992tj}. The true SM prediction is then very sensitive to the precise values of the two isospin contributions \cite{Bertolini:1998vd,Hambye:1999yy}. Small corrections to any of the two amplitudes get strongly amplified in \eqn{eq:cp1} because they destroy the accidental numerical cancellation.

Isospin violation plays a very important role in $\varepsilon'/\varepsilon$ due to the large ratio $1/\omega$. Small isospin-violating corrections to the dominant decay amplitude $A_0$ generate very sizeable contributions to $A_2$, which have obviously a direct impact on $\varepsilon'/\varepsilon$. Isospin-breaking effects in $K\to\pi\pi$ decays have been systematically analysed in Refs.~\cite{Cirigliano:2003nn,Cirigliano:2003gt,Cirigliano:2009rr}, including corrections from electromagnetic interactions, at NLO in $\chi$PT.
To first order in isospin violation, $\varepsilon'$ can be expressed in the following form, which makes explicit all sources of isospin breaking:
\begin{equation}
\varepsilon' \, = \, - \frac{i}{\sqrt{2}}\:  \e^{i ( \chi_2 - \chi_0 )}\:
\omega_+ \;   \left[
\frac{\mathrm{Im} A_{0}^{(0)}}{ \mathrm{Re} A_{0}^{(0)}}\,
\left( 1 + \Delta_0 + f_{5/2}\right)  -  \frac{\mathrm{Im} A_{2}}{ \mathrm{Re}
  A_{2}^{(0)}} \right] .
\label{eq:cpiso}
\end{equation}
From the (isospin conserving) phenomenological fit in Eq.~\eqn{eq:isoamps}, one actually extracts $\omega_+ = \mathrm{Re} A_{2}^{+}/\mathrm{Re} A_{0}$, which differs from $\omega$ by a pure $\Delta I = \frac{5}{2}$ correction induced by the electromagnetic interaction at NLO, {\it i.e.}, at $\cO (e^2 p^2)$ \cite{Cirigliano:2011ny,Cirigliano:2003nn,Cirigliano:2003gt,Cirigliano:2009rr},
\be
f_{5/2} \; =\; \frac{\mathrm{Re} A_{2}}{\mathrm{Re} A_{2}^{+}} - 1 \; =\;
\left( 8.44\pm 0.02_{\mathrm{exp}}\pm 2.5_{\mathrm{th}}\right)\cdot 10^{-2} \, .
\ee
The superscript $(0)$ on the amplitudes denotes the isospin limit and \cite{Cirigliano:2003nn,Cirigliano:2003gt}
\be
\Delta_0 \; =\; \frac{\mathrm{Im} A_0}{\mathrm{Im} A_0^{(0)}}\;
\frac{\mathrm{Re} A_0^{(0)}}{\mathrm{Re}A_0}  - 1
\; =\; (8.4\pm 3.6)\cdot 10^{-2} 
\label{eq:epspdefs}
\ee
includes corrections of $\cO [(m_u-m_d) p^2,e^2 p^2]$. The final numerical result for $\Delta_0$ is governed to a large extent by the electromagnetic penguin contribution to $\mathrm{Im} A_0$.

The expression \eqn{eq:cpiso} takes already into account that $\mathrm{Im} A_2$ is itself of first order in isospin violation. It is convenient to separate the leading contribution of the electromagnetic penguin operator from the isospin-breaking effects generated by other four-quark operators:
\begin{equation}
\mathrm{Im} A_2\; =\; \mathrm{Im} A_2^{\rm emp} \, + \, \mathrm{Im}  A_2^{\rm non-emp}\, .
\label{eq:empsep}
\end{equation}
This separation is renormalization-scheme dependent,\footnote{
The renormalization-scheme ambiguity is only present in the electromagnetic contribution. The splitting between $\mathrm{Im} A_2^{\rm emp}$ and $\mathrm{Im}  A_2^{\rm non-emp}$ has been performed (in the $\overline{\mathrm{MS}}$ scheme with naive dimensional regularization) matching the short-distance Hamiltonian and $\chi$PT  effective descriptions (see the next sections) at leading order in $1/N_C$.}
but allows one to identify those isospin-violating contributions which are enhanced by the ratio $1/\omega$ and write them explicitly as corrections to the $I=0$ side through the parameter \cite{Cirigliano:2003nn,Cirigliano:2003gt}
\be\label{eq:omegaIB}
\Omega_{\rm IV} \; =\;  \frac{\mathrm{Re} A_0^{(0)} }
{ \mathrm{Re} A_2^{(0)} } \; \frac{\mathrm{Im} A_2^{\rm non-emp} }
{ \mathrm{Im} A_0^{(0)} } 
\; =\; (22.7 \pm 7.6)\cdot 10^{-2}\, .
\ee
This quantity includes a sizeable contribution from $\pi^0$--$\eta$ mixing \cite{Ecker:1999kr} which dominates the full NLO correction from strong isospin violation: $\Omega_{\rm IV}^{\alpha=0} = (15.9\pm 4.5)\cdot 10^{-2}$. Electromagnetic contributions are responsible for the numerical difference with the value in Eq.~\eqn{eq:omegaIB}.

The phenomenological analysis of $\varepsilon'/\varepsilon$ can then be more easily performed with the expression
\be\label{eq:epsp_simp}
\mathrm{Re}(\varepsilon'/\varepsilon) \; = \; - \frac{\omega_+}{\sqrt{2}\, |\varepsilon|}  \; \left[
\frac{\mathrm{Im} A_{0}^{(0)} }{ \mathrm{Re} A_{0}^{(0)} }\;
\left( 1 - \Omega_{\rm eff} \right) - \frac{\mathrm{Im} A_{2}^{\rm emp}}{ \mathrm{Re}
  A_{2}^{(0)} } \right] \, ,
\ee
with \cite{Cirigliano:2003nn,Cirigliano:2003gt}
\be
\Omega_{\rm eff} \; =\; \Omega_{\rm IV} - \Delta_0 - f_{5/2}
\; =\; (6.0\pm 7.7) \cdot 10^{-2}\, .
\ee

Notice that there is a large numerical cancellation among the different isospin-breaking corrections. Although the separate strong and electromagnetic contributions are sizeable, they interfere destructively leading to a final isospin-violation correction of moderate size.

\section{Short-distance contributions}
\label{sec:SD}

\begin{figure}[t]

\centering
\hbox{  
  \subcaptionbox{\label{fig1:a}}{\tikzset{
    vector/.style={decorate, decoration={snake}, draw},
      gluon1/.style={decorate, draw=blue,
        decoration={snake}, draw},
    fermion/.style={draw=black, postaction={decorate},
        decoration={markings,mark=at position .55 with {\arrow[draw=black]{>}}}},
}

\begin{tikzpicture}[line width=1.5 pt, node distance=2cm and 2cm ]
	
\coordinate[label=below:\scriptsize{$ $}] (v1);
\coordinate[above=of v1,label=above:\scriptsize{$ $}] (v2);
\coordinate[right=of v2,label=above:\scriptsize{$ $}] (v3);
\coordinate[below= of v3,label=below:\scriptsize{$ $}] (v4);
\coordinate[above right= 2cm and .2cm of v4,label=right:$s$] (e2);
\coordinate[below right= 2cm and .2cm of v3,label=right:$u$] (f2);
\coordinate[above left = 2cm and .2cm of v1,label=left :$u$] (f1);
\coordinate[below left = 2cm and .2cm of v2,label=left :$d$] (e1);
\coordinate[above left=1cm and  -0.9cm of v1,label=left :$W$] (a1);
\coordinate[below right=0.35cm and  -0.8cm of v4,label=left :$ $] (a2);
\coordinate[above right=2cm and  -1cm of v4,label=left :$ $] (o1);   
\coordinate[above right=0cm and  -1cm of v4,label=left :$ $] (o2);

		\draw[fermion] (e1) -- (o2);
		\draw[fermion] (f1) -- (o1);
		\draw[fermion] (o1) -- (e2);
		\draw[fermion] (o2) -- (f2);
	    \draw[vector] (o1) -- (o2);
	    \fill[black] (o1) circle (.1cm);
	    \fill[black] (o2) circle (.1cm);
	    
\end{tikzpicture}
	
	\tikzset{
    vector/.style={decorate, decoration={snake}, draw},
      gluon/.style={decorate, draw=red,
        decoration={coil,amplitude=4pt, segment length=5pt}}, 
    fermion/.style={draw=black, postaction={decorate},
        decoration={markings,mark=at position .55 with {\arrow[draw=black]{>}}}},
}

\hskip .6cm
\begin{tikzpicture}[line width=1.5 pt, node distance=2cm and 2cm ]
	
\coordinate[label=below:\scriptsize{$ $}] (v1);
\coordinate[above=of v1,label=above:\scriptsize{$ $}] (v2);
\coordinate[right=of v2,label=above:\scriptsize{$ $}] (v3);
\coordinate[below= of v3,label=below:\scriptsize{$ $}] (v4);
\coordinate[above right= 2cm and 1.1cm of v4,label=right:$s$] (e2);
\coordinate[below right= 2cm and 1.1cm of v3,label=right:$u$] (f2);
\coordinate[above left = 2cm and 1.1cm of v1,label=left :$u$] (f1);
\coordinate[below left = 2cm and 1.1cm of v2,label=left :$d$] (e1);
\coordinate[above left=1cm and  0.2cm of v1,label=left :$\color{red}{g}$] (a1);
\coordinate[above right=0.3cm and  1.2cm of v2,label=left :$ $] (a2);
\coordinate[below right=1cm and  0.8cm of v3,label=left :$W$] (a3);
\coordinate[below right=0.35cm and  -0.8cm of v4,label=left :$ $] (a2);
\coordinate[above right=0.3cm and 0cm of v1,label=right:\tiny{$ $}];
\coordinate[below right=0.3cm and 0cm of v2,label=right:\tiny{$ $}];
\coordinate[below right=0.3cm and 0cm of v3,label=left:\tiny{$ $}];
\coordinate[above right=0.3cm and 0cm of v4,label=left:\tiny{$ $}];

		\draw[fermion] (e1) -- (v1);
		\draw[fermion] (v2) -- (v3);
		\draw[fermion] (f1) -- (v2);
		\draw[fermion] (v3) -- (e2);
		\draw[fermion] (v1) -- (v4);
		\draw[fermion] (v4) -- (f2);
		\draw[gluon] (v1) -- (v2);
		\draw[vector] (v3) -- (v4);
		
	    \fill[blue] (v1) circle (.1cm);
	    \fill[blue] (v2) circle (.1cm);
	    \fill[black] (v3) circle (.1cm);
	    \fill[black] (v4) circle (.1cm);
\end{tikzpicture}}

\subcaptionbox{\label{fig1:b}}{\tikzset{
    vector/.style={decorate, decoration={snake}, draw},
      gluon/.style={decorate, draw=red,
        decoration={coil,amplitude=4pt, segment length=5pt}}, 
    fermion/.style={draw=black, postaction={decorate},
        decoration={markings,mark=at position .55 with {\arrow[draw=black]{>}}}},
}

\hskip .6cm
\begin{tikzpicture}[line width=1.5 pt, node distance=2cm and 2cm ]
	
\coordinate[label=below:\scriptsize{$ $}] (v1);
\coordinate[above=of v1,label=above:\scriptsize{$ $}] (v2);
\coordinate[right=of v2,label=above:\scriptsize{$ $}] (v3);
\coordinate[below= of v3,label=below:\scriptsize{$ $}] (v4);
\coordinate[above right= 2cm and 1.1cm of v4,label=right:$s$] (e2);
\coordinate[below right= 2cm and 1.1cm of v3,label=right:$q$] (f2);
\coordinate[above left = 2cm and 1.1cm of v1,label=left :$d$] (f1);
\coordinate[below left = 2cm and 1.1cm of v2,label=left :$q$] (e1);
\coordinate[above left=1.3cm and  -0.5cm of v1,label=left :$u\text{,}\:c\text{,}\:t$] (a1);
\coordinate[above right=0.4cm and  1.4cm of v2,label=left :$W$] (a2);
\coordinate[below right=0.7cm and  0.8cm of v3,label=left :$u\text{,}\:c\text{,}\:t$] (a3);
\coordinate[below right=0.35cm and  -0.8cm of v4,label=left :$ $] (a2);
\coordinate[below right=1.5cm and  -0.3cm of v3,label=left :$\color{red}{g}$];
\coordinate[above right=1cm and  -1cm of v4,label=left :$ $] (o1);   
\coordinate[above right=0cm and  -1cm of v4,label=left :$ $] (o2);

		\draw[fermion] (e1) -- (o2);
		\draw[vector] (v3) -- (v2);
		\draw[fermion] (f1) -- (v2);
		\draw[fermion] (v3) -- (e2);
		\draw[fermion] (o2) -- (f2);
		\draw[fermion] (v2) -- (o1);
		\draw[fermion] (o1) -- (v3);
	    \draw[gluon] (o1) -- (o2);
		
	    \fill[black] (v2) circle (.1cm);
	    \fill[black] (v3) circle (.1cm);
	    \fill[blue] (o1) circle (.1cm);
	    \fill[blue] (o2) circle (.1cm);
	    
\end{tikzpicture}}
}
\vskip .5cm
	\subcaptionbox{\label{fig1:c}}{\tikzset{
    vector/.style={decorate, decoration={snake}, draw},
      gluon1/.style={decorate, draw=blue,
        decoration={snake}, draw},
    fermion/.style={draw=black, postaction={decorate},
        decoration={markings,mark=at position .55 with {\arrow[draw=black]{>}}}},
}

\begin{tikzpicture}[line width=1.5 pt, node distance=2cm and 2cm ]
	
\coordinate[label=below:\scriptsize{$ $}] (v1);
\coordinate[above=of v1,label=above:\scriptsize{$ $}] (v2);
\coordinate[right=of v2,label=above:\scriptsize{$ $}] (v3);
\coordinate[below= of v3,label=below:\scriptsize{$ $}] (v4);
\coordinate[above right= 2cm and 1.1cm of v4,label=right:$s$] (e2);
\coordinate[below right= 2cm and 1.1cm of v3,label=right:$q$] (f2);
\coordinate[above left = 2cm and 1.1cm of v1,label=left :$d$] (f1);
\coordinate[below left = 2cm and 1.1cm of v2,label=left :$q$] (e1);
\coordinate[above left=1.3cm and  -0.5cm of v1,label=left :$u\text{,}\:c\text{,}\:t$] (a1);
\coordinate[above right=0.4cm and  1.4cm of v2,label=left :$W$] (a2);
\coordinate[below right=0.7cm and  0.8cm of v3,label=left :$u\text{,}\:c\text{,}\:t$] (a3);
\coordinate[below right=0.35cm and  -0.8cm of v4,label=left :$ $] (a2);
\coordinate[below right=1.5cm and  0.2cm of v3,label=left :$\color{blue}{\gamma\text{,}\:Z}$];
\coordinate[above right=1cm and  -1cm of v4,label=left :$ $] (o1);   
\coordinate[above right=0cm and  -1cm of v4,label=left :$ $] (o2);

		\draw[fermion] (e1) -- (o2);
		\draw[vector] (v3) -- (v2);
		\draw[fermion] (f1) -- (v2);
		\draw[fermion] (v3) -- (e2);
		\draw[fermion] (o2) -- (f2);
		\draw[fermion] (v2) -- (o1);
		\draw[fermion] (o1) -- (v3);
	    \draw[gluon1] (o1) -- (o2);
		
	    \fill[black] (v2) circle (.1cm);
	    \fill[black] (v3) circle (.1cm);
	    \fill[red] (o1) circle (.1cm);
	    \fill[red] (o2) circle (.1cm);
	    
\end{tikzpicture}
\hskip .6cm	
	\tikzset{
    vector/.style={decorate, decoration={snake}, draw},
      gluon1/.style={decorate, draw=blue,
        decoration={snake}, draw},
    fermion/.style={draw=black, postaction={decorate},
        decoration={markings,mark=at position .55 with {\arrow[draw=black]{>}}}},
}

\begin{tikzpicture}[line width=1.5 pt, node distance=2cm and 2cm ]
	
\coordinate[label=below:\scriptsize{$ $}] (v1);
\coordinate[above=of v1,label=above:\scriptsize{$ $}] (v2);
\coordinate[right=of v2,label=above:\scriptsize{$ $}] (v3);
\coordinate[below= of v3,label=below:\scriptsize{$ $}] (v4);
\coordinate[above right= 2cm and 1.1cm of v4,label=right:$s$] (e2);
\coordinate[below right= 2cm and 1.1cm of v3,label=right:$q$] (f2);
\coordinate[above left = 2cm and 1.1cm of v1,label=left :$d$] (f1);
\coordinate[below left = 2cm and 1.1cm of v2,label=left :$q$] (e1);
\coordinate[above left=1.3cm and  -0.4cm of v1,label=left :$W$] (a1);
\coordinate[above right=0.3cm and  1.7cm of v2,label=left :$u\text{,}\:c\text{,}\:t$] (a2);
\coordinate[below right=0.7cm and  0.3cm of v3,label=left :$W$] (a3);
\coordinate[below right=0.35cm and  -0.8cm of v4,label=left :$ $] (a2);
\coordinate[below right=1.5cm and  0.2cm of v3,label=left :$\color{blue}{\gamma\text{,}\:Z}$];
\coordinate[above right=1cm and  -1cm of v4,label=left :$ $] (o1);   
\coordinate[above right=0cm and  -1cm of v4,label=left :$ $] (o2);

		\draw[fermion] (e1) -- (o2);
		\draw[fermion] (v2) -- (v3);
		\draw[fermion] (f1) -- (v2);
		\draw[fermion] (v3) -- (e2);
		\draw[fermion] (o2) -- (f2);
		\draw[vector] (v2) -- (o1);
		\draw[vector] (o1) -- (v3);
	    \draw[gluon1] (o1) -- (o2);
		
	    \fill[black] (v2) circle (.1cm);
	    \fill[black] (v3) circle (.1cm);
	    \fill[green] (o1) circle (.1cm);
	    \fill[red] (o2) circle (.1cm);
	    
\end{tikzpicture}}
	
\caption{SM Feynman diagrams contributing to $\DS$ transitions:
current--current (a), QCD penguin (b) and electroweak penguin (c) topologies.}	
\end{figure}
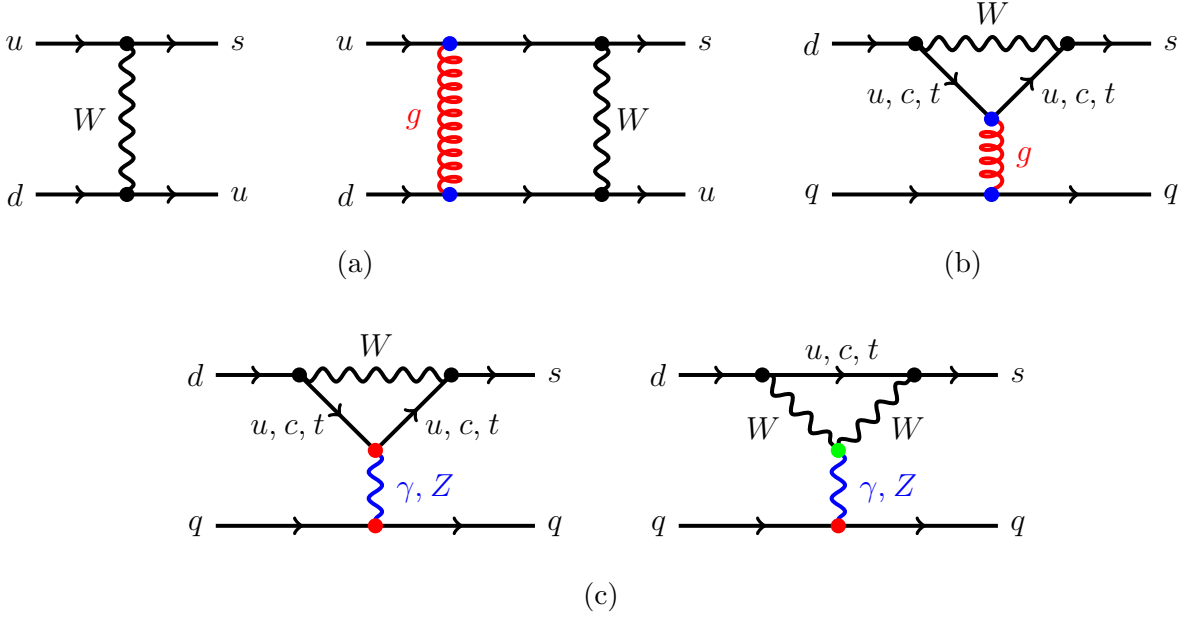

In the SM, the flavour-changing $\Delta S=1$ transition proceeds through the exchange of a $W$ boson between two weak charged currents. Since $M_K\ll M_W$, the heavy $W$ boson can be integrated out and the effective interaction reduces to a local four-fermion operator, $[\bar{s}\gamma^\mu(1-\gamma_5 )u]\, [\bar{u}\gamma_\mu(1-\gamma_5) d]$, multiplied by the Fermi coupling $G_F/\sqrt{2} = g^2/(8 M_W^2)$ and the relevant CKM factors $V_{ud}^{\phantom{*}} V_{us}^*$. The inclusion of gluonic corrections generates additional four-fermion operators, which mix under renormalization
\cite{Altarelli:1974exa,Gaillard:1974nj,Vainshtein:1975sv,Shifman:1975tn,Gilman:1979bc}:
\begin{align}
Q_{1}\, &=\, \left( \bar s_{\alpha} u_{\beta}  \right)_{\rm V-A}
            \left( \bar u_{\beta}  d_{\alpha} \right)_{\rm V-A}\, ,
\qquad 
&Q_{2}\, =\, \left( \bar s u \right)_{\rm V-A}
            \left( \bar u d \right)_{\rm V-A}\, , \qquad\qquad\;
\no\\[5pt]
Q_{3} \, &=\, \left( \bar s d \right)_{\rm V-A}
\sum_{q=u,d,s} \left( \bar q q \right)_{\rm V-A}\, , 
\qquad
&Q_{4} \, =\, \left( \bar s_{\alpha} d_{\beta}  \right)_{\rm V-A}
   \sum_{q=u,d,s} \left( \bar q_{\beta}  q_{\alpha} \right)_{\rm V-A}\, ,  
\nn
Q_{5} \, &=\,  \left( \bar s d \right)_{\rm V-A}
   \sum_{q=u,d,s} \left( \bar q q \right)_{\rm V+A}\, , 
\qquad
&Q_{6} \, =\,  \left( \bar s_{\alpha} d_{\beta}  \right)_{\rm V-A}
   \sum_{q=u,d,s} \left( \bar q_{\beta}  q_{\alpha} \right)_{\rm V+A}\, , 
\end{align}
where $V \pm A$ indicates the Lorentz structure $\gamma_{\mu} (1 \pm \gf)$ and
$\alpha$, $\beta$ denote color indices. When colour labels are not explicit, colour-singlet currents are understood  ($\bar{q} \Gamma  q\equiv \bar{q}_\alpha \Gamma q_\alpha$).
The first two operators originate in the $W$-exchange topology of figure~\ref{fig1:a}, while the QCD penguin diagram in figure~\ref{fig1:b} gives rise to $Q_{3,4,5,6}$.

Four additional four-quark operators appear when one-loop electroweak corrections are incorporated. The electroweak penguin diagrams in figure~\ref{fig1:c} generate the structures  \cite{Bijnens:1983ye,Buras:1987wc,Sharpe:1987cx,Lusignoli:1988fz,Flynn:1989iu}
\begin{align}
Q_{7}\, & =\,  \frac{3}{2}\, \left( \bar s d \right)_{\rm V-A}\,
         \sum_{q=u,d,s} e_{q} \left( \bar q q \right)_{\rm V+A}
\, , \qquad
&Q_{8}\, =\,  \frac{3}{2}\, \left( \bar s_{\alpha} d_{\beta} \right)_{\rm V-A}\,
         \sum_{q=u,d,s} e_{q} \left( \bar q_{\beta}  q_{\alpha}\right)_{\rm V+A}
\, ,  
\nn
Q_{9}\, & =\,  \frac{3}{2}\, \left( \bar s d \right)_{\rm V-A}\,
         \sum_{q=u,d,s} e_{q} \left( \bar q q \right)_{\rm V-A}
\, , \qquad
&Q_{10}\,  =\,  \frac{3}{2}\, \left( \bar s_{\alpha} d_{\beta} \right)_{\rm V-A}\,
         \sum_{q=u,d,s} e_{q} \left( \bar q_{\beta}  q_{\alpha}\right)_{\rm V-A}
\, ,\label{penguin2}
\end{align}
where $e_q$ denotes the corresponding quark charge in units of $e=\sqrt{4\pi\alpha}$.

The presence of very different mass scales ($M_\pi < M_K \ll M_W$) amplifies the gluonic corrections to the $K\to\pi\pi$ amplitudes with large logarithms that can be summed up all the way down from $M_W$ to scales $\mu < m_c$, using the Operator Product Expansion (OPE) and the renormalization group. One finally gets a short-distance effective Lagrangian \cite{Buchalla:1995vs},
\be\label{eq:Leff}
\cL_{\mathrm{eff}}^{\Delta S=1}\, =\, - \frac{G_F}{\sqrt{2}}\,
 V_{ud}^{\phantom{*}}V^*_{us}\,  \sum_{i=1}^{10}
 C_i(\mu) \, Q_i (\mu)\, ,
\ee
defined in the three-flavour theory, with the different local operators modulated by Wilson coefficients $C_i(\mu)$ that are functions of the heavy masses ($M_Z, M_W, m_t, m_b, m_c >\mu$) and CKM parameters:
\be\label{eq:tau}
C_i(\mu)\, =\, z_i(\mu)\, +\,\tau \:y_i(\mu)\, ,
\qquad\qquad 
\tau\,\equiv\, -\frac{V_{td}^{\phantom{*}} V^{*}_{ts}}{V_{ud}^{\phantom{*}} V^{*}_{us}}\, .
\ee
For convenience, the global normalization in Eq.~\eqn{eq:Leff} incorporates the tree-level dependence on CKM factors, so that $C_i = \delta_{i2}$ at lowest order (LO), and the unitarity of the CKM matrix has been used to remove the dependences on $V_{cd}^{\phantom{*}} V^{*}_{cs}$.

The Wilson coefficients $C_i(\mu)$ are known at the next-to-leading logarithmic 
order \cite{Buras:1991jm,Buras:1992tc,Buras:1992zv,Ciuchini:1993vr}. This includes all corrections of $\cO(\alpha_s^n t^n)$ and $\cO(\alpha_s^{n+1} t^n)$, where
$t\equiv\log{(M_1/M_2)}$ refers to the logarithm of any ratio of
heavy mass scales $M_1,M_2\geq\mu$. Moreover, the full $m_t/M_W$ dependence (at LO in $\alpha_s$) is taken into account. Some next-to-next-to-leading-order (NNLO) corrections are already known \cite{Buras:1999st,Gorbahn:2004my} and efforts towards a complete short-distance calculation at the NNLO are currently under way \cite{Cerda-Sevilla:2016yzo}.

\begin{table}[t]
\renewcommand*{\arraystretch}{1.1}
\begin{center}
\begin{tabular}{|c|c|c|c|}
\hline
   & NDR scheme & HV scheme & $\alpha_s(M_\tau)$ error \\
\hline\hline
$z_1$ & $-0.4967$ & $-0.6238$ &  $\pm\:\text{0.04\phantom{00}}$\\
$z_2$ & $\phantom{-}1.2697$  & $\phantom{-}1.3587$  &  $\pm\:\text{0.03\phantom{00}}$\\
\hline\hline
$z_3$ & $\phantom{-}0.0115$  & $\phantom{-}0.0064$ &  $\pm\:\text{0.002\phantom{0}}$\\
$z_4$ & $-0.0321$ & $-0.0143$ & $\pm\:\text{0.006\phantom{0}}$\\
$z_5$ & $\phantom{-}0.0073$  & $\phantom{-}0.0034$  &  $\pm\:\text{0.0007}$\\
$z_6$ & $-0.0318$ & $-0.0124$  &   $\pm\:\text{0.006\phantom{0}}$\\
\hline\hline
$z_7/\alpha$ & $\phantom{-}0.0107$ & $-0.0017$ &  $\pm\:\text{0.003\phantom{0}}$\\
$z_8/\alpha$ & $\phantom{-}0.0121$ & $\phantom{-}0.0082$ &  $\pm\:\text{0.003\phantom{0}}$\\
$z_9/\alpha$ & $\phantom{-}0.0169$ & $\phantom{-}0.0037$ &  $\pm\:\text{0.004\phantom{0}}$ \\
$z_{10}/\alpha$ & $-0.0072$ & $-0.0082$ &  $\pm\:\text{0.001\phantom{0}}$ \\
\hline\hline
$y_3$ & $\phantom{-}0.0318$ & $\phantom{-}0.0367$  & $\pm\:\text{0.003\phantom{0}}$\\
$y_4$ & $-0.0575$ & $-0.0607$ &  $\pm\:\text{0.004\phantom{0}}$\\
$y_5$ & $\phantom{-}0.0000$ & $\phantom{-}0.0161$  & $\pm\:\text{0.003\phantom{0}}$ \\
$y_6$ & $-0.1081$  & $- 0.0948$  &  $\pm\:\text{0.02\phantom{00}}$ \\
\hline\hline
$y_7/\alpha$ & $-0.0364$ & $-0.0349$ & $\pm\:\text{0.0004}$ \\
$y_8/\alpha$ & $\phantom{-}0.1605$ & $\phantom{-}0.1748$  & $\pm\:\text{0.02\phantom{00}}$ \\
$y_9/\alpha$ & $-1.5087$ & $-1.5103$ & $\pm\:\text{0.04\phantom{00}}$ \\
$y_{10}/\alpha$ & $\phantom{-}0.6464$ & $\phantom{-}0.6557$ & $\pm\:\text{0.06\phantom{00}}$ \\
\hline
\end{tabular}
\caption{ $\DS$ Wilson coefficients at $\mu=1$~GeV ($y_1 = y_2 = 0$). }
\label{tab:WilsonCoeff}
\end{center}
\end{table}

In table~\ref{tab:WilsonCoeff} we provide the numerical values of the Wilson coefficients, computed at the NLO with a renormalization scale $\mu=1$~GeV. The results are displayed in the $\overline{\mathrm{MS}}$ renormalization scheme and for two different definitions of $\gamma_5$ within dimensional regularization: the Naive Dimensional Regularization (NDR) and 't Hooft-Veltman (HV) schemes \cite{tHooft:1972tcz,Breitenlohner:1977hr,Breitenlohner:1975hg,Breitenlohner:1976te}. The inputs adopted for the relevant SM parameters are detailed in the appendix~\ref{sec:inputs}, in table~\ref{tab:inputs}. The dependence of the Wilson coefficients on the renormalization scheme and scale should cancel with a corresponding dependence on the hadronic matrix elements of the four-quark operators. However, given their non-perturbative character, a rigorous evaluation of these matrix elements, keeping full control of the QCD renormalization conventions, is a very challenging task. As shown in table~\ref{tab:WilsonCoeff}, the Wilson coefficients have a sizeable sensitivity to the chosen scheme for $\gamma_5$, which limits the currently achievable precision. The table illustrates also their variation with the input value of the strong coupling, which has been taken in the range $\alpha_s^{(n_f=3)}(m_\tau) = 0.325\pm 0.015$ \cite{Olive:2016xmw,Pich:2016yfh}.

To generate CP-violating effects, the SM requires at least three fermion families so that the CKM matrix incorporates a measurable complex phase. For the $K\to 2\pi$ transitions, this implies that direct violations of CP can only originate from penguin diagrams where the three generations play an active role. Thus, the CP-violating parts of the decay amplitudes, $\mathrm{Im} A_I$, are proportional to the $y_i(\mu)$ components of the Wilson coefficients, which are only non-zero for $i>2$. 
In the Wolfenstein parametrization \cite{Wolfenstein:1983yz}, $\mathrm{Im}\,\tau\approx -\lambda^4 A^2\eta \sim -6\cdot 10^{-4}$, exhibiting the strong suppression of these effects in the SM.

\section{Hadronic matrix elements}
\label{sec:hadronic}

Symmetry considerations allow us to better understand the dynamical role of the different four-quark operators. The difference $Q_-\equiv Q_2 - Q_1$ and the QCD penguin operators $Q_{3,4,5,6}$ transform as $(8_L,1_R)$ under chiral $SU(3)_L\otimes SU(3)_R$ transformations in the flavour space, and induce pure $\Delta I=\frac{1}{2}$ transitions. Thus, they do not contribute to the $A_2$ amplitude if isospin is conserved. $\Delta I = \frac{3}{2}$ transitions can only be generated through the complementary combination $Q^{(27)}\equiv 2 Q_2 + 3 Q_1 - Q_3$, which transforms as a $(27_L,1_R)$ operator and can also give rise to processes with $\Delta I = \frac{1}{2}$. Owing to their explicit dependence on the quark charges $e_q$, the electroweak penguin operators do not have definite chiral and isospin quantum numbers. The operators $Q_7$ and $Q_8$ contain $(8_L,1_R)$ and $(8_L,8_R)$ components, while $Q_9$ and $Q_{10}$ are combinations of $(8_L,1_R)$ and $(27_L,1_R)$ pieces.

The chiralities of the different $Q_i$ operators play also a very important dynamical role. Making a Fierz rearrangement, one can rewrite all operators in terms of colour-singlet quark currents. While the $(V-A)\otimes (V-A)$ operators remain then with a similar Lorentz structure,
\be\label{eq:Fierz1}
Q_{4} \, =\, \sum_{q=u,d,s}  \left( \bar s q  \right)_{\rm V-A}
\left( \bar q  d \right)_{\rm V-A}\, ,
\qquad\qquad
Q_{10}\,  =\,  \frac{3}{2}\, \sum_{q=u,d,s} e_{q} \left( \bar s q \right)_{\rm V-A}
         \left( \bar q d\right)_{\rm V-A} \, ,
\ee
the two $(V-A)\otimes (V+A)$ operators transform into a product of scalar/pseudoscalar currents,
\be\label{eq:Fierz2}
Q_{6} \, =\,  -8\, \sum_{q=u,d,s} \left( \bar s_L q_R  \right)
  \left( \bar q_R  d_L \right)\, , 
\qquad\qquad
Q_{8}\, =\,  -12\, \sum_{q=u,d,s} e_{q} \left( \bar s_L q_R \right)
          \left( \bar q_R  d_L\right) \, .
\ee
For light quarks, the hadronic matrix elements of this type of operators turn out to be much larger than the $(V-A)\otimes (V-A)$
ones. 

This chiral enhancement can be easily estimated in the limit of a large number of QCD colours \cite{tHooft:1973alw,Witten:1979kh}, because the product of two colour-singlet quark currents factorizes at the hadron level into two current matrix elements:
\be\label{eq:factorization}
\langle J\cdot J\rangle\, =\,  \langle J\rangle\,\langle J\rangle\,
\left\{ 1 + \cO(1/N_C)\right\}\, .
\ee
Thus, when $N_C\to\infty$,
\beqn\label{eq:V-A_hme}
\langle\pi^+\pi^-|(\bar s_L\gamma^\mu u_L) (\bar u_L\gamma_\mu d_L)|K^0\rangle
& = & 
\langle\pi^+|\bar u_L\gamma_\mu d_L|0\rangle\,
\langle \pi^-|\bar s_L\gamma^\mu u_L|K^0\rangle
\no\\[5pt]
& = &
\frac{i\sqrt{2}}{4}\, F_\pi\, (M_K^2-M_\pi^2)\,\left\{ 1 + \cO\!\left(\frac{M_\pi^2}{F_\pi^2}\right)\right\} ,
\eeqn
while
\beqn\label{eq:S-P_hme}
\langle\pi^+\pi^-|(\bar s_L u_R) (\bar u_R d_L)|K^0\rangle
& = & 
\langle\pi^+|\bar u_R d_L|0\rangle\,
\langle \pi^-|\bar s_L u_R|K^0\rangle
\no\\[5pt]
& = &
\frac{i\sqrt{2}}{4}\, F_\pi\, \left[\frac{M_K^2}{m_d(\mu) + m_s(\mu)}\right]^2\,\left\{ 1 + \cO\!\left(\frac{M_K^2}{F_\pi^2}\right)\right\} ,
\eeqn
where $F_\pi = 92.1$~MeV is the pion decay constant.
Notice the $\mu$ dependence of this last matrix element, which arises because the scalar and pseudoscalar currents get renormalized, but keeping the products $m_q\, \bar q (1,\gamma_5) q$ invariant under renormalization.  On the other side, the vector and axial currents are renormalization invariant, since they are protected by chiral symmetry. This different short-distance behaviour has important consequences in the analysis of $\varepsilon'/\varepsilon$. At $\mu=1$~GeV, the relative ratio between the matrix elements in Eqs.~\eqn{eq:S-P_hme} and \eqn{eq:V-A_hme} is a large factor $M_K^2/[m_s(\mu)+ m_d(\mu)]^2\sim 14$.

Owing to their chiral enhancement, the operators $Q_6$ and $Q_8$ dominate the CP-odd amplitudes $\mathrm{Im} A_{0}^{(0)}$ and $\mathrm{Im} A_{2}^{\rm emp}$, respectively, in Eq.~\eqn{eq:epsp_simp}. As shown in table~\ref{tab:WilsonCoeff}, $Q_6$ has in addition the largest Wilson coefficient $y_i(\mu)$. Ignoring all other contributions to the CP-violating decay amplitudes, one can then make a rough estimate of $\varepsilon'/\varepsilon$ with their matrix elements \cite{Buras:1987wc,Buras:1985yx}:
\beqn \label{eq:A0Q6}
\left.\mathrm{Im} A_{0}\right|_{Q_6} & =& 
 \frac{G_F}{\sqrt{2}}\, A^2\lambda^5\eta
 \; y_6(\mu)\; 4\sqrt{2}\,\left(F_K - F_\pi\right) \;  \left[\frac{M_K^2}{m_d(\mu) + m_s(\mu)}\right]^2\; B_6^{(1/2)}\, ,
\\[5pt]\label{eq:A2Q8}
\left.\mathrm{Im} A_{2}\right|_{Q_8} & =&
 -  \frac{G_F}{\sqrt{2}}\, A^2\lambda^5\eta
\; y_8(\mu)\; 2\,
F_\pi \;  \left[\frac{M_K^2}{m_d(\mu) + m_s(\mu)}\right]^2\; B_8^{(3/2)}\, ,
\eeqn
where $F_K = (1.193\pm 0.003)\, F_\pi$ \cite{Aoki:2016frl} is the kaon decay constant and the factors $B_6^{(1/2)}$ and $B_8^{(3/2)}$ parametrize the deviations of the true hadronic matrix elements from their large-$N_C$ approximations; {\it i.e.},  
$B_6^{(1/2)} = B_8^{(3/2)} = 1$ at $N_C\to\infty$.\footnote{
Actually, the expressions~\eqn{eq:A0Q6} and \eqn{eq:A2Q8} receive small chiral corrections even at $N_C\to\infty$. We will take them later into account, using an appropriate effective field theory framework.}
The renormalization-scale dependence of $y_6(\mu)$ and $y_8(\mu)$ is cancelled to a large extent by the running quark masses, leaving a very soft residual dependence on $\mu$ for the unknown parameters $B_6^{(1/2)}$ and $B_8^{(3/2)}$. This very fortunate fact originates in the large-$N_C$ structure of the anomalous dimension matrix $\gamma_{ij}$ of the four-quark operators $Q_i$. At $N_C\to\infty$, all entries of this matrix are zero, except $\gamma_{66}$ and $\gamma_{88}$ \cite{Bardeen:1986uz}. This just reflects the factorization property in Eq.~\eqn{eq:factorization} and the fact that the product $m_q \bar q q$ is renormalization invariant.

Inserting these two matrix elements in Eq.~\eqn{eq:epsp_simp} and taking the experimental values for all the other inputs, one finds
\be\label{eq:epsilonp_naive}
\mathrm{Re}(\varepsilon'/\varepsilon) \; \approx \; 
2.2\cdot 10^{-3}\;
\left\{ B_6^{(1/2)} \left( 1 - \Omega_{\rm eff} \right) - 0.48\; B_8^{(3/2)}
\right\} \, .
\ee
With $B_6^{(1/2)} = B_8^{(3/2)} = 1$ and $\Omega_{\rm eff}= 0.06$, this gives
$\mathrm{Re}(\varepsilon'/\varepsilon) \approx 1.0\cdot 10^{-3}$ as the expected order of magnitude for the SM prediction. However, there is a subtle cancellation among the three terms in Eq.~\eqn{eq:epsilonp_naive}, making the final number very sensitive to the exact values of these three inputs. For instance, with the inputs adopted in Ref.~\cite{Buras:2015yba}, $B_6^{(1/2)} = 0.57$, $B_8^{(3/2)} = 0.76$ and $\Omega_{\rm eff}= 0.15$, one finds instead $\mathrm{Re}(\varepsilon'/\varepsilon) \approx 2.6\cdot 10^{-4}$, which is nearly one order of magnitude smaller and in clear conflict with the experimental value in Eq.~\eqn{eq:exp}. Which such a choice of inputs, the cancellation is so strong that contributions from other four-quark operators become then sizeable. 
On the other side, a moderate increase of $B_6^{(1/2)}$ over its large-$N_C$ prediction, {\it i.e.}, $B_6^{(1/2)}>1$, gets amplified in \eqn{eq:epsilonp_naive}, which results in much larger values of $\varepsilon'/\varepsilon$.

The crucial observation made in Refs.~\cite{Pallante:1999qf,Pallante:2000hk,Pallante:2001he} is that the chiral dynamics of the final-state pions generates large logarithmic corrections to the two relevant decay amplitudes,
$\left. A_{0}\right|_{Q_6}$ and $\left. A_{2}\right|_{Q_8}$, which are of NLO in $1/N_C$. These logarithmic corrections can be rigorously computed with standard $\chi$PT methods and are tightly related to the large phase-shift difference in Eq.~\eqn{eq:isoamps}. They turn out to be positive for $\left. A_{0}\right|_{Q_6}$ and negative for $\left. A_{2}\right|_{Q_8}$, destroying the numerical cancellation in Eq.~\eqn{eq:epsilonp_naive} and bringing, therefore, a sizeable enhancement of the SM prediction for $\varepsilon'/\varepsilon$, in good agreement with its experimental value.

\section{Effective field theory description}
\label{sec:eft}

Effective field theory provides the appropriate framework to address the multi-scale dynamics involved in kaon decays. While the short-distance electroweak transitions occur at the $W$ mass scale, kaons and pions are the lightest particles in the QCD spectrum and their dynamics is governed by the non-perturbative regime of the strong interaction. A proper description of non-leptonic kaon decays requires then a good theoretical control of short-distance and long-distance contributions, through a combined application of perturbative and non-perturbative techniques.

We have already displayed in Eq.~\eqn{eq:Leff} the relevant short-distance effective Lagrangian at scales $\mu$ just below the charm mass, where perturbation theory remains still valid. This Lagrangian corresponds to an effective field theory description with all heavy ($M> \mu$) fields integrated out. Only the three light quarks (and $e$, $\mu$, $\nu_i$, $\gamma$, $G_a$) are kept as explicit dynamical fields. All informations on the heavy fields that are no longer in the effective theory are captured by the Wilson coefficients $C_i(\mu)$, which can be conveniently calculated with the OPE and renormalization-group methods.

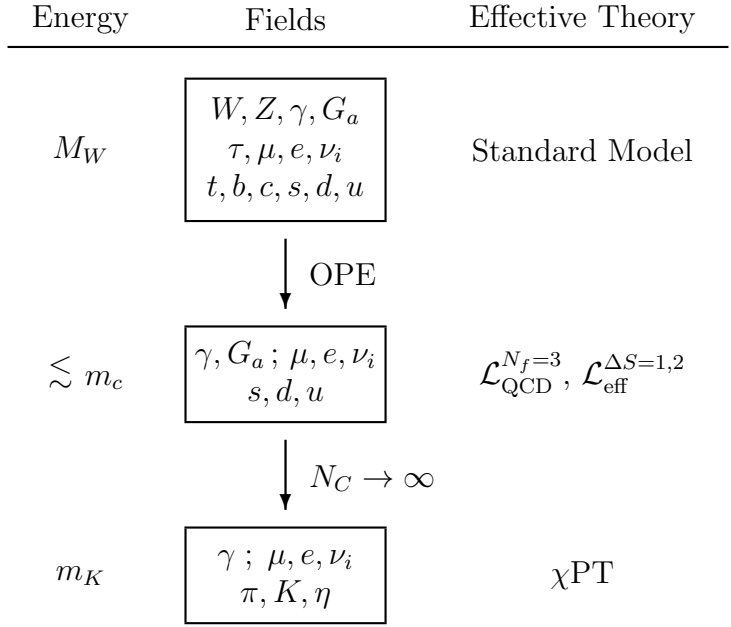
\begin{figure}[t]\centering
\setlength{\unitlength}{0.62mm} 
\begin{picture}(163,133)
\put(0,0){\makebox(163,133){}}
\thicklines
\put(8,124){\makebox(25,10){Energy}}
\put(43,124){\makebox(42,10){Fields}}
\put(101,124){\makebox(52,10){Effective Theory}}
\put(5,123){\line(1,0){153}} {
\put(8,86){\makebox(25,30){$M_W$}}
\put(43,86){\framebox(42,30){
   $\ba W, Z, \gamma, G_a \\  \tau, \mu, e, \nu_i \\ t, b, c, s, d, u \ea $}}
\put(101,86){\makebox(52,30){Standard Model}}

\put(8,43){\makebox(25,20){$\lsim m_c$}}
\put(43,43){\framebox(42,20){
 $\ba  \gamma, G_a  \, ;\, \mu ,  e, \nu_i \\ s, d, u \ea $}}
\put(101,43){\makebox(52,20){$\cL_{\mathrm{QCD}}^{N_f=3}$,
             $\cL_{\mathrm{eff}}^{\Delta S=1,2}$}}

\put(8,0){\makebox(25,20){$m_K$}}
\put(43,0){\framebox(42,20){
 $\ba\gamma \; ;\; \mu , e, \nu_i  \\  \pi, K,\eta  \ea $}}
\put(101,0){\makebox(52,20){$\chi$PT}}
\linethickness{0.3mm}
\put(64,39){\vector(0,-1){15}}
\put(64,82){\vector(0,-1){15}}
\put(69,72){OPE}
\put(69,29){$N_C\to\infty $}}    
\end{picture}
\vskip -.1cm\mbox{}
\caption{Evolution from $M_W$ to the kaon mass scale.
  \label{fig:eff_th}}
\end{figure}

Chiral symmetry considerations allow us to formulate another effective field theory that is valid at the kaon mass scale where perturbation theory can no longer be applied. Since kaons and pions are the Goldstone modes of the QCD chiral symmetry breaking, their dynamics is highly constrained by chiral symmetry, which provides a very powerful tool to describe kaon decays in a rigorous way \cite{Cirigliano:2011ny}. Figure~\ref{fig:eff_th} shows schematically the chain of effective theories entering the analysis of the kaon decay dynamics.

\subsection{Chiral perturbation theory}
\label{subsec:chpt}

At very low energies, below the $\rho$ mass scale, the hadronic spectrum only contains the pseudoscalar meson octet; {\it i.e.}, the Goldstone modes $\phi_a$ associated with the dynamical breaking of chiral symmetry by the QCD  vacuum, which are conveniently parametrized through the $3\times 3$ unitary matrix \cite{Pich:1995bw}
\be
U(\phi)\,\equiv\, \exp{\{ i\sqrt{2}\,\Phi(x)/F\}}\, ,
\ee
where
\be
\Phi(x)\,\equiv\, \sum_{a=1}^8\frac{\lambda^a}{\sqrt{2}}\, \phi_a(x)
\, =\,
\begin{pmatrix}
{1\over\sqrt 2}\pi^0 \, +
\, {1\over\sqrt 6}\eta_8
 & \pi^+ & K^+ \cr
\pi^- & - {1\over\sqrt 2}\pi^0 \, + \, {1\over\sqrt 6}\eta_8
 & K^0 \cr K^- & \bar K^0 & - {2 \over\sqrt 6}\eta_8
\end{pmatrix}
\, .
\ee
Under a chiral transformation $(g_L,g_R)\in SU(3)_L\otimes SU(3)_R$ in the  flavour space $q\equiv (u,d,s)^T$, $q_L\to g_L\, q_L ,\; q_R\to g_R\, q_R$, $U(\phi)$ transforms as $g_R\, U(\phi)\, g_L^\dagger$, inducing a non-linear transformation on the Goldstone fields $\phi_a(x)$.

The low-energy effective realization of QCD is obtained by writing the most general Lagrangian involving the matrix $U(\phi)$ that is consistent with chiral symmetry \cite{Weinberg:1978kz}. The Lagrangian can be organised through an expansion in powers of momenta (derivatives) and explicit breakings of chiral symmetry (light quark masses, electromagnetic coupling, etc.):
\be
\cL_{\mathrm{eff}} \, =\, \cL_2 + \cL_4 + \cL_6 + \cdots
\ee
Parity conservation requires the number of derivatives to be even, and a minimum of two derivatives is needed to generate non-trivial interactions because $U\, U^\dagger = 1$. 
The terms with a minimum number of derivatives will dominate at low energies. To lowest order, $\cO(p^2)$, the effective Lagrangian is given by \cite{Gasser:1984gg}:
\be
\cL_2\, =\, \frac{F^2}{4}\; \langle D_\mu U^\dagger D^\mu U + U^\dagger\chi + \chi^\dagger U\rangle\, ,
\ee
where $\langle \cdots\rangle$ denotes the three-dimensional flavour trace,
$D_\mu U = \partial_\mu U - i r_\mu U + i U \ell_\mu$ is the covariant chiral derivative, in the presence of arbitrary right-handed and left-handed (matrix-valued) external sources $r_\mu$ and $\ell_\mu$, and $\chi\equiv 2 B_0 (s+ip)$ with $s$ and $p$ external scalar and pseudoscalar sources, respectively. Taking $s=\cM$, $p=0$, and 
$r_\mu = \ell_\mu = e \cQ A_\mu$ allows one to incorporate the explicit chiral symmetry breakings generated by the non-zero quark masses and electric charges:
\be 
\cM\, =\, \mathrm{diag}(m_u, m_d, m_s)\, ,
\qquad\qquad\qquad
\cQ \, =\, \frac{1}{3}\;\mathrm{diag}(2, -1, -1)\, .
\ee
Moreover, taking derivatives with respect to the external sources one can easily obtain the effective realization of the QCD quark currents in terms of the Goldstone bosons \cite{Pich:1995bw}. One then finds that $F$ is the pion decay constant in the chiral limit ($m_q=0$), while the constant $B_0$ is related to the quark condensate.

While only two low-energy constants (LECs) appear at $\cO(p^2)$, $F$ and $B_0$, ten additional couplings $L_i$ characterize the $\cO(p^4)$ $\chi$PT Lagrangian \cite{Gasser:1984gg}, 
\beqn
\cL_4  & = &
L_1 \,\langle D_\mu U^\dagger D^\mu U\rangle^2 \, + \,
L_2 \,\langle D_\mu U^\dagger D_\nu U\rangle\,
   \langle D^\mu U^\dagger D^\nu U\rangle \, + \,
L_3 \,\langle D_\mu U^\dagger D^\mu U D_\nu U^\dagger
D^\nu U\rangle   
\no\\[4pt]  &+&
L_4 \,\langle D_\mu U^\dagger D^\mu U\rangle\,
   \langle U^\dagger\chi +  \chi^\dagger U \rangle \, + \,
L_5 \,\langle D_\mu U^\dagger D^\mu U \left( U^\dagger\chi +
\chi^\dagger U
\right)\rangle
\no\\[4pt]  &+&
L_6 \,\langle U^\dagger\chi +  \chi^\dagger U \rangle^2\, + \,
L_7 \,\langle U^\dagger\chi -  \chi^\dagger U \rangle^2\,
+ \, L_8 \,\langle\chi^\dagger U \chi^\dagger U
+ U^\dagger\chi U^\dagger\chi\rangle
\no\\[4pt]  &-&
i L_9 \,\langle F_R^{\mu\nu} D_\mu U D_\nu U^\dagger +
     F_L^{\mu\nu} D_\mu U^\dagger D_\nu U\rangle\,
+ \, L_{10} \,\langle U^\dagger F_R^{\mu\nu} U F_{L\mu\nu} \rangle\, ,
\label{eq:l4}
\eeqn
with $F_R^{\mu\nu} =\partial^\mu r^\nu - \partial^\nu r^\mu -i\, [r^\mu,r^\nu]$
and $F_L^{\mu\nu} =\partial^\mu \ell^\nu - \partial^\nu \ell^\mu -i\, [\ell^\mu,\ell^\nu]$,
and 90 LECs more would be needed to compute corrections of $\cO(p^6)$:\footnote{
There are, in addition, 2 contact terms without Goldstone bosons at $\cO(p^4)$, and 4 more at $\cO(p^6)$, which are only needed for renormalization. The $\cO(p^6)$ LECs are usually denoted $C_i\equiv F^{-2} X_i$. We have changed the notation to avoid possible confusions with the short-distance Wilson coefficients. The $\chi$PT Lagrangian contains also the $\cO(p^4)$ Wess-Zumino-Witten term that has no free parameters and accounts for the QCD chiral anomaly \cite{Wess:1971yu,Witten:1983tw}.
}
%
\be 
\cL_6 \, =\, F^{-2}\;\sum_{i=1}^{90} X_i\; O_i^{p^6}\, .
\ee
The explicit form of the 
$\cO(p^6)$ operators can be found in Ref.~\cite{Bijnens:1999sh}. The current knowledge on all these LECs has been summarized in Ref.~\cite{Bijnens:2014lea}.

Quantum loops with Goldstone boson propagators in the internal lines generate non-polynomial contributions, with logarithms and threshold factors as required by unitarity. Each loop increases the chiral dimension by two powers of momenta \cite{Weinberg:1978kz}. Thus, to achieve an $\cO(p^4)$ accuracy  one needs to compute tree-level contributions with a single insertion of $\cL_4$ plus one-loop graphs with only $\cL_2$ vertices. These chiral one-loop corrections are then fully predicted in terms of $F_\pi$ and the meson masses. 
Two-loop corrections with only $\cL_2$ vertices contribute at $\cO(p^6)$, together with one-loop graphs with a single insertion of $\cL_4$ and tree-level diagrams with one insertion of $\cL_6$.

The ultraviolet divergences generated by quantum loops get reabsorbed by the corresponding LECs contributing to the same order in momenta. This induces an explicit dependence of the renormalized LECs on the chiral renormalization scale $\nu_\chi$: 
\be 
L_i\, =\, L_i^r(\nu_\chi)\:+\Gamma_i\:\Lambda(\nu_\chi)\, ,
\ee
with
\be
\Lambda(\nu_\chi)\, =\, \frac{{\nu_\chi}^{d-4}}{16\pi^2}\:\left\lbrace\frac{1}{d-4}\:-\:\frac{1}{2}\left[\log{(4 \pi)} + \Gamma'(1) + 1\right]\right\rbrace 
\ee
the divergent subtraction constant in the usual $\chi$PT renormalization scheme. Similar expressions apply for the other $\cO(p^4)$ and $\cO(e^2p^2)$ LECs ($K_i$, $N_i$, $D_i$, $Z_i$) that will be discussed next, while the $\cO(p^6)$ LECs $X_i$ require a two-loop subtraction \cite{Bijnens:1999hw}. The divergent parts of all these $\chi$PT couplings are fully known \cite{Gasser:1984gg,Kambor:1989tz,Ecker:1992de,Urech:1994hd,Ecker:2000zr,Bijnens:1999hw}.

In order to include loop corrections with virtual photon propagators, one needs to consider also the electromagnetic Lagrangian \cite{Ecker:1988te,Urech:1994hd,Knecht:1999ag}
\be \label{eq:emL}
\cL_{\mathrm{em}}\, =\, e^2 Z F^4\; \langle \cQ U^\dagger \cQ U\rangle +
e^2 F^2 \sum_{i=1}^{14} K_i\; O_i^{e^2p^2} + \cO(e^2p^4)\, .
\ee
The presence of the quark charge matrix allows for a chiral structure without derivatives. The corresponding LEC of this $\cO(e^2p^0)$ term is determined by the electromagnetic pion mass difference \cite{Ecker:1988te}
\be 
Z \, =\, \frac{1}{8\pi\alpha F^2}\,\left( M_{\pi^\pm}^2 - M_{\pi^0}^2 \right)\,\approx\, 0.8\, .
\ee
%

\subsection[Chiral realization of the $\Delta S=1$ effective Lagrangian]{Chiral realization of the {\boldmath $\Delta S=1$} effective Lagrangian}
\label{subsec:chpt-weak}

Strangeness-changing weak interactions with $\Delta S=1$ are incorporated in the low-energy theory as a perturbation to the strong Lagrangian. At LO, the most general effective Lagrangian with the same transformation properties as the short-distance Lagrangian \eqn{eq:Leff} contains three terms \cite{Cronin:1967jq,Bijnens:1983ye,Grinstein:1985ut}:
\beqn\label{eq:L2_weak}
\cL_2^{\Delta S=1}& =& G_8 F^4\;\langle\lambda D^\mu U^\dagger D_\mu U\rangle
+ G_{27} F^4 \,\left( L_{\mu 23} L^\mu_{11} + \frac{2}{3}\, L_{\mu 21} L^\mu_{13}\right)
\no\\
&+& e^2 G_8\, g_{\mathrm{ewk}} F^6\;\langle\lambda U^\dagger \cQ U\rangle\, ,
\eeqn
where $\lambda = (\lambda_6 - i\lambda_7)/2$ projects into the $\bar s\to\bar d$ transition and $L_\mu = i\, U^\dagger D_\mu U$ represents the octet of $V-A$ currents to lowest order in derivatives. Under chiral transformations, these three terms transform  as $(8_L,1_R)$, $(27_L,1_R)$ and $(8_L,8_R)$, respectively. To simplify notation, we have reabsorbed the Fermi coupling and the CKM factors into effective LECs:
\be 
G_{8,27}\,\equiv\, - \frac{G_F}{\sqrt{2}}\; V_{ud}^{\phantom{*}} V_{us}^*\; g_{8,27}\, ,
\ee
where $g_{8}$, $g_{27}$ and $g_{\mathrm{ewk}}$ are dimensionless couplings.

The $G_8$ and $G_{27}$ chiral operators contain two derivatives and, therefore, lead to amplitudes that vanish at zero momenta. However, the electromagnetic penguin operator has a chiral realization at $\cO(e^2 p^0)$, given by the term proportional to $G_8\, g_{\mathrm{ewk}}$. The absence of derivatives implies a chiral enhancement that we have already seen before in Eq.~\eqn{eq:S-P_hme}. 

The corresponding NLO effective Lagrangians have been worked out in Refs.~\cite{Kambor:1989tz,Ecker:1992de,Ecker:2000zr} and include $\chi$PT operators of $\cO(p^4)$ and $\cO(e^2 p^2)$:
\beqn 
\cL_4^{\Delta S=1}& =& G_8 F^2\;\sum_{i=1}^{22} N_i\; O_i^8 + G_{27} F^2\;\sum_{i=1}^{28} D_i\; O_i^{27} + e^2 G_8 F^4\;\sum_{i=1}^{14} Z_i\; O_i^{\mathrm{EW}}\, .
\eeqn

\section{Matching}
\label{sec:matching}

In principle, the chiral LECs could be computed through a matching calculation between the three-flavour quark effective theory and $\chi$PT. This is however a formidable non-perturbative task. Therefore, one needs to resort to phenomenological determinations, using the available hadronic data \cite{Bijnens:2014lea}. Nevertheless, a very good understanding of the strong LECs has been achieved in the large-$N_C$ limit, where the meson scattering amplitudes reduce to tree-level diagrams with physical hadrons exchanged \cite{Pich:2002xy}. The contributions from tree-level meson resonance exchanges have been shown to saturate the phenomenologically known LECs at $\nu_\chi\sim M_\rho$ \cite{Ecker:1988te,Ecker:1989yg,Pich:2002xy,Cirigliano:2006hb}. Nowadays, lattice simulations are also able to provide quantitative values of some of the $\cO(p^4)$ $L_i$ couplings \cite{Aoki:2016frl}.

In the limit $N_C\to\infty$, thanks to the factorization property in Eq.~\eqn{eq:factorization}, the electroweak $\chi$PT couplings can be related to strong 
LECs because the QCD currents have well-known chiral realizations. The quark currents are obtained as functional derivatives with respect to the appropriate external sources of the QCD generating functional $Z[v_\mu,a_\mu,s,p]$, defined via the path integral formula
%
\beqn 
\exp{\{i Z\}} &  = &  \int  {\cal D}q \, {\cal D} \bar q
\, {\cal D}G_\mu \;
\exp{\left\{i \int d^4x\, \left[ {\cal L}_{QCD}
+ \bar q \gamma^\mu (v_\mu + \gamma_5 a_\mu ) q -
\bar q (s - i \gamma_5 p) q \right] \right\}}
\no\\[5pt]
&  = &
\int  {\cal D}U \; \exp{\left\{i \int d^4x\, {\cal L}_{\mathrm{eff}}\right\}} \, .
\label{eq:generatingfunctional}
\eeqn
The corresponding derivatives of the $\chi$PT generating functional determine the chiral expressions of the QCD currents. 

At $N_C\to\infty$, the generating functional reduces to the classical action because quantum loops are suppressed by powers of $1/N_C$. The left and right vector currents are then easily computed, by taking derivatives of $\cL_{\mathrm{eff}}$ with respect to
$\ell_\mu^{ji}\equiv (v_\mu - a_\mu)^{ji}$ and $r_\mu^{ji}\equiv (v_\mu + a_\mu)^{ji}$, respectively \cite{Pich:1995bw}. One easily finds:
\beqn\label{eq:LLcurrent}
\bar q^j_L\gamma^\mu q^i_L &\dot =& \frac{i}{2}\, \Big\{
D^\mu U^\dagger U \left[  F^2 + 8 L_1\,\langle D_\alpha U^\dagger D^\alpha U\rangle 
\right]
  + 4 L_2\,D_\alpha U^\dagger U \, \langle D^\alpha U^\dagger D^\mu U + D^\mu U^\dagger D^\alpha U \rangle
\nn &&\hskip .3cm\mbox{}  
  + 4 L_3\, \left\{ D^\mu U^\dagger U\, ,\,  D_\alpha U^\dagger D^\alpha U\right\}
+ 2 L_5\, \left\{ D^\mu U^\dagger U\,  ,\, \left( U^\dagger\chi + \chi^\dagger U\right)\right\}
\nn &&\hskip .3cm\mbox{}
+ 2 L_9\, \partial_\alpha\left(  D^\alpha U^\dagger D^\mu U - D^\mu U^\dagger D^\alpha U\right)
+ 2 i L_9\,\left(\ell_\alpha  D^\mu U^\dagger D^\alpha U - D^\alpha U^\dagger D^\mu U \ell_\alpha\right)\Big\}^{ij}
\nn &&\hskip .3cm\mbox{}
+\; \cO(p^5 N_C,p^3 N_C^0)\, ,
\\[8pt]
\bar q^j_R\gamma^\mu q^i_R &\dot =& \frac{i}{2}\, \Big\{
D^\mu U U^\dagger \left[F^2 + 8 L_1\,\langle D_\alpha U^\dagger D^\alpha U\rangle 
\right] 
  + 4 L_2\,D_\alpha U U^\dagger \, \langle D^\alpha U D^\mu U^\dagger + D^\mu U D^\alpha U^\dagger \rangle
\nn &&\hskip .3cm\mbox{}  
  + 4 L_3\, \left\{ D^\mu U U^\dagger\, ,\,  D_\alpha U D^\alpha U^\dagger\right\}
+ 2 L_5\, \left\{ D^\mu U U^\dagger\,  ,\, \left( \chi U^\dagger + U \chi^\dagger \right)\right\}
\nn &&\hskip .3cm\mbox{} 
+ 2 L_9\, \partial_\alpha\left(  D^\alpha U D^\mu U^\dagger - D^\mu U D^\alpha U^\dagger\right)
+ 2 i L_9\,\left( r_\alpha  D^\mu U D^\alpha U^\dagger - D^\alpha U D^\mu U^\dagger r_\alpha\right)\Big\}^{ij}
\nn &&\hskip .3cm\mbox{} 
+\; \cO(p^5 N_C,p^3 N_C^0)\, .
\label{eq:RRcurrent}\eeqn
We have made explicit the $\cO(p)$ contributions from the LO Lagrangian $\cL_2$, which are proportional to $F^2\sim \cO(N_C)$, and those $\cO(p^3)$ contributions from $\cL_4$ that are of $\cO(N_C)$. 
Taking derivatives with respect to $-(s-ip)^{ji}$ and $-(s+ip)^{ji}$, one obtains the scalar bilinears
\beqn\label{eq:LRcurrent}
\bar q^j_L q^i_R &\dot =& - \frac{B_0}{2}\,\Big\{ U \left[
F^2 
+ 4 L_5\, D_\alpha U^\dagger D^\alpha U
- 8 L_7\, \langle U^\dagger\chi - \chi^\dagger U\rangle
+ 8 L_8\, \chi^\dagger U\right]\Big\}^{ij}
\nn &&\hskip .9cm\mbox{} +\; \cO(p^4 N_C,p^2 N_C^0)\, ,
\\[5pt]
\bar q^j_R q^i_L &\dot =& - \frac{B_0}{2}\,\Big\{ U^\dagger \left[
F^2 
+ 4 L_5\, D_\alpha U D^\alpha U^\dagger
+ 8 L_7\, \langle U^\dagger\chi - \chi^\dagger U\rangle
+ 8 L_8\, \chi U^\dagger\right]\Big\}^{ij}
\nn &&\hskip .9cm\mbox{} +\; \cO(p^4 N_C,p^2 N_C^0)\, .
\label{eq:RLcurrent}\eeqn
The vacuum expectation value of the last two equations relates the coupling $B_0$ with the quark vacuum condensate: $\langle 0 |\bar q^j q^i |0\rangle = -F^2B_0\,\delta^{ij}$, at LO.

Inserting these expressions into the four-quark operators in Eq.~\eqn{eq:Leff}, using the factorization property \eqn{eq:factorization}, one finds the $\chi$PT realization of 
$\cL_{\mathrm{eff}}^{\Delta S=1}$ in the large-$N_C$ limit. At $\cO(p^2)$, the three chiral structures in Eq.~\eqn{eq:L2_weak} are generated, with the following large-$N_C$ values for their electroweak LECs \cite{Pallante:2001he,Cirigliano:2003gt}:
%
\begin{align}
g_8^\infty\:&=\:-\:\frac{2}{5}\, C_1(\mu)\:+\:\frac{3}{5}\, C_2(\mu)\:+\:C_4(\mu)\:-\:16\:L_5\,B(\mu)\, C_6(\mu)\, ,\quad
\nonumber\\
g_{27}^\infty\:&=\:\frac{3}{5} \left[C_1(\mu)\:+\:C_2(\mu)\right],
\label{largeNccouplings}\\
(e^2\:g_8\:g_{\text{ewk}})^\infty\:&=\:-\:3\, B(\mu)\, C_8(\mu)
\:-\:\frac{16}{3}\, B(\mu)\, C_6(\mu)\, e^2\, (K_9\:-2\, K_{10})
\, .
\nonumber
\end{align}
In the last line, we have also taken into account the contribution to $g_8\, g_{\mathrm{ewk}}$ from electromagnetic corrections to $Q_6$ \cite{Cirigliano:2003gt}.

The LO terms in Eqs.~\eqn{eq:LRcurrent} and \eqn{eq:RLcurrent} give rise to the $\cO(p^0)$ electroweak chiral structure $Q_8\dot= -3 B_0^2 F^4\,\langle\lambda U^\dagger \cQ U \rangle$. An analogous $\cO(p^0)$ contribution is absent for $Q_6$ because $\langle\lambda U^\dagger U \rangle = \langle\lambda \rangle =0$. Therefore, the $\chi$PT realization of the penguin operator $Q_6$ starts at $\cO(p^2)$, giving rise to the same octet structure as the $(V-A)\otimes (V-A)$ operators. The only difference is that
$Q_{1,2,4}$ generate this structure with the LO terms in Eqs.~\eqn{eq:LLcurrent} and \eqn{eq:RRcurrent}, while in the $Q_6$ case it originates from the interference of the $\cO(p^0)$ and $\cO(p^2)$ terms in Eqs.~\eqn{eq:LRcurrent} and \eqn{eq:RLcurrent}. This is the reason why the $C_6$ contribution to $g_8^\infty$ appears multiplied by the strong LEC $L_5$, reducing the expected chiral enhancement in a very significant way.

There are no $\cO(p^2)$ contributions from the operators $Q_3$ and $Q_5$, at large-$N_C$, because they are proportional to the flavour trace of the left and right currents, respectively, which vanish identically at LO. The operators $Q_{7,9,10}$ start to contribute at $\cO(e^2p^2)$.

The dependence on the short-distance renormalization scale of the 
Wilson coefficients $C_i(\mu)$ is governed by the anomalous dimension matrix $\gamma_{ij}$ of the four-quark operators $Q_i$, which vanishes at $N_C\to\infty$, except for the non-zero entries $\gamma_{66}$ and $\gamma_{88}$. Thus, the $\mu$ dependence 
of $C_i(\mu)$ with $i\not= 6,8$  disappears when $N_C\to\infty$, while that of $C_{6,8}(\mu)$ is exactly cancelled by the factor
\be\label{eq:Bmu}
B(\mu)\equiv \left(\frac{B_0^2}{F^2}\right)^\infty
\, =\, \left\{ \frac{M_K^2}{[m_s(\mu) + m_d(\mu)]\, F_\pi}\right\}^2\; 
\left[ 1 - \frac{16 M_K^2}{F_\pi^2}\, (2 L_8-L_5)  
  + \frac{8 M_\pi^2}{ F_\pi^2}\,  L_5 \right]\, .
\ee
Thus, the computed LECs in Eq.~\eqn{largeNccouplings} do not depend on $\mu$, as it should. Notice, however, that the numerical values of the Wilson coefficients in table~\ref{tab:WilsonCoeff} do include those $1/N_C$ corrections responsible for the QCD running and, when inserted in Eq.~\eqn{largeNccouplings}, will generate a residual $\mu$ dependence at NLO in $1/N_C$.
At $N_C\to\infty$, the strong LECs are also independent on the $\chi$PT renormalization scale $\nu_\chi$ because chiral loops are suppressed by a factor $1/N_C$. The renormalization of the LECs $L_i$ and their corresponding $\nu_\chi$ dependences can only appear at NLO in the $1/N_C$ expansion.

The $\cO(p^4)$ strong LEC $L_5$ plays a very important role in the $\varepsilon'/\varepsilon$ prediction because it appears as a multiplicative factor in the $C_6(\mu)$ contribution to $g_8^\infty$. Its large-$N_C$ value can be determined from resonance exchange, using the single-resonance approximation (SRA) \cite{Ecker:1988te,Pich:2002xy}:
\be 
L_5^\infty\, =\, \frac{F^2}{4\, M_S^2}\, .
\ee
The numerical result is, however, very sensitive to the chosen value for the scalar resonance mass. Taking $F = F_\pi$ and $M_S = 1.48$~GeV, as advocated in Ref \cite{Cirigliano:2003yq}, one gets $L_5^\infty = 1.0\cdot 10^{-3}$ \cite{Cirigliano:2003gt}, while $M_S= 1.0$~GeV would imply $L_5^\infty = 2.1\cdot 10^{-3}$.
An independent determination can be obtained from the pion and kaon decay constants, ignoring the $1/N_C$ suppressed loop contributions~\cite{Pallante:2001he}:
%
%
%
\be \label{eq:L5_FKpi}
L_5^\infty\, \approx\, \frac{F_\pi^2}{4\, (M_K^2-M_\pi^2)}\;\left(\frac{F_K}{F_\pi}-1\right)\, =\, 1.8\cdot 10^{-3}\, .
\ee
This procedure, which has actually been used in the rough estimate of the $Q_6$ matrix element in Eq.~\eqn{eq:A0Q6}, is also subject to large uncertainties because the $SU(3)$-breaking difference $F_K-F_\pi$ is very sensitive to logarithmic chiral corrections that are no longer present when $N_C\to\infty$.

Quantitative values for the $\chi$PT coupling $L_5$ have been also extracted through lattice simulations. The most recent determination has been obtained by the HPQCD collaboration \cite{Dowdall:2013rya}, analysing $F_K$ and $F_\pi$ at different quark masses with $N_f=2+1+1$ dynamical flavours, and is the result advocated in the current FLAG compilation \cite{Aoki:2016frl}:
\be\label{eq:L5Latt}
L_5^r(M_\rho)\, =\, (1.19\pm 0.25)\cdot 10^{-3}\, .
\ee
We will adopt this number in our analysis and will comment later on the sensitivity to this parameter of the $\varepsilon'/\varepsilon$ prediction.

The combination $2 L_8 - L_5$ can also be estimated in the large-$N_C$ limit, through the SRA, and it gets determined by $L_5$ \cite{Pich:2002xy}:
\be 
(2 L_8 - L_5)^\infty\, =\, -\frac{1}{4}\, L_5^\infty
\, .
\ee
This relation is well satisfied by the current lattice results that find 
$(2 L^r_8 - L^r_5)(M_\rho)= (-0.10\pm 0.20)\cdot 10^{-3}$ 
\cite{Aoki:2016frl,Dowdall:2013rya}.

The electromagnetic LECs $K_i$ can be expressed as convolutions of QCD correlators with a photon propagator \cite{Moussallam:1997xx}, and their evaluation involves an integration over the virtual photon momenta. In contrast to the strong LECs $L_i$, the $K_i$ couplings have then an explicit dependence on the $\chi$PT renormalization scale $\nu_\chi$ already at LO in $1/N_C$. Moreover, they also depend on the short-distance renormalization scale $\mu$ and the gauge parameter $\xi$. Those dependences cancel in the physical decay amplitudes with photon-loop contributions. In order to fix the combination $K_9 -2 K_{10}$ that enters $g_{\mathrm{ewk}}$ in Eq.~\eqn{largeNccouplings}, we follow Ref.~\cite{Cirigliano:2003gt} and adopt the value 
\cite{Moussallam:1997xx,Bijnens:1996kk}
\be 
(K^r_9 -2 K^r_{10})(M_\rho)\, =\, - (9.3\pm 4.6)\cdot 10^{-3}\, ,
\ee
which refers to the renormalized parameter at $\nu_\chi = M_\rho$, in the Feynman gauge $\xi=1$ and with a short-distance scale $\mu=1$~GeV.

Expanding the products of chiral currents to NLO, one obtains the large-$N_C$ predictions for the $\cO(p^4)$ and $\cO(e^2p^2)$ LECs $N_i$, $D_i$ and $Z_i$. The explicit expressions 
can be found in section 5.2 of Ref.~\cite{Cirigliano:2003gt}.

\begin{table}[t] 
\renewcommand*{\arraystretch}{1.3}
\begin{center}
\begin{tabular}{|c|c|c|c|}\hline 
Scheme & $\mathrm{Re}(g_8)$  & $g_{27}$  & $\mathrm{Re}(g_8\, g_{ewk})$
\\
\hline   
\hline 
NDR  & $1.22 \pm 0.13_{\mu} \pm 0.06_{L_i} \, {{}^{+\, 0.03}_{ -\, 0.02}}_{m_s}$   & 
$0.46 \pm 0.02_\mu$ &
$-2.24 \pm 1.44_{\mu} \pm 0.38_{K_i}\, {{}^{+\, 0.19}_{ -\, 0.21}}_{m_s}$
\\
\hline 
HV & $1.16 \pm 0.20_{\mu} \pm 0.02_{L_i}\pm 0.01_{m_s}$   &
$0.44 \pm 0.02_\mu$ &   
$-1.29 \pm 1.04_{\mu} \pm 0.15_{K_i}\, {{}^{+\, 0.11}_{-\, 0.13}}_{m_s}$  
\\
\hline\hline
NDR\,+\,HV & $1.19 \pm 0.17_{\mu} \pm 0.04_{L_i} \pm 0.02_{m_s}$   &
$0.45 \pm 0.02_\mu$ &
$-1.77 \pm 1.22_{\mu} \pm 0.26_{K_i}\, {{}^{+\, 0.15}_{-\, 0.17}}_{m_s}$   
\\
\hline   
\end{tabular}
\caption{Large-$N_C$ predictions for the CP-even parts of the LO electroweak LECs.}  
\label{tab:TableG8even} 
\end{center} 
\end{table}
\begin{table}[t] 
\renewcommand*{\arraystretch}{1.3}
\begin{center}
\begin{tabular}{|c|c|c|}\hline 
Scheme & $\mathrm{Im}(g_8)/\mathrm{Im}(\tau)$  & $\mathrm{Im}(g_8\, g_{ewk})/\mathrm{Im}(\tau)$
\\
\hline   
\hline 
NDR  & $\: 0.93 \pm 0.22_{\mu} \pm 0.21_{L_i}\, {{}^{+\, 0.10}_{-\, 0.08}}_{m_s}\;$   & 
$-22.2 \pm 5.0_{\mu} \pm 1.3_{K_i}\, {{}^{+\, 1.9}_{-\, 2.1}}_{m_s}$
\\
\hline 
HV & $0.81 \pm 0.23_{\mu} \pm 0.18_{L_i}\, {{}^{+\, 0.08}_{-\, 0.07}}_{m_s}$   &
$-23.6 \pm 5.0_{\mu} \pm 1.1_{K_i}\, {{}^{+\, 2.0}_{-\, 2.2}}_{m_s}$  
\\
\hline\hline
NDR\,+\,HV & $0.87 \pm 0.22_{\mu} \pm 0.20_{L_i}\, {{}^{+\, 0.09}_{-\, 0.08}}_{m_s}$   &
$-22.9 \pm 5.0_{\mu} \pm 1.2_{K_i}\, {{}^{+\, 1.9}_{-\, 2.2}}_{m_s}$   
\\
\hline   
\end{tabular}
\caption{Large-$N_C$ predictions for the CP-odd parts of the LO electroweak LECs.}  
\label{tab:TableG8odd} 
\end{center} 
\end{table}

Tables~\ref{tab:TableG8even} and \ref{tab:TableG8odd} show the numerical predictions obtained for the CP-even and CP-odd parts, respectively, of the LO electroweak LECs $g_8$, $g_{27}$ and $g_8\, g_{ewk}$. The large-$N_C$ limit has been only applied to the matching between the two EFTs. The full evolution from the electroweak scale down to $\mu<m_c$ has been taken into account without making any unnecessary expansion in powers of $1/N_C$; otherwise, one would miss the large short-distance corrections encoded in the Wilson coefficients $C_i(\mu)$ with $i\not= 6,8$. 

The central values quoted in the tables have been obtained at $\mu= 1$~GeV. The first uncertainty has been estimated by varying the short-distance renormalization scale $\mu$ between $M_\rho$ and $m_c$, taking into account the large-$N_C$ running for the factor $(m_s+m_d)(\mu)$ in $B(\mu)$. The current uncertainties on the strong and electromagnetic LECs that appear in Eqs.~\eqn{largeNccouplings} and  \eqn{eq:Bmu} is reflected in the second error, while the third one corresponds to the uncertainty from the input quark masses given in table~\ref{tab:inputs}. To better assess the perturbative uncertainties, the Wilson coefficients have been evaluated in two different schemes for $\gamma_5$, and an educated average of the two results is displayed in the tables.

It is important to realize the different levels of reliability of these predictions. 
The large-$N_C$ matching is only able to capture the anomalous dimensions of the operators $Q_6$ and $Q_8$. In fact, $\gamma_{66}$ and $\gamma_{88}$ are very well approximated by their leading estimates in $1/N_C$. Therefore, the contributions of these two operators to the electroweak LECs are quite robust, within the estimated uncertainties. This implies that the predicted CP-odd components of $g_8$ and $g_8\, g_{ewk}$ ($g_{27}$ is CP even) are very reliable. However, this is no longer true for their CP-even parts because the anomalous dimensions of the relevant operators are completely missed at large-$N_C$. The parametric errors quoted in table~\ref{tab:TableG8even} are probably underestimating the actual uncertainties of the calculated CP-even components of the electroweak LECs. An accurate estimate of these components would require to perform the matching calculation at the NLO in $1/N_C$.

\section{$\boldsymbol{\chi}$PT determination of the {\boldmath $K\to\pi\pi$} amplitudes}
\label{sec:K2pi-chpt}

The evaluation of the kaon decay amplitudes is a straightforward perturbative calculation within the $\chi$PT framework. To lowest order in the chiral expansion, one only needs to consider tree-level Feynman diagrams with one insertion of $\cL_2^{\Delta S=1}$. In the limit of isospin conservation, the $\cA_{\Delta I}$ amplitudes defined in Eq.~\eqn{eq:2pipar} are given by \cite{Pallante:2001he,Cirigliano:2003gt}
\begin{eqnarray}
\cA_{1/2} & = & \!\mbox{} - \sqrt{2}\, G_8 F\, \Big[  \left( M_{K}^2 - M_{\pi}^2 \right)
- {2 \over 3}\, F^2  e^2 g_{\rm ewk} \Big]
- {\sqrt{2} \over 9}\, G_{27} F \left( M_{K}^2 - M_{\pi}^2 \right)  ,
\nn
\cA_{3/2} & = & \!\mbox{} -
{10 \over 9}\,  G_{27} F \left( M_{K}^2 - M_{\pi}^2 \right)
+  {2 \over 3}\, G_8 F^3 e^2   g_{\rm ewk}\, ,
\label{eq:LOamp}
\end{eqnarray}
and $\cA_{5/2} = 0$.
From the measured values of the decay amplitudes in Eq.~\eqn{eq:isoamps}, one gets the tree-level determinations $g_8 = 5.0$ and $g_{27} = 0.25$ for the octet and 27-plet chiral couplings. The large numerical difference between these two LECs just reflects the small experimental value of the ratio $\omega$ in Eq.~\eqn{eq:omega}. Moreover, the sizeable difference between these LO phenomenological determinations and the large-$N_C$ estimates in table~\ref{tab:TableG8even} makes evident that the neglected $1/N_C$ corrections are numerically important.

Inserting in~\eqn{eq:LOamp} the large-$N_C$ predictions for the electroweak LECs given in Eq.~\eqn{largeNccouplings} and taking for $L_5$ the value in Eq.~\eqn{eq:L5_FKpi}, one immediately gets the $Q_6$ and $Q_8$ CP-odd amplitudes estimated before in Eqs.~\eqn{eq:A0Q6} and \eqn{eq:A2Q8}, 
with $B_6^{(1/2)} = B_8^{(3/2)} = 1$, including in addition some small chiral corrections that were still missing there. Eq.~\eqn{eq:LOamp} contains, moreover, the $\cO(p^2)$ contributions from all other four-quark operators.

Note, however, that the $\pi\pi$ phase shifts are predicted to be zero at LO in $\chi$PT, since phase shifts are generated by absorptive contributions in quantum loop diagrams. We know experimentally that the phase-shift difference $\chi_0-\chi_2 = 47.5^\circ$ is large, which implies that chiral loop corrections are very sizeable. Chiral loops bring a $1/N_C$ suppression but they get enhanced by large logarithms. The large absorptive contributions originate in those logarithmic corrections that are related with unitarity. A large absorptive contribution implies, moreover, a large dispersive correction because they are related by analyticity. A proper understanding of the kaon dynamics cannot be achieved without the inclusion of these $1/N_C$ suppressed contributions.

At the NLO in $\chi$PT the $\cA_{\Delta I}$ amplitudes can be expressed as \cite{Cirigliano:2003gt}
\be
\mathcal{A}_{\Delta I} \, = \mbox{} -
G_8\, F_\pi\;\Bigl\{ (M_K^2-M_\pi^2)\;\mathcal{A}_{\Delta I}^{(8)} -e^2\: F_{\pi}^2 \:g_{\text{ewk}}\:\mathcal{A}_{\Delta I}^{(g)}\Bigr\}
- G_{27}\:F_\pi\:(M_K^2-M_\pi^2)\:\mathcal{A}_{\Delta I}^{(27)}\, ,
\label{amplitudegeneral}
\ee
where $\mathcal{A}_{\Delta I}^{(8)}$ and $\mathcal{A}_{\Delta I}^{(27)}$ represent the octet and 27-plet components, and $\mathcal{A}_{\Delta I}^{(g)}$ contains the electroweak penguin contributions.
Each of these amplitudes can be decomposed in the form
\be 
\cA_{\Delta I}^{(X)}\, =\, a_{\Delta I}^{(X)} \left[ 1 + \Delta_L\cA_{\Delta I}^{(X)} + \Delta_C\cA_{\Delta I}^{(X)}\right]\, ,
\ee
with
\begin{align} 
a_{1/2}^{(8)}\, &=\, \sqrt{2}\, ,
\qquad &
a_{1/2}^{(g)}\, &=\, \frac{2\sqrt{2}}{3}\, ,
\qquad &
a_{1/2}^{(27)}\, &=\,  \frac{\sqrt{2}}{9}\, ,
\nn
a_{3/2}^{(8)} \, &=\,  0\, ,
\qquad &
a_{3/2}^{(g)} \, &=\,  \frac{2}{3}\, ,
\qquad &
a_{3/2}^{(27)} \, &=\,  \frac{10}{9}\, ,
\end{align}
parametrizing the corresponding tree-level contributions, $\Delta_L\cA_{\Delta I}^{(X)}$ the one-loop chiral corrections and $\Delta_C\cA_{\Delta I}^{(X)}$ the NLO local corrections from $\cL_4^{\Delta S=1}$. Since we are not considering electromagnetic corrections, $\cA_{5/2} = 0$.  

A small part of the $\cO(p^4)$ corrections has been reabsorbed into the physical pion decay constant $F_\pi$, which appears explicitly in the three terms of Eq.~\eqn{amplitudegeneral}. The NLO relation between $F_\pi$ and the Lagrangian parameter $F$ is given by \cite{Gasser:1984gg}
\begin{align}
F\, =\, F_\pi \; &\bigg\{ 1 - \frac{4}{F^2} \left[   
\left( M_\pi^2 + 2 M_K^2 \right) L_{4}^{r} (\nu_\chi) + M_\pi^2\; L_{5}^{r}(\nu_\chi)   \right] 
\nonumber\\[4pt] 
&\mbox{} +  
\frac{1}{2 (4 \pi)^2 F^2} \left[ 2 M_\pi^2\; \log{\left(\frac{M_\pi^2}
{\nu_\chi^2}\right)} + 
M_K^2\; \log{\left(\frac{M_K^2}{\nu_\chi^2}\right)} \right] 
\bigg\} \ .
\end{align}
The chiral logarithmic corrections are obviously suppressed by the geometrical loop factor $(4\pi)^{-2}$ and two powers of the Goldstone scale $F\sim\sqrt{N_C}$. Thus, these contributions and the corresponding dependence of the renormalized $L_i^{r}(\nu_\chi)$ couplings on the $\chi$PT renormalization scale $\nu_\chi$ are of $\cO(1/N_C)$.
The $\nu_\chi$-independent parts of the LECs have a different scaling with $1/N_C$: while $L_4/F^2 \sim \cO(1/N_C)$, $L_5/F^2$ is a leading correction of $\cO(1)$. This implies the large-$N_C$ result for $L_5$ given in Eq.~\eqn{eq:L5_FKpi} (the $L_5$ contribution to $F_K$ is multiplied by $M_K^2$ instead of $M_\pi^2$).

The numerical values of the different $\cA_{1/2}^{(X)}$ and $\cA_{3/2}^{(X)}$ components are given in tables~\ref{tab:Table12} and \ref{tab:Table32}, respectively.
We comment next on the most important features of the different contributions.

\begin{table}[h!] 
\renewcommand*{\arraystretch}{1.3}
\begin{center}
\begin{tabular}{|c|c|c|c|c|}\hline 
X & $a_{1/2}^{(X)}$ &
$\Delta_{L} \mathcal{A}_{1/2}^{(X)} $ &
$[\Delta_{C} \mathcal{A}_{1/2}^{(X)}]^+ $  & 
$[\Delta_{C} \mathcal{A}_{1/2}^{(X)}]^- $ \\
\hline   
\hline 
8  & $\sqrt{2}$   & $0.27 +  0.47\;  i$   &
$\phantom{-}0.01 \pm 0.05$   &  $\phantom{-}0.02 \pm 0.05$  
\\
\hline 
g  & $ \frac{2 \sqrt{2}}{3}$  &  $0.27 +  0.47\; i$ & 
$-0.19 \pm 0.01$  & $-0.19 \pm 0.01$ 
\\
\hline 
27  &  $\frac{\sqrt{2}}{9}$   &  $1.03 + 0.47\; i$    & 
$\phantom{-}0.01 \pm 0.63$ & $\phantom{-}0.01  \pm 0.63$ 
\\
\hline   
\end{tabular}
\caption{Numerical predictions for the $\mathcal{A}_{1/2}$ components:  
$a_{1/2}^{(X)}$, 
$\Delta_{L} \mathcal{A}_{1/2}^{(X)} $, 
$\Delta_{C} \mathcal{A}_{1/2}^{(X)} $. The local NLO correction to the CP-even ($[\Delta_{C} \mathcal{A}_{1/2}^{(X)}]^+$) and CP-odd ($[\Delta_{C} \mathcal{A}_{1/2}^{(X)}]^-$) amplitudes is only different in the octet case.} 
\label{tab:Table12} 
\end{center} 
\end{table}


\begin{table}[h!] 
\renewcommand*{\arraystretch}{1.3}
\begin{center}
\begin{tabular}{|c|c|c|c|}\hline 
X & $a_{3/2}^{(X)}$ &
$\Delta_{L} \mathcal{A}_{3/2}^{(X)} $ & $\Delta_{C} \mathcal{A}_{3/2}^{(X)} $
\\
\hline   
\hline 
g  & $ \frac{2}{ 3}$  &  $-0.50 -  0.21\; i$  & $-0.19 \pm 0.19$      
\\
\hline 
27  &  $\frac{10}{9}$  &  $-0.04 - 0.21\; i$    & 
$\phantom{-}0.01 \pm 0.05$    
\\
\hline   
\end{tabular}
\caption{Numerical predictions for the $\mathcal{A}_{3/2}$ components:  
$a_{3/2}^{(X)}$, 
$\Delta_{L} \mathcal{A}_{3/2}^{(X)} $, 
$\Delta_{C} \mathcal{A}_{3/2}^{(X)} $ }  
\label{tab:Table32} 
\end{center} 
\end{table}

\subsection{Chiral loop corrections}

\begin{figure}[t]\centering
\begin{center}
\begin{minipage}[t]{0.5\textwidth}
\centering
\tikzset{
  goldstone/.style={draw=black}
}

\NewDocumentCommand\semiloop{O{black}mmmO{}O{above}}
{
\draw[#1] let \p1 = ($(#3)-(#2)$) in (#3) arc (#4:({#4+180}):({0.5*veclen(\x1,\y1)})node[midway, #6] {#5};)
}

\begin{tikzpicture}[line width=1.5 pt,node distance=1cm and 1cm]
\coordinate[ ] (v1);
\coordinate[right=1.5cm of v1] (v4);
\coordinate[above right=of v4,label=right:$\pi$] (f1);
\coordinate[below right=of v4,label=right:$\pi$] (f2);
\coordinate[ left =of v1,label=left :$K$] (e2);
\draw[goldstone] (v1) -- (e2);
\draw[goldstone] (v4) -- (f1);
\draw[goldstone] (f2) -- (v4);
\semiloop[goldstone]{v1}{v4}{0}[$ $];
\semiloop[goldstone]{v4}{v1}{180}[$ $][below];
\coordinate[above right=1.0cm and -0.5cm of v4,label=left:$i$];
\coordinate[above right=-1.0cm and -0.5cm of v4,label=left:$j$];
\fill[red] (v1) circle (.1cm);
\end{tikzpicture}
\end{minipage}\hfill
\begin{minipage}[t]{0.5\textwidth}
\centering
\tikzset{
  goldstone/.style={draw=black}
}

\NewDocumentCommand\semiloop{O{black}mmmO{}O{above}}
{
\draw[#1] let \p1 = ($(#3)-(#2)$) in (#3) arc (#4:({#4+180}):({0.5*veclen(\x1,\y1)})node[midway, #6] {#5};)
}

\begin{tikzpicture}[line width=1.5 pt,node distance=1.2cm and 1.7cm]
\coordinate[ ] (v1);
\coordinate[ right=of v1] (v4);
\coordinate[above right=1.5cm of v4,label=right:$\pi$] (f1);
\coordinate[below right=1.5cm of v4,label=right:$\pi$] (f2);
\coordinate[below left =1.0cm of v1,label=left :$K$] (e2);
\coordinate[above right=0.7cm and -0.6cm of v4](g1);
\coordinate[above right=-0.7cm and -0.6cm of v4](g2);
\draw[goldstone] (g2) -- (e2);
\draw[goldstone] (g1) -- (f1);
\draw[goldstone] (f2) -- (g2);
\semiloop[goldstone]{g1}{g2}{270}[$ $];
\semiloop[goldstone]{g2}{g1}{90}[$ $][below];
\coordinate[above right=0.7cm and -0.6cm of v4] (v5);
\coordinate[above right=-0.7cm and -0.6cm of v4] (v6);
\fill[red] (v5) circle (.1cm);
\coordinate[above right=0.0cm and -1.3cm of v4,label=left:$i$];
\coordinate[above right=0.0cm and 0.6cm of v4,label=left:$j$];
\end{tikzpicture}
\end{minipage}\hfill
\vskip .5cm

\begin{minipage}[t]{0.5\textwidth}
\centering
\tikzset{
  goldstone/.style={draw=black}
}

\NewDocumentCommand\semiloop{O{black}mmmO{}O{above}}
{
\draw[#1] let \p1 = ($(#3)-(#2)$) in (#3) arc (#4:({#4+180}):({0.5*veclen(\x1,\y1)})node[midway, #6] {#5};)
}

\begin{tikzpicture}[line width=1.5 pt,node distance=1.5cm and 1.5cm]
\coordinate[ ] (v1);
\coordinate[above right=of v1,label=right:$\pi$] (f1);
\coordinate[right=of v1,label=right:$\pi$] (f2);
\coordinate[below =0.6cm of v1] (f3);
\coordinate[below =1.8cm of v1] (f4);
\coordinate[ left =of v1,label=left :$K$] (e2);
\draw[goldstone] (v1) -- (e2);
\draw[goldstone] (v1) -- (f1);
\draw[goldstone] (f2) -- (v1);
\draw[goldstone] (f3) -- (v1);
\semiloop[goldstone]{f3}{f4}{270}[$ $];
\semiloop[goldstone]{f4}{f3}{90}[$ $][below];
\coordinate[above right=0.8cm and -0.6cm of v4];
\coordinate[above right=-1.2cm and -0.6cm of v4,label=left:$i$];
\fill[red] (f3) circle (.1cm);
\end{tikzpicture}
\end{minipage}\hfill
\begin{minipage}[t]{0.5\textwidth}
\centering
\tikzset{
  goldstone/.style={draw=black}
}

\NewDocumentCommand\semiloop{O{black}mmmO{}O{above}}
{
\draw[#1] let \p1 = ($(#3)-(#2)$) in (#3) arc (#4:({#4+180}):({0.5*veclen(\x1,\y1)})node[midway, #6] {#5};)
}

\begin{tikzpicture}[line width=1.5 pt,node distance=1.2cm and 1.7cm]
\coordinate[ ] (v1);
\coordinate[ ,  right=of v1] (v4);
\coordinate[above right=-0.5cm and 1.4cm of v4,label=right:$\pi$] (f1);
\coordinate[below right=1.9cm of v4,label=right:$\pi$] (f2);
\coordinate[below left =1.0cm of v1,label=left :$K$] (e2);
\coordinate[above right=0.7cm and -0.6cm of v4](g1);
\coordinate[above right=-0.7cm and -0.6cm of v4](g2);
\draw[goldstone] (g2) -- (e2);
\draw[goldstone] (g2) -- (f1);
\draw[goldstone] (f2) -- (g2);
\semiloop[goldstone]{g1}{g2}{270}[$ $];
\semiloop[goldstone]{g2}{g1}{90}[$ $][below];
\coordinate[above right=0.0cm and -1.3cm of v4,label=left:$i$];
\fill[red] (g2) circle (.1cm);
\end{tikzpicture}
\end{minipage}
\end{center}
\caption{One-loop topologies contributing to $K\rightarrow\pi\pi$. The filled red circles indicate LO $\Delta S=1$ vertices. The labels $i$ and $j$ represent the Goldstone bosons inside the loop. Wave-function renormalization topologies are not shown.}
\label{fig:1-loop}
\end{figure}
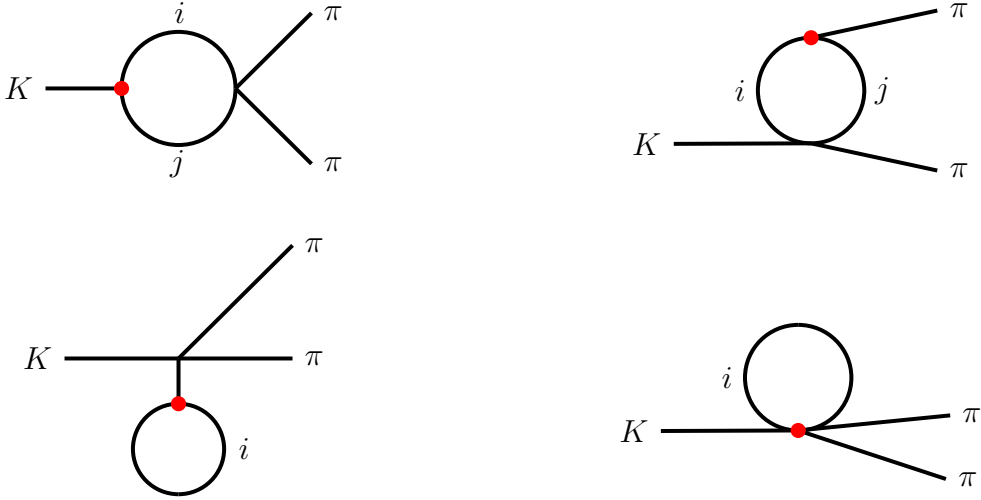

The one-loop chiral corrections are generated through the Feynman topologies depicted in figure~\ref{fig:1-loop} \cite{Kambor:1989tz,Kambor:1991ah,Bijnens:1998mb,Pallante:1998gk,Pallante:1999qf,Pallante:2000hk,Pallante:2001he,Cirigliano:2003nn,Cirigliano:2003gt,Cirigliano:2009rr}. They include one insertion of the LO weak Lagrangian $\cL_2^{\Delta S=1}$ (filled red vertices), and the first two diagrams contain also interaction vertices from the $\cO(p^2)$ strong Lagrangian $\cL_2$. 
The resulting $\Delta_L\cA_{1/2}^{(X)}$ and $\Delta_L\cA_{3/2}^{(X)}$ corrections given in tables~\ref{tab:Table12} and \ref{tab:Table32}, respectively, exhibit a very clear pattern. The one-loop chiral corrections are always positive for all $\Delta I=1/2$ amplitudes and negative for $\Delta I=3/2$. Moreover, the absorptive contributions (the imaginary parts of $\Delta_L\cA_{\Delta I}^{(X)}$) only depend on the isospin of the final 2-pion state. 
The elastic final-state interaction of the two pions induces a very large and positive absorptive correction when $I=0$, while this contribution becomes much smaller and negative when $I=2$. 

The absorptive contribution fully originates in the first topology of figure~\ref{fig:1-loop}, since it is the only one where the two intermediate pions can be put on their mass-shell \cite{Pallante:2001he}: 
\beqn
\Delta_L\cA_{1/2}^{(X)} & = & \frac{M_K^2}{(4\pi F_\pi)^2}\;
\left(1 - \frac{M_\pi^2}{2 M_K^2}\right)\; \widetilde B(M_\pi^2,M_\pi^2,M_K^2) + \cdots\, ,
\nn
\Delta_L\cA_{3/2}^{(X)} & = & -\frac{1}{2}\,\frac{M_K^2}{(4\pi F_\pi)^2}\;
\left(1 - \frac{2 M_\pi^2}{M_K^2}\right)\; \widetilde B(M_\pi^2,M_\pi^2,M_K^2) + \cdots\, ,
\eeqn
where
\be \label{eq:2pionintegral}
\widetilde B(M_\pi^2,M_\pi^2,M_K^2)\, =\, 
\sigma_\pi\,\left[ \log{\left(\frac{1-\sigma_\pi}{1+\sigma_\pi}\right)} + i \pi\right]
+ \log{\left(\frac{\nu_\chi^2}{M_\pi^2}\right)}+1
\ee
is the renormalized one-loop scalar integral with two pion propagators and $q^2 = M_K^2$, and $\sigma_\pi \equiv\sqrt{1-4 M_\pi^2/M_K^2}$. These results reproduce the LO $\chi$PT values for the strong $\pi\pi$ scattering phase shifts with $J=0$ and $I=0,2$, at $s=M_K^2$:
\be 
\tan{\delta_0(M_K^2)}\, =\, \frac{\sigma_\pi}{32\pi F_\pi^2}\,\left( 2 M_K^2 - M_\pi^2\right)\, ,
\qquad\quad
\tan{\delta_2(M_K^2)}\, =\, \frac{\sigma_\pi}{32\pi F_\pi^2}\,\left( 2 M_\pi^2 - M_K^2\right)\, .
\ee
The predicted phase-shift difference, $\delta_0(M_K^2)-\delta_2(M_K^2) = 37^\circ$, is somewhat lower than its experimental value showing that higher-order rescattering contributions are numerically relevant.
The one-loop integral in Eq.~\eqn{eq:2pionintegral} contains, in addition, a large chiral logarithm of the ultraviolet scale $\nu_\chi$ over the infrared scale $M_\pi$, which enhances significantly the dispersive component of the $I=0$ amplitude and suppresses the $I=2$ one.

The complete analytical expressions for the one-loop corrections $\Delta_L\cA_{\Delta I}^{(X)}$ can be found in Refs.~\cite{Pallante:2001he,Cirigliano:2003gt}.
The absorptive contributions are finite and, therefore, do not depend on the chiral renormalization scale. An explicit dependence on $\nu_\chi$ is, however, present on the dispersive components. The numbers quoted in tables~\ref{tab:Table12} and \ref{tab:Table32} have been obtained at $\nu_\chi = M_\rho = 0.77$~GeV. The dependence on $\nu_\chi$ is of course exactly cancelled by the local counterterm contributions.

One observes in table~\ref{tab:Table12} a huge ($\sim 100\%$) dispersive one-loop correction to the $\cA^{27}_{1/2}$ amplitude. Fortunately, since the 27-plet contribution is a very small part of the total $\Delta I = \frac{1}{2}$ amplitude, this does not introduce any significant uncertainty in the final numerical value of $A_0$.
Moreover, the 27-plet components do not contribute to the CP-odd amplitudes we interested in, because Im$(g_{27})=0$.

The corrections relevant for $\varepsilon'/\varepsilon$ are
the octet contribution to the isoscalar amplitude and the electroweak-penguin
contribution to $A_2$. The first one generates a very sizeable enhancement of Im$A_0$ by a factor $|1 +\Delta_L\cA_{1/2}^{(8)}|\approx 1.35$, while the second one induces a strong suppression of Im$A_2^{\mathrm{emp}}$ with a factor $|1 +\Delta_L\cA_{3/2}^{(g)}|\approx 0.54$. Looking to the simplified formula in Eq.~\eqn{eq:epsilonp_naive}, one immediately realizes the obvious impact of these chiral corrections on the final value predicted for 
$\varepsilon'/\varepsilon$, since they destroy completely the accidental numerical cancellation between the $Q_6$ and $Q_8$ contributions.

\subsection[Local $\cO(p^4)$ contributions]{Local {\boldmath $\cO(p^4)$} contributions}

Explicit expressions for the local $\Delta_C\cA_{\Delta I}^{(X)}$ corrections in terms of the $\cO(p^4)$ electroweak LECs can be found in Refs.~\cite{Pallante:2001he,Cirigliano:2003gt}. In the large-$N_C$ limit, the local contribution to the $\cA_{1/2}^{(8)}$ amplitude takes the form
\beqn
g_8^\infty\left[1 + \Delta_C\cA_{1/2}^{(8)}\right]^\infty & = &
\left[ -\frac{2}{5}\, C_1(\mu)+\frac{3}{5}\, C_2(\mu)+C_4(\mu)\right]\,
\left\{ 1 + \frac{4 M_\pi^2}{F^2}\, L_5\right\}
\nn & - & 16\, B(\mu)\, C_6(\mu) \,\left\{ L_5 + 
\frac{4 M_K^2}{F^2}\, \delta_8^K
+ \frac{4 M_\pi^2}{F^2}\, \delta_8^\pi
\right\} . 
\eeqn
The NLO corrections $\delta^P_8$ depend on some $\cO(p^6)$ LECs $X_i$ that are not very well known. The relevant combinations can be estimated with the SRA, up to unknown contributions from couplings with two resonance fields \cite{Cirigliano:2006hb}:
\beqn
\delta_8^K & = & L_5\, (2 L_8 - L_5)  + \frac{1}{4}\, (2 X_{14} + X_{34})
\, \approx\, \frac{1}{2}\, L_5^2\, ,
\nn
\delta_8^\pi & = & (8 L_8^2-3 L_5^2)
+ X_{12} + X_{14} + X_{17} - 3 X_{19} -4 X_{20} - X_{31}
\, \approx\, -\frac{15}{8}\, L_5^2\, .
\eeqn
Since the contributions from $Q_6$ and the other four-quark operators get different NLO corrections, the $\cO(p^4)$ corrections to the CP-even and CP-odd octet amplitudes, 
\be 
\mathrm{Re}(g_8\,\Delta_C\cA_{1/2}^{(8)}) + i\:\mathrm{Im}(g_8\,\Delta_C\cA_{1/2}^{(8)})
\,\equiv\,
\mathrm{Re}(g_8)\; [\Delta_C\cA_{1/2}^{(8)}]^+ + i\:
\mathrm{Im}(g_8)\; [\Delta_C\cA_{1/2}^{(8)}]^-\, ,
\ee
are also different. The predicted numerical values for the separate corrections $[\Delta_C\cA_{1/2}^{(8)}]^+$ and $[\Delta_C\cA_{1/2}^{(8)}]^-$  are given in table~\ref{tab:Table12}.

The $\cO(p^4)$ local corrections to the other $\cA_{\Delta I}^{(X)}$ amplitudes only depend on $L_5$, in the large-$N_C$ limit:
\beqn
\left.\Delta_C\cA_{1/2}^{(27)}\right|^\infty & =& \left.\Delta_C\cA_{3/2}^{(27)}\right|^\infty \; =\;\frac{4 M_\pi^2}{F^2}\, L_5\, ,
\nn
\left.\Delta_C\cA_{1/2}^{(g)}\right|^\infty & =& \left.\Delta_C\cA_{3/2}^{(g)}\right|^\infty \; =\; - \frac{4}{F^2}\, (M_K^2 + 5 M_\pi^2)\, L_5\, .
\eeqn
Thus, all local NLO contributions are finally determined by the input value of $L_5^\infty$.

The numerical predictions for the different local corrections $\Delta_C\cA_{\Delta I}^{(X)}$ are shown in tables~\ref{tab:Table12} and \ref{tab:Table32}. The main uncertainty originates in their dependence on the chiral renormalization scale $\nu_\chi$, which is totally missed by the large-$N_C$ approximation. We take the large-$N_C$ results as our numerical estimates at $\nu_\chi = M_\rho$. The errors have been estimated varying $\nu_\chi$ between 0.6 and 1 GeV in the corresponding loop contributions $\Delta_L\cA_{\Delta I}^{(X)}$. We have also varied the short-distance renormalization scale $\mu$ between $M_\rho$ and $m_c$, but the impact on the $\Delta_C\cA_{\Delta I}^{(X)}$ corrections is negligible compared with the $\nu_\chi$ uncertainty.

The relevant corrections for our determination of $\varepsilon'/\varepsilon$ are
$[\Delta_C\cA_{1/2}^{(8)}]^- = 0.02\pm 0.05$ and
$\Delta_C\cA_{3/2}^{(g)} = -0.19\pm 0.19$. They are much smaller than the corresponding loop contributions, which is also reflected in the large relative uncertainties induced by the $\nu_\chi$ variation.

\section{\boldmath The SM prediction for $\varepsilon'/\varepsilon$}
\label{sec:result}

Putting all computed corrections together in Eq.~\eqn{eq:epsp_simp}, we obtain the updated SM prediction
\beqn\label{eq:finalRes}
\mbox{Re}\left(\varepsilon'/\varepsilon\right)& =&
\left(15\pm 2_{\mu}\pm 2_{m_s} \pm 2_{\Omega_\text{eff}}\pm 6_{1/N_C}\right) \times 10^{-4}
\no\\[5pt] & =& 
\left(15\pm 7\right) \times 10^{-4}\, .
\eeqn
The input values adopted for the relevant SM parameters are given in table~\ref{tab:inputs}, in the appendix. We have only calculated theoretically the CP-odd amplitudes $\mathrm{Im}A_I$. For their CP-even counterparts (and $|\varepsilon|$) the experimental values have been taken instead.

We display explicitly the four main sources of errors. The first one reflects the fluctuations under changes of the short-distance renormalization scale $\mu$ in the range between $M_\rho$ and $m_c$, and the choice of scheme for $\gamma_5$. The second uncertainty shows the sensitivity to variations of the input quark masses within their currently allowed ranges, while the third one displays the uncertainty from the isospin-breaking parameter $\Omega_{\text{eff}}$. The fourth error accounts for the sum of squared uncertainties from the input value of $L_5$ ($\pm \:5\cdot 10^{-4}$) and the $\chi$PT scale $\nu_\chi$  that is varied between 0.6 and 1~GeV ($\pm \:3\cdot 10^{-4}$). This fourth error is by far the dominant one and reflects our current ignorance about $1/N_C$-suppressed contributions in the matching process.

\begin{figure}[t]\centering
\includegraphics[scale=.8]{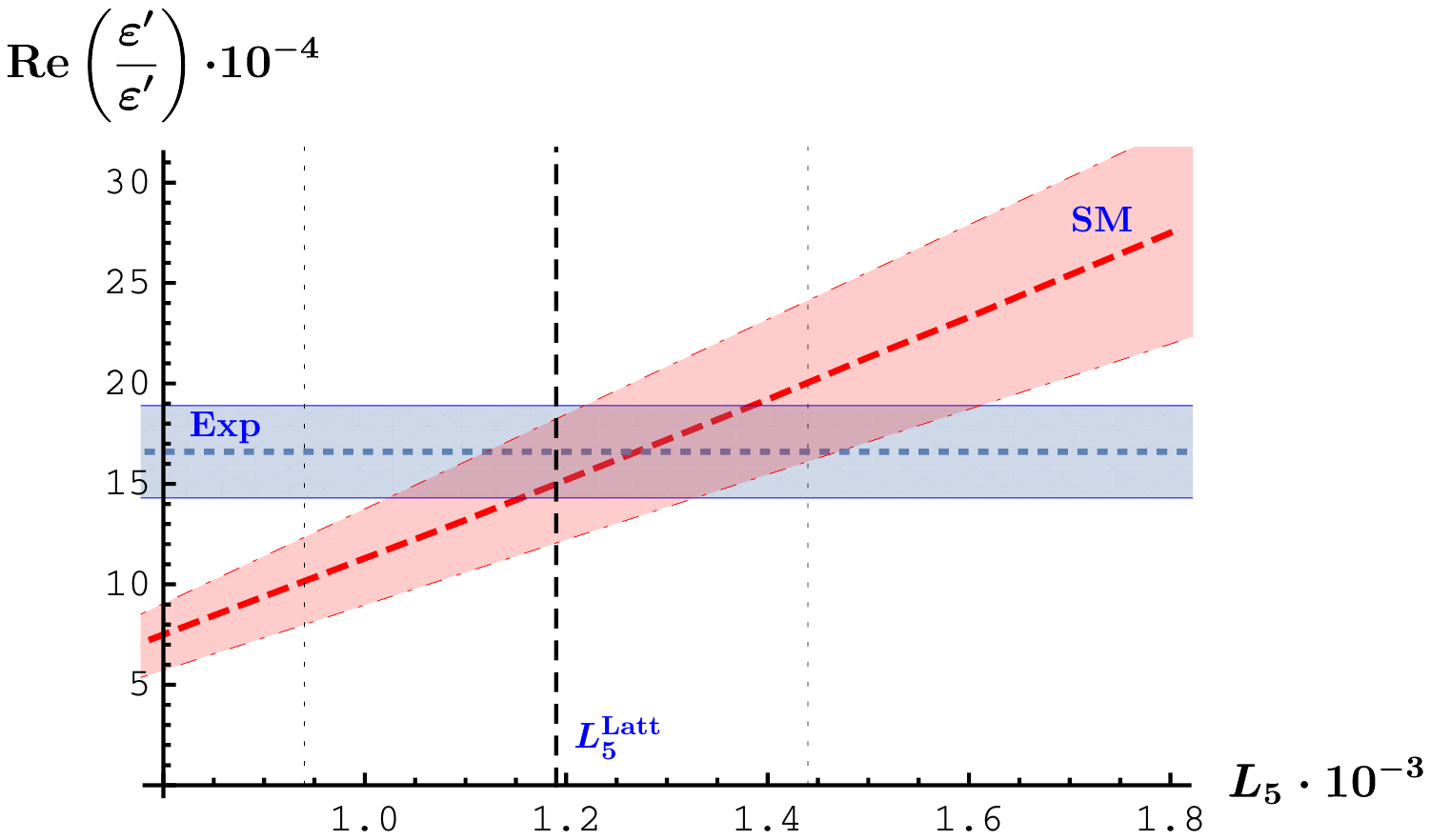}
\caption{SM prediction for $\varepsilon'/\varepsilon$ as function of $L_5$
(red dashed line) with $1\,\sigma$ errors (oblique band).
The horizontal blue band displays the experimentally measured value with $1\,\sigma$ error bars. The dashed vertical line shows the current lattice determination of $L_5^r(M_\rho)$.}
\label{fig:epsp}
\end{figure}

In figure~\ref{fig:epsp}, we show the prediction for $\varepsilon'/\varepsilon$ as function of the input value of $L_5$. The strong dependence on this important parameter is evident from the plot. The experimental $1\,\sigma$ range is indicated by the horizontal band, while the dashed vertical lines display the current lattice 
determination of $L_5^r(M_\rho)$. The measured value of $\varepsilon'/\varepsilon$ is nicely reproduced with the preferred lattice inputs.

In order to better appreciate the dynamical contributions that have been included in Eq.~\eqn{eq:finalRes}, it is worth to go back to the schematic description of EFTs displayed in figure~\ref{fig:eff_th}. Starting with the SM at the electroweak scale, where the underlying $\Delta S=1$ transitions take place, we have first used the full machinery of the short-distance OPE to determine the effective Lagrangian $\cL_{\mathrm{eff}}^{\Delta S=1}$, defined in the three-flavour quark theory at scales just below the charm mass. We have included all NLO contributions to the Wilson coefficients $C_i(\mu)$, without making any large-$N_C$ approximation. The OPE sums up large logarithmic QCD corrections, but most of these logarithms are suppressed by  factors of $1/N_C$ because the anomalous dimension matrix $\gamma_{ij}$ of the $Q_i$ operators vanishes at $N_C\to \infty$, except for $\gamma_{66}$ and  
$\gamma_{88}$ which remain non-zero. Since
\be 
\frac{1}{N_C}\:\log{(M_W^2/\mu_0^2)}\, =\, 2.9
\ee
at $\mu=\mu_0 = 1~\mathrm{GeV}$,
it would not make much sense to neglect ``subleading'' corrections in $1/N_C$.

At the kaon mass scale, we have made use of a different effective field theory that
takes advantage of the chiral symmetry properties of QCD to constrain the pseudoscalar Goldstone dynamics. $\chi$PT is the appropriate tool to describe rigorously the physics of kaons and pions, through a low-energy expansion in powers of momenta and quark masses \cite{Cirigliano:2011ny}. Chiral symmetry determines the effective realization of $\cL_{\mathrm{eff}}^{\Delta S=1}$ at the hadronic mass scale; {\it i.e.}, the most general form of the low-energy $\chi$PT structures with the same symmetry properties than the four-quark operators $Q_i$, at a given order in momenta. All short-distance information is encoded in LECs that are not fixed by symmetry considerations.
The $K\to\pi\pi$ amplitudes can then be easily predicted in terms of those LECs. The $\chi$PT predictions include unambiguous quantum corrections, which comply with the requirements of unitarity and analyticity. 

The so-called chiral logarithmic corrections are also suppressed by factors of $1/N_C$ (quantum loops are absent from the $N_C\to\infty$ mesonic world \cite{Pich:2002xy}), but they cannot be ignored since they are responsible for the large $\pi\pi$ phase shifts that originate in their absorptive contributions. Moreover, the dispersive logarithmic corrections are also large when the two pions are in a $J=0$ state.  Once again, a $1/N_C$ suppression gets compensated by a large logarithm: at $\nu_\chi = \nu_0 = 1~\mathrm{GeV}$,
\be 
\frac{1}{N_C}\:\log{(\nu_0^2/M_\pi^2)}\, =\, 1.3\, .
\ee
The measured kaon decay amplitudes cannot be understood without the inclusion of these large, but $1/N_C$ suppressed, contributions. Our SM prediction in Eq.~\eqn{eq:finalRes} includes of course the full $\cO(p^4)$ $\chi$PT results, without any unnecessary $1/N_C$ approximation.

The limit of a large number of QCD colours has been only used to perform the matching between the two EFTs; {\it i.e.}, to evaluate the numerical values of the $\chi$PT LECs. Thanks to the factorization property in Eq.~\eqn{eq:factorization}, the hadronic matrix elements of the four-quark operators can be reduced to matrix elements of QCD currents at $N_C\to\infty$. Since these currents have well-known $\chi$PT realizations at low energies, the electroweak LECs can be easily determined in the large-$N_C$ limit. The intrinsic uncertainty of this determination is of $\cO(1/N_C)$, but it is not enhanced by any large logarithm of two widely separated mass scales.

The large-$N_C$ structure of the anomalous dimension matrix $\gamma_{ij}$ allows us to better assess the quality of our matching procedure. At $N_C\to\infty$,  the only non-zero entries are $\gamma_{66}$ and $\gamma_{88}$, which, moreover, are well approximated by their large-$N_C$ estimates. Thus, the short-distance properties of $Q_6$ and $Q_8$ are very efficiently incorporated into the corresponding $\chi$PT couplings through the large-$N_C$ matching. In fact, the leading renormalization-scale dependence of $C_6(\mu)$ and $C_8(\mu)$ cancels exactly with the running of the light quark masses appearing through the $\chi$PT factor $B(\mu)$ in Eq.~\eqn{eq:Bmu}. Fortunately, $Q_6$ and $Q_8$ are precisely the only two operators that really matter for the numerical prediction of $\varepsilon'/\varepsilon$. 

This is no longer true for the other four-quark operators because their anomalous dimensions are lost at $N_C\to\infty$. The $\mu$ dependence of their Wilson coefficients cannot be compensated in the large-$N_C$ matching process, which indicates the relevance
of the missing $1/N_C$ contributions. The bulk of the $\Delta I=1/2$ enhancement is associated with the octet operator $Q_-$, while $\Delta I=3/2$ transitions originate from $Q^{(27)}$. Since the $1/N_C$-suppressed anomalous dimensions of these two operators are a crucial ingredient of the $K\to\pi\pi$ dynamics, an accurate prediction of the CP-even  decay amplitudes will only become possible with a matching calculation at NLO in $1/N_C$   \cite{Pich:1995qp,Pich:1990mw,Jamin:1994sv}.

\section{Discussion and outlook}
\label{sec:outlook}

The SM prediction for $\varepsilon'/\varepsilon$, given in Eq.~\eqn{eq:finalRes}, is in perfect agreement with the measured experimental value \eqn{eq:exp}. 
The final result emerges from a delicate balance among several contributions, where the chiral dynamics of the two final pions plays a very crucial role. The $\pi\pi$ rescattering corrections destroy the naive cancellation between the $Q_6$ and $Q_8$ terms in Eq.~\eqn{eq:epsilonp_naive}, enhancing the positive $Q_6$ contribution and suppressing the negative contribution from $Q_8$. The small corrections from other four-quark operators to $\mathrm{Im} A_0^{(0)}$ and $\mathrm{Im} A_2^{\mathrm{emp}}$ are not important numerically, once the cancellation is no longer operative.

The low values of $\varepsilon'/\varepsilon$ claimed in some recent references
\cite{Buras:2015xba,Buras:2016fys,Buras:2015yba} originate in simplified estimates of the relevant $K\to\pi\pi$ amplitudes that neglect the long-distance contributions from pion loops. Following a $1/N_C$-inspired approach \cite{Buras:2014maa}, Ref.~\cite{Buras:2015xba} has advocated the inequality  $B_6^{(1/2)} \le B_8^{(3/2)} < 1$, which has been later adopted in subsequent works. However, this very questionable result is obtained within a chiral model that only includes the $\cO(p^4)$ $L_5$ structure, neglecting all other terms in Eq.~\eqn{eq:l4}. Moreover, the only computed $1/N_C$ corrections correspond to some non-factorizable divergent contributions of the form $\log{(\Lambda^2/M^2)}$, with $\Lambda$ an UV cut-off that is identified with the short-distance renormalization scale $\mu$, and $M$ a badly-defined infrared scale of $\cO(M_K)$. All other quantum corrections (including the important absorptive contributions) are just ignored.
Notice also that in order to properly define the parameter $B_6^{(1/2)}$, one needs first to specify $L_5^\infty$, which in Ref.~\cite{Buras:2015xba} is fixed at the large value shown in Eq.~\eqn{eq:L5_FKpi}.

It has been well known for many years that the elastic scattering of two pions with zero relative angular momentum is very strong and generates a large phase-shift difference between the $I=0$ and $I=2$ states \cite{Weinberg:1966kf}. This important dynamical effect is well understood and has been rigorously predicted within the $\chi$PT framework. The relevant quantum corrections have been computed by many groups for $K\to\pi\pi$ \cite{Kambor:1989tz,Kambor:1991ah,Bijnens:1998mb,Pallante:1998gk,Pallante:1999qf,Pallante:2000hk,Pallante:2001he,Cirigliano:2003nn,Cirigliano:2003gt,Cirigliano:2009rr,Cirigliano:1999pv,Cirigliano:2001hs}, $K_{\ell4}$ \cite{Bijnens:1989mr,Riggenbach:1990zp,Bijnens:1994ie,Amoros:1999qq,Amoros:2000mc} and $\pi\pi\to\pi\pi$ \cite{Gasser:1983yg,Gasser:1984gg,Bijnens:1995yn,Bijnens:1997vq}, reaching a two-loop precision in the last two cases. Higher-loop effects have been also estimated with dispersive methods and many successful phenomenological analyses of the relevant data have been put forward along the years \cite{Ananthanarayan:2000ht,Colangelo:2000jc,Colangelo:2001df,Caprini:2003ta,Caprini:2011ky,DescotesGenon:2001tn,GarciaMartin:2011cn,Colangelo:2015kha}. 
The inclusion of all known $\chi$PT corrections is a compulsory requirement for a reliable prediction of the kaon decay amplitudes \cite{Cirigliano:2011ny}.

The recent RBC-UKQCD lattice calculation~\cite{Bai:2015nea,Blum:2015ywa}, which also finds a low central value for $\varepsilon'/\varepsilon$, follows the Lellouch-L\"uscher prescription \cite{Lellouch:2000pv} in order to incorporate the Minkowskian pion dynamics into the numerical simulation. Their results look quite encouraging, since it is the first time that a clear signal of the $\Delta I=1/2$ enhancement seems to emerge from lattice data \cite{Boyle:2012ys}. However, the value obtained for the isoscalar $\pi\pi$ phase shift disagrees with the experimental determination by $2.9\,\sigma$. This discrepancy is larger than the one quoted for $\varepsilon'/\varepsilon$
($2.1\,\sigma$), indicating that these results are still in a very premature stage and  improvements are clearly needed. In addition, the current lattice result does not take into account any isospin-breaking effects, which are a very important ingredient of the $\varepsilon'/\varepsilon$ prediction \cite{Cirigliano:2003nn,Cirigliano:2003gt,Cirigliano:2009rr}.
The inclusion of electromagnetic corrections in lattice simulations of the $K\to\pi\pi$ amplitudes looks difficult, but proposals to face some of the technical problems involved are already being considered \cite{Christ:2017pze}.

The quoted uncertainty of the SM prediction of $\varepsilon'/\varepsilon$ in Eq.~\eqn{eq:finalRes} is three times larger than the current experimental error.
This leaves ample margin to speculate with hypothetical new-physics contributions, but prevents us from making a precise test of the SM mechanism of CP violation that could give significant constraints on the CKM parameters \cite{Lehner:2015jga}.

In order to achieve a better theoretical accuracy, the different ingredients entering the calculation must be substantially refined. Improvements look possible in the near future through a combination of analytical calculations, numerical simulations and data analyses:
\begin{itemize}
\item A NNLO computation of the Wilson coefficients is currently being performed \cite{Cerda-Sevilla:2016yzo}. The known NNLO corrections to the electroweak penguin operators \cite{Buras:1999st} reduce the scheme dependence of $C_8(\mu)$ in a quite significant way and slightly increase the negative $Q_8$ contribution to $\varepsilon'/\varepsilon$. A complete NNLO calculation should allow for a similar reduction of the $y_6(\mu)$ uncertainty. Since the quark-mass anomalous dimension is already known with a much better $\cO(\alpha_s^5)$ precision \cite{Baikov:2014qja}, the large-$N_C$ matching could be trivially promoted to NNLO accuracy in $\alpha_s$, once the Wilson coefficients are determined at this order.

\item The isospin-breaking correction $\Omega_{\mathrm{eff}}$ plays a quite important role in the $\varepsilon'/\varepsilon$ prediction. A complete re-analysis with updated inputs would be highly welcome because the moderate value of $\Omega_{\mathrm{eff}}$ results from a large numerical cancellation among different electromagnetic and strong contributions.
Work in this direction is currently under way \cite{VincenzoEtAl}.

\item Applying soft-pion techniques, the $\cO(e^2p^0)$ coupling $g_8\, g_{ewk}$ can be related to a dispersive integral over the hadronic vector and axial-vector spectral functions \cite{Donoghue:1993xd,Knecht:1998nn,Donoghue:1999ku,Knecht:2001bc}. This makes possible to perform a phenomenological estimate of this LEC with $\tau$ decay data. The published analyses, using the $\tau$ spectral functions measured at LEP, agree reasonably well with the large-$N_C$ determination, but their errors are rather large \cite{Narison:2000ys,Cirigliano:2001qw,Bijnens:2001ps,Cirigliano:2002jy}. A new phenomenological analysis, using the recently updated and more precise ALEPH $\tau$ data is close to being finalized \cite{Antonio}. Several lattice calculations of the matrix element $\langle\pi\pi|Q_8|K\rangle$ have been also published (some of them in the chiral limit) \cite{Bai:2015nea,Blum:2015ywa,Noaki:2001un,Ishizuka:2015oja,Blum:2001xb,Boucaud:2004aa}.

\item A matching calculation of the weak LECs at NLO in $1/N_C$ is a very challenging task that so far remains unsolved. Several analytical approaches have been pursued in the past to estimate the hadronic matrix elements of the $Q_i$ operators beyond the large-$N_C$ approximation \cite{Bertolini:1995tp,Bertolini:1997nf,Bertolini:1998vd,Bijnens:2000im,Hambye:2003cy,Pich:1995qp,Pich:1990mw,Jamin:1994sv,Bertolini:1997ir,Antonelli:1995nv,Antonelli:1995gw,Hambye:1998sma,Bardeen:1986vz,Buras:2014maa,Bijnens:1998ee}. A fresh look to these pioneering attempts from a modern perspective could bring new enlightenment and, perhaps, could suggest ways to implement some of these methods within a well-defined EFT framework where a proper NLO matching calculation could be accomplished.

\item The dominant two-loop $\chi$PT corrections originate from large chiral logarithms, either associated with unitarity contributions or infrared singularities of the massless Goldstone theory \cite{Pallante:1999qf,Pallante:2000hk,Pallante:2001he,Buchler:2001nm,Buchler:2005xn}. A reliable estimate of these two-loop contributions should be feasible.

\item In the next few years, lattice simulations are expected to provide new data on $K\to\pi\pi$ transitions, with improved methods and higher statistics \cite{Feng:2017voh}. Combined with appropriate $\chi$PT techniques, a better control of systematic uncertainties could be achieved. Moreover, analysing the sensitivity of the lattice results to several input parameters, such as quark masses and/or the electromagnetic coupling, one could try to disentangle the different contributions to the decay amplitudes and get a better understanding of the underlying dynamics. Improved lattice determinations of the strong LECs are also needed, in particular of the crucial $L_5$ parameter.

\end{itemize}

At present, the SM prediction of $\varepsilon'/\varepsilon$ agrees well with the measured value and provides a qualitative confirmation of the SM mechanism of CP violation. The theoretical error is still large, but the prospects to achieve a better accuracy in the next few years are good. A significant step forward in our theoretical understanding of the kaon dynamics would allow us to perform a precise test of the electroweak theory, giving complementary and very relevant information on the CKM matrix structure in the kaon sector.

\appendix

\section{Inputs}
\label{sec:inputs}

Table~\ref{tab:inputs} shows the inputs values adopted for the different SM parameters entering the $\varepsilon'/\varepsilon$ computation. 

\begin{table}[ht]
\renewcommand*{\arraystretch}{1.2}
\begin{center}
\begin{tabular}{|c|c|c|}
\hline
Parameter & Value & Ref. \\
\hline
\hline
$\lambda$ & $0.22506 \pm 0.00050$ & \cite{Olive:2016xmw} \\
$A$ & $0.811 \pm 0.026$ & \cite{Olive:2016xmw}  \\
$\rho$ & $0.124 \pm 0.019$ & \cite{Olive:2016xmw}  \\
$\eta$ & $0.356 \pm 0.011$ & \cite{Olive:2016xmw}  \\
\hline\hline
$\alpha^{-1}(M_Z^2)$ & $128.947 \pm 0.012$ & \cite{Davier:2017zfy} \\
$\alpha_s^{(n_f=3)}(M_\tau)$ & $0.325 \pm 0.015$  & \cite{Olive:2016xmw,Pich:2016yfh} \\
$\sin^2{\theta}_W(M_Z)_{\overline{\mathrm{MS}}}$ & $0.23129 \pm 0.00005$  & \cite{Olive:2016xmw}  \\
$M_W$ & $(80.385 \pm 0.015)\:\text{GeV}$ & \cite{Olive:2016xmw}  \\
$M_\tau$ & $(1.77686 \pm 0.00012)\:\text{GeV}$ & \cite{Olive:2016xmw} \\
\hline\hline
$\overline{m}_u(2\:\mathrm{GeV})$ & $(2.36\pm 0.24)\:\text{MeV}$ & \cite{Aoki:2016frl}  \\
$\overline{m}_d(2\:\mathrm{GeV})$ & $(5.03\pm 0.26)\:\text{MeV}$ & \cite{Aoki:2016frl}  \\
$\overline{m}_s(2\:\mathrm{GeV})$ & $(93.9\pm 1.1)\:\text{MeV}$ & \cite{Aoki:2016frl}  \\
$\overline{m}_c(\overline{m}_c)$ & $(1.286 \pm 0.030)\:\text{GeV}$ & \cite{Aoki:2016frl}  \\
$\overline{m}_b(\overline{m}_b)$ & $(4.190 \pm 0.021)\:\text{GeV}$  & \cite{Aoki:2016frl}    \\
$\overline{m}_t(\overline{m}_t)$ & $(165.9 \pm 2.1)\:\text{GeV}$ & \cite{Fuster:2017rev,Aad:2015waa} \\
\hline
\end{tabular}
\caption{Input values adopted for the relevant SM parameters.}
\label{tab:inputs}
\end{center}
\end{table}

\section*{Acknowledgements}
We have benefited from useful discussions with Vincenzo Cirigliano, Gerhard Ecker, Elvira Gámiz, Helmut Neufeld, Jorge Portolés and Antonio Rodríguez Sánchez.
This work has been supported in part by the Spanish Government and ERDF funds from the EU Commission [Grant FPA2014-53631-C2-1-P], by Generalitat Valenciana [Grant Prometeo/2017/053] and by the Spanish
Centro de Excelencia Severo Ochoa Programme [Grant SEV-2014-0398]. 
The work of H.G. is supported by a FPI doctoral contract [BES-2015-073138], funded by the Spanish Ministry of Economy, Industry and Competitiveness.


\begin{thebibliography}{99}

\bibitem{Cirigliano:2011ny}
  V.~Cirigliano, G.~Ecker, H.~Neufeld, A.~Pich and J.~Portol\'es,
  ``Kaon Decays in the Standard Model'',
  {\it Rev. Mod. Phys.}  {\bf 84} (2012) 399.

\bibitem{Glashow:1961tr}
  S.~L.~Glashow,
  ``Partial Symmetries of Weak Interactions'',
  Nucl.\ Phys.\  {\bf 22} (1961) 579.

\bibitem{Weinberg:1967tq}
  S.~Weinberg,
  ``A Model of Leptons'',
  Phys.\ Rev.\ Lett.\  {\bf 19} (1967) 1264.

\bibitem{Salam:1968rm}
  A.~Salam,
  ``Weak and Electromagnetic Interactions'',
  in Proc. $8^{\mathrm{th}}$ Nobel Symposium,
  N. Svartholm, ed. (Almqvist and Wiksell, Stockholm, 1968), 367.

\bibitem{GellMann:1953zza}
  M.~Gell-Mann,
  ``Isotopic Spin and New Unstable Particles'',
  Phys.\ Rev.\  {\bf 92} (1953) 833.

\bibitem{Pais:1952zz}
  A.~Pais,
  ``Some Remarks on the V-Particles'',
  Phys.\ Rev.\  {\bf 86} (1952) 663.

\bibitem{Dalitz:1954cq}
  R.~H.~Dalitz,
  ``Decay of tau mesons of known charge'',
  Phys.\ Rev.\  {\bf 94} (1954) 1046.
  
\bibitem{Lee:1956qn}
  T.~D.~Lee and C.~N.~Yang,
  ``Question of Parity Conservation in Weak Interactions'',
  Phys.\ Rev.\  {\bf 104} (1956) 254.
  
\bibitem{GellMann:1955jx}
  M.~Gell-Mann and A.~Pais,
  ``Behavior of neutral particles under charge conjugation'',
  Phys.\ Rev.\  {\bf 97} (1955) 1387.
  
\bibitem{Cabibbo:1963yz}
  N.~Cabibbo,
  ``Unitary Symmetry and Leptonic Decays'',
  Phys.\ Rev.\ Lett.\  {\bf 10} (1963) 531.

\bibitem{Glashow:1970gm}
  S.~L.~Glashow, J.~Iliopoulos and L.~Maiani,
  ``Weak Interactions with Lepton-Hadron Symmetry'',
  Phys.\ Rev.\ D {\bf 2} (1970) 1285.

\bibitem{Christenson:1964fg}
  J.~H.~Christenson, J.~W.~Cronin, V.~L.~Fitch and R.~Turlay,
  ``Evidence for the $2\pi$ Decay of the $K_2^0$ Meson'',
  Phys.\ Rev.\ Lett.\  {\bf 13} (1964) 138.

\bibitem{Kobayashi:1973fv}
  M.~Kobayashi and T.~Maskawa,
  ``CP Violation in the Renormalizable Theory of Weak Interaction'',
  Prog.\ Theor.\ Phys.\  {\bf 49} (1973) 652.

\bibitem{Rochester:1947mi}
  G.~D.~Rochester and C.~C.~Butler,
  ``Evidence for the Existence of New Unstable Elementary Particles'',
  Nature {\bf 160} (1947) 855.
  

\bibitem{Gaillard:1974hs}
  M.~K.~Gaillard and B.~W.~Lee,
  ``Rare Decay Modes of the K-Mesons in Gauge Theories'',
  Phys.\ Rev.\ D {\bf 10} (1974) 897.

\bibitem{Buras:1992uf}
  A.~J.~Buras and M.~K.~Harlander,
  ``A Top quark story: Quark mixing, CP violation and rare decays in the standard model'',
  Adv.\ Ser.\ Direct.\ High Energy Phys.\  {\bf 10} (1992) 58.

\bibitem{Olive:2016xmw}
  C.~Patrignani {\it et al.} [Particle Data Group],
  ``Review of Particle Physics'',
  Chin.\ Phys.\ C {\bf 40} (2016) no.10,  100001.

\bibitem{Burkhardt:1988yh}
  H.~Burkhardt {\it et al.}  [NA31 Collaboration],
  ``First Evidence for Direct CP Violation'',
  Phys.\ Lett.\ B
  {\bf 206} (1988) 169.

\bibitem{Barr:1993rx}
  G.~D.~Barr {\it et al.}  [NA31 Collaboration],
  ``A New measurement of direct CP violation in the neutral kaon system'',
  {\it Phys. Lett.} B {\bf 317} (1993) 233.

\bibitem{Fanti:1999nm}
  V.~Fanti {\it et al.}  [NA48 Collaboration],
  ``A New measurement of direct CP violation in two pion decays of the neutral kaon'',
  Phys.\ Lett.\ B
  {\bf 465} (1999) 335
  [hep-ex/9909022].

\bibitem{Lai:2001ki}
  A.~Lai {\it et al.}  [NA48 Collaboration],
  ``A Precise measurement of the direct CP violation parameter $\mathrm{Re}(\varepsilon'/\varepsilon)$'',
  {\it Eur. Phys. J.} C {\bf 22} (2001) 231
  [hep-ex/0110019].
  
\bibitem{Batley:2002gn}
  J.~R.~Batley {\it et al.}  [NA48 Collaboration],
  ``A Precision measurement of direct CP violation in the decay of neutral kaons into two pions'',
  {\it Phys. Lett.} B {\bf 544} (2002) 97
  [hep-ex/0208009].

\bibitem{Gibbons:1993zq} L.~K.~Gibbons {\it et al.} [E731 Collaboration],
  ``Measurement of the CP violation parameter $\mathrm{Re} (\varepsilon'/\varepsilon)$'',
  {\it Phys. Rev. Lett.}  {\bf 70} (1993) 1203.

\bibitem{AlaviHarati:1999xp}
  A.~Alavi-Harati {\it et al.}  [KTeV Collaboration],
  ``Observation of direct CP violation in $K_{S,L}\to\pi\pi$ decays'',
  {\it Phys. Rev. Lett.}  {\bf 83} (1999) 22
  [hep-ex/9905060].

\bibitem{AlaviHarati:2002ye}
  A.~Alavi-Harati {\it et al.}  [KTeV Collaboration],
  ``Measurements of direct CP violation, CPT symmetry, and other parameters in the neutral kaon system'',
  Phys.\ Rev.\ D
{\bf 67} (2003) 012005
   [{\it Erratum-ibid.} D {\bf 70} (2004) 079904]
  [hep-ex/0208007].

\bibitem{Abouzaid:2010ny}
  E.~Abouzaid {\it et al.}  [KTeV Collaboration],
  ``Precise Measurements of Direct CP Violation, CPT Symmetry, and Other Parameters in the Neutral Kaon System'',
  {\it Phys. Rev.} D {\bf 83} (2011) 092001
  [arXiv:1011.0127 [hep-ex]].

\bibitem{Gilman:1978wm}
  F.~J.~Gilman and M.~B.~Wise,
  ``The $\Delta I = 1/2$ Rule and Violation of CP in the Six Quark Model'',
  Phys.\ Lett.\  {\bf 83B} (1979) 83.

\bibitem{Buchalla:1989we}
  G.~Buchalla, A.~J.~Buras and M.~K.~Harlander,
  ``The Anatomy of $\varepsilon' / \varepsilon$ in the Standard Model'',
  Nucl.\ Phys.\ B {\bf 337} (1990) 313.
  
\bibitem{Buras:1993dy}
  A.~J.~Buras, M.~Jamin and M.~E.~Lautenbacher,
  ``The Anatomy of $\varepsilon' / \varepsilon$ beyond leading logarithms with improved hadronic matrix elements'',
  Nucl.\ Phys.\ B {\bf 408} (1993) 209
  [hep-ph/9303284].

\bibitem{Buras:1996dq}
  A.~J.~Buras, M.~Jamin and M.~E.~Lautenbacher,
  ``A 1996 analysis of the CP violating ratio $\varepsilon' / \varepsilon$'',
  {\it Phys. Lett.} B {\bf 389} (1996) 749
  [hep-ph/9608365].

\bibitem{Bosch:1999wr}
  S.~Bosch, A.~J.~Buras, M.~Gorbahn, S.~Jager, M.~Jamin, M.~E.~Lautenbacher and L.~Silvestrini,
  ``Standard model confronting new results for $\varepsilon' / \varepsilon$'',
  Nucl.\ Phys.\ B {\bf 565} (2000) 3
  [hep-ph/9904408].

\bibitem{Buras:2000qz}
  A.~J.~Buras, P.~Gambino, M.~Gorbahn, S.~Jager and L.~Silvestrini,
  ``$\varepsilon' / \varepsilon$ and rare K and B decays in the MSSM'',
  Nucl.\ Phys.\ B {\bf 592} (2001) 55
  [hep-ph/0007313].



\bibitem{Ciuchini:1995cd}  
  M.~Ciuchini, E.~Franco, G.~Martinelli, L.~Reina and L.~Silvestrini,
  ``An Upgraded analysis of $\varepsilon' / \varepsilon$ at the next-to-leading order'',
  {\it Z. Phys.} C {\bf 68} (1995) 239
  [hep-ph/9501265].

\bibitem{Ciuchini:1992tj}
  M.~Ciuchini, E.~Franco, G.~Martinelli and L.~Reina,
  ``$\varepsilon' / \varepsilon$ at the Next-to-leading order in QCD and QED'',
  {\it Phys. Lett.} B {\bf 301} (1993) 263
  [hep-ph/9212203].

\bibitem{Bertolini:1995tp}
  S.~Bertolini, J.~O.~Eeg and M.~Fabbrichesi,
  ``A New estimate of $\varepsilon'/\varepsilon$'',
  Nucl.\ Phys.\ B {\bf 476} (1996) 225
  [hep-ph/9512356].

\bibitem{Bertolini:1997nf}
  S.~Bertolini, J.~O.~Eeg, M.~Fabbrichesi and E.~I.~Lashin,
  ``$\varepsilon'/\varepsilon$ at $O(p^4)$ in the chiral expansion'',
  Nucl.\ Phys.\ B {\bf 514} (1998) 93
  [hep-ph/9706260].
      
\bibitem{Bertolini:1998vd}
  S.~Bertolini, M.~Fabbrichesi and J.~O.~Eeg,
  ``Theory of the CP violating parameter $\varepsilon' / \varepsilon$'',
  {\it Rev. Mod. Phys.} {\bf 72} (2000) 65
  [hep-ph/9802405].


\bibitem{Bertolini:2000dy}
  S.~Bertolini, J.~O.~Eeg and M.~Fabbrichesi,
  ``An Updated analysis of $\varepsilon'/\varepsilon$ in the standard model with hadronic matrix elements from the chiral quark model'',
  Phys.\ Rev.\ D {\bf 63} (2001) 056009
  [hep-ph/0002234].





\bibitem{Hambye:1999yy} 
  T.~Hambye, G.~O.~Kohler, E.~A.~Paschos and P.~H.~Soldan,
  ``Analysis of $\varepsilon' / \varepsilon$ in the $1 / N_c$ expansion'',
  {\it Nucl. Phys.} B {\bf 564} (2000) 391
  [hep-ph/9906434].


\bibitem{Pallante:1999qf}
  E.~Pallante and A.~Pich,
  ``Strong enhancement of $\varepsilon' / \varepsilon$ through final state interactions'',
  {\it Phys. Rev. Lett.}  {\bf 84} (2000) 2568
  [hep-ph/9911233].

\bibitem{Pallante:2000hk}
  E.~Pallante and A.~Pich,
  ``Final state interactions in kaon decays'',
  {\it Nucl. Phys.} B {\bf 592} (2001) 294
  [hep-ph/0007208].

\bibitem{Pallante:2001he}
  E.~Pallante, A.~Pich and I.~Scimemi,
  ``The Standard model prediction for $\varepsilon' / \varepsilon$'',
  {\it Nucl. Phys.} B {\bf 617} (2001) 441
 [hep-ph/0105011].

\bibitem{Bijnens:2000im}
  J.~Bijnens and J.~Prades,
  ``$\varepsilon'_K / \varepsilon_K$ in the chiral limit'',
  JHEP {\bf 0006} (2000) 035
  [hep-ph/0005189].
  
\bibitem{Hambye:2003cy}
  T.~Hambye, S.~Peris and E.~de Rafael,
  ``$\Delta I = 1/2$ and $\varepsilon' / \varepsilon$ in large $N_C$ QCD'',
  JHEP {\bf 0305} (2003) 027
  [hep-ph/0305104].

\bibitem{Buras:2003zz}
  A.~J.~Buras and M.~Jamin,
  ``$\varepsilon' / \varepsilon$ at the NLO: 10 years later'',
  JHEP {\bf 0401} (2004) 048
  [hep-ph/0306217].

\bibitem{Pich:2004ee}
  A.~Pich,
 ``$\varepsilon'/\varepsilon$ in the standard model: Theoretical update'',
  hep-ph/0410215.

\bibitem{Maiani:1990ca}
  L.~Maiani and M.~Testa,
  ``Final state interactions from Euclidean correlation functions'',
  Phys.\ Lett.\ B {\bf 245} (1990) 585.

\bibitem{Pekurovsky:1998jd}
  D.~Pekurovsky and G.~Kilcup,
  ``Matrix elements relevant for $\Delta I = 1/2$ rule and $\epsilon' / \epsilon$ from lattice QCD with staggered fermions'',
  Phys.\ Rev.\ D {\bf 64} (2001) 074502
  [hep-lat/9812019].
  
\bibitem{Noaki:2001un}
  J.~I.~Noaki {\it et al.} [CP-PACS Collaboration],
  ``Calculation of nonleptonic kaon decay amplitudes from $K\to\pi$ matrix elements in quenched domain wall QCD'',
  Phys.\ Rev.\ D {\bf 68} (2003) 014501
  [hep-lat/0108013].
  
\bibitem{Blum:2001xb}
  T.~Blum {\it et al.} [RBC Collaboration],
  ``Kaon matrix elements and CP violation from quenched lattice QCD: 1. The three flavor case'',
  Phys.\ Rev.\ D {\bf 68} (2003) 114506
  [hep-lat/0110075].

\bibitem{Bhattacharya:2004qu}
  T.~Bhattacharya, G.~T.~Fleming, R.~Gupta, G.~Kilcup, W.~Lee and S.~R.~Sharpe,
  ``Calculating $\epsilon' / \epsilon$ using HYP staggered fermions'',
  Nucl.\ Phys.\ Proc.\ Suppl.\  {\bf 140} (2005) 369
  [hep-lat/0409046].

\bibitem{Blum:2011ng} T.~Blum {\it et al.},
  T.~Blum {\it et al.}  [RBC and UKQCD Collaborations],
  ``The $K\to(\pi\pi)_{I=2}$ Decay Amplitude from Lattice QCD'',
  Phys.\ Rev.\ Lett.\
  {\bf 108} (2012) 141601;
  [arXiv:1111.1699 [hep-lat]];

\bibitem{Blum:2012uk}
T.~Blum {\it et al.} [RBC and UKQCD Collaborations],
  ``Lattice determination of the $K \to (\pi\pi)_{I=2}$ Decay Amplitude $A_2$'',
  {\it Phys. Rev.} D {\bf 86} (2012) 074513
  [arXiv:1206.5142 [hep-lat]].

\bibitem{Blum:2015ywa}
  T.~Blum {\it et al.},
  ``$K \rightarrow \pi\pi$ $\Delta I=3/2$ decay amplitude in the continuum limit'',
  Phys.\ Rev.\ D {\bf 91} (2015) no.7,  074502
  [arXiv:1502.00263 [hep-lat]].

\bibitem{Boyle:2012ys}
  P.~A.~Boyle {\it et al.}  [RBC and UKQCD Collaborations],
  ``Emerging understanding of the $\Delta I = 1/2$ Rule from Lattice QCD'',
  {\it Phys. Rev. Lett.}  {\bf 110} (2013) 152001
  [arXiv:1212.1474 [hep-lat]].

\bibitem{Pich:1995qp}
  A.~Pich and E.~de Rafael,
  ``Weak K amplitudes in the chiral and $1/N_c$ expansions'',
  {\it Phys. Lett.} B {\bf 374} (1996) 186
  [hep-ph/9511465].

\bibitem{Pich:1990mw}
  A.~Pich and E.~de Rafael,
  ``Four quark operators and nonleptonic weak transitions'',
  {\it Nucl. Phys.} B {\bf 358} (1991) 311.

\bibitem{Jamin:1994sv}
  M.~Jamin and A.~Pich,
  ``QCD corrections to inclusive $\Delta S = 1,2$ transitions at the next-to-leading order'',
  {\it Nucl. Phys.} B {\bf 425} (1994) 15
  [hep-ph/9402363].

\bibitem{Bertolini:1997ir}
  S.~Bertolini,  J.~O.~Eeg, M.~Fabbrichesi and E.~I.~Lashin,
  ``The $\Delta I = 1/2$ rule and $B_K$ at $O (p^4)$ in the chiral expansion'',
  {\it Nucl. Phys.} B {\bf 514} (1998) 63
  [hep-ph/9705244].

\bibitem{Antonelli:1995nv}
  V.~Antonelli, S.~Bertolini, J.~O.~Eeg, M.~Fabbrichesi and E.~I.~Lashin,
  ``The $\Delta S = 1$ weak chiral lagrangian as the effective theory of the chiral quark model'',
  Nucl.\ Phys.\ B {\bf 469} (1996) 143
   [hep-ph/9511255];

\bibitem{Antonelli:1995gw}
  V.~Antonelli, S.~Bertolini, M.~Fabbrichesi and E.~I.~Lashin,
  ``The $\Delta I = 1/2$ selection rule'',
  Nucl.\ Phys.\ B {\bf 469} (1996) 181
  [hep-ph/9511341].

\bibitem{Hambye:1998sma}
  T.~Hambye, G.~O.~Kohler, E.~A.~Paschos, P.~H.~Soldan and W.~A.~Bardeen,
  ``$1 / N_c$ corrections to the hadronic matrix elements of $Q_6$ and $Q_8$ in $K \to \pi \pi$ decays'',
  Phys.\ Rev.\ D {\bf 58} (1998) 014017
  [hep-ph/9802300].
  
\bibitem{Bardeen:1986vz}
  W.~A.~Bardeen, A.~J.~Buras and J.~M.~Gerard,
  ``A Consistent Analysis of the $\Delta I = 1/2$ Rule for K Decays'',
  {\it Phys. Lett.} B {\bf 192} (1987) 138.

\bibitem{Buras:2014maa}
  A.~J.~Buras, J.~M.~Gerard and W.~A.~Bardeen,
  ``Large $N$ Approach to Kaon Decays and Mixing 28 Years Later: $\Delta I = 1/2$ Rule, $\hat B_K$ and $\Delta M_K$'',
  {\it Eur. Phys. J.} C {\bf 74} (2014) 2871
  [arXiv:1401.1385 [hep-ph]].

\bibitem{Bijnens:1998ee}
  J.~Bijnens and J.~Prades,
  ``The Delta I = 1/2 rule in the chiral limit'',
  {\it JHEP} {\bf 9901} (1999) 023
  [hep-ph/9811472].

\bibitem{Ishizuka:2015oja}
  N.~Ishizuka, K.-I.~Ishikawa, A.~Ukawa and T.~Yoshié,
  ``Calculation of $K\to\pi\pi$ decay amplitudes with improved Wilson fermion action in lattice QCD'',
  Phys.\ Rev.\ D {\bf 92} (2015) no.7,  074503
  [arXiv:1505.05289 [hep-lat]].

\bibitem{Bai:2015nea}
  Z.~Bai {\it et al.} [RBC and UKQCD Collaborations],
  ``Standard Model Prediction for Direct CP Violation in $K\to\pi\pi$ Decay'',
  Phys.\ Rev.\ Lett.\  {\bf 115} (2015) no.21,  212001
  [arXiv:1505.07863 [hep-lat]].


\bibitem{Buras:2015xba}
  A.~J.~Buras and J.~M.~Gérard,
  ``Upper bounds on $\varepsilon'/\varepsilon$  parameters B$_{6}^{(1/2)}$ and B$_{8}^{(3/2)}$ from large N QCD and other news'',
  JHEP {\bf 1512} (2015) 008
  [arXiv:1507.06326 [hep-ph]].

\bibitem{Buras:2016fys}
  A.~J.~Buras and J.~M.~Gerard,
  ``Final state interactions in $K\rightarrow \pi \pi $ decays: $\Delta I=1/2$ rule vs. $\varepsilon '/\varepsilon $'',
  Eur.\ Phys.\ J.\ C {\bf 77} (2017) no.1,  10
  [arXiv:1603.05686 [hep-ph]].

\bibitem{Buras:2015yba}
  A.~J.~Buras, M.~Gorbahn, S.~Jäger and M.~Jamin,
 ``Improved anatomy of $\varepsilon '/\varepsilon $ in the Standard Model'',
  JHEP {\bf 1511} (2015) 202
  [arXiv:1507.06345 [hep-ph]].

\bibitem{Buras:2014sba}
  A.~J.~Buras, F.~De Fazio and J.~Girrbach,
  ``$\Delta I=1/2$ rule, $\varepsilon'/\varepsilon $ and $K\rightarrow \pi \nu \bar{\nu }$ in $Z' (Z)$ and $G' $ models with FCNC quark couplings'',
  Eur.\ Phys.\ J.\ C {\bf 74} (2014) no.7,  2950
  [arXiv:1404.3824 [hep-ph]].

\bibitem{Buras:2015yca}
  A.~J.~Buras, D.~Buttazzo and R.~Knegjens,
  ``$ K\to \pi \nu \overline{\nu} $ and $\varepsilon'/\varepsilon $ in simplified new physics models'',
  JHEP {\bf 1511} (2015) 166
  [arXiv:1507.08672 [hep-ph]].
  
\bibitem{Blanke:2015wba}
  M.~Blanke, A.~J.~Buras and S.~Recksiegel,
  ``Quark flavour observables in the Littlest Higgs model with T-parity after LHC Run 1'',
  Eur.\ Phys.\ J.\ C {\bf 76} (2016) no.4,  182
  [arXiv:1507.06316 [hep-ph]].

\bibitem{Buras:2015kwd}
  A.~J.~Buras and F.~De Fazio,
  ``$\varepsilon'/\varepsilon$ in 331 Models'',
  JHEP {\bf 1603} (2016) 010
  [arXiv:1512.02869 [hep-ph]].

\bibitem{Buras:2016dxz}
  A.~J.~Buras and F.~De Fazio,
  ``331 Models Facing the Tensions in $\Delta F=2$ Processes with the Impact on $\varepsilon^\prime/\varepsilon$, $B_s\to\mu^+\mu^-$ and $B\to K^*\mu^+\mu^-$'',
  JHEP {\bf 1608} (2016) 115
  [arXiv:1604.02344 [hep-ph]].

\bibitem{Buras:2015jaq}
  A.~J.~Buras,
  ``New physics patterns in $\varepsilon^\prime/\varepsilon$ and $\varepsilon_K$ with implications for rare kaon decays and $\Delta M_K$'',
  JHEP {\bf 1604} (2016) 071
  [arXiv:1601.00005 [hep-ph]].
      
\bibitem{Kitahara:2016otd}
  T.~Kitahara, U.~Nierste and P.~Tremper,
  ``Supersymmetric Explanation of CP Violation in $K\to \pi\pi$ Decays'',
  Phys.\ Rev.\ Lett.\  {\bf 117} (2016) no.9,  091802
  [arXiv:1604.07400 [hep-ph]].

\bibitem{Kitahara:2016nld}
  T.~Kitahara, U.~Nierste and P.~Tremper,
  ``Singularity-free next-to-leading order $\Delta$S = 1 renormalization group evolution and $\epsilon_K'/\epsilon_K$ in the Standard Model and beyond'',
  JHEP {\bf 1612} (2016) 078
  [arXiv:1607.06727 [hep-ph]].

\bibitem{Endo:2016aws}
  M.~Endo, S.~Mishima, D.~Ueda and K.~Yamamoto,
  ``Chargino contributions in light of recent $\epsilon'/\epsilon$'',
  Phys.\ Lett.\ B {\bf 762} (2016) 493
  [arXiv:1608.01444 [hep-ph]].

\bibitem{Endo:2016tnu}
  M.~Endo, T.~Kitahara, S.~Mishima and K.~Yamamoto,
  ``Revisiting Kaon Physics in General $Z$ Scenario'',
  Phys.\ Lett.\ B {\bf 771} (2017) 37
  [arXiv:1612.08839 [hep-ph]].

\bibitem{Cirigliano:2016yhc}
  V.~Cirigliano, W.~Dekens, J.~de Vries and E.~Mereghetti,
  ``An $\epsilon'$ improvement from right-handed currents'',
  Phys.\ Lett.\ B {\bf 767} (2017) 1
  [arXiv:1612.03914 [hep-ph]].

\bibitem{Alioli:2017ces}
  S.~Alioli, V.~Cirigliano, W.~Dekens, J.~de Vries and E.~Mereghetti,
  ``Right-handed charged currents in the era of the Large Hadron Collider'',
  JHEP {\bf 1705} (2017) 086
  [arXiv:1703.04751 [hep-ph]].
  
\bibitem{Bobeth:2016llm}
  C.~Bobeth, A.~J.~Buras, A.~Celis and M.~Jung,
  ``Patterns of Flavour Violation in Models with Vector-Like Quarks'',
  JHEP {\bf 1704} (2017) 079
  [arXiv:1609.04783 [hep-ph]].
    
\bibitem{Bobeth:2017xry}
  C.~Bobeth, A.~J.~Buras, A.~Celis and M.~Jung,
  ``Yukawa enhancement of $Z$-mediated new physics in $\Delta S = 2$ and $\Delta B = 2$ processes'',
  JHEP {\bf 1707} (2017) 124
  [arXiv:1703.04753 [hep-ph]].

\bibitem{Crivellin:2017gks}
  A.~Crivellin, G.~D'Ambrosio, T.~Kitahara and U.~Nierste,
  ``$K\to \pi \nu\overline{\nu}$ in the MSSM in light of the $\epsilon^{\prime}_K/\epsilon_K$ anomaly'',
  Phys.\ Rev.\ D {\bf 96} (2017) no.1,  015023
  [arXiv:1703.05786 [hep-ph]].

\bibitem{Chobanova:2017rkj}
  V.~Chobanova, G.~D'Ambrosio, T.~Kitahara, M.~Lucio Martinez, D.~Martinez Santos, I.~S.~Fernandez and K.~Yamamoto,
  ``Probing SUSY effects in $K_S^0\rightarrow\mu^+\mu^-$'',
  arXiv:1711.11030 [hep-ph].

\bibitem{Bobeth:2017ecx}
  C.~Bobeth and A.~J.~Buras,
  ``Leptoquarks meet $\varepsilon'/\varepsilon$ and rare Kaon processes'',
  arXiv:1712.01295 [hep-ph].

\bibitem{Endo:2017ums}
  M.~Endo, T.~Goto, T.~Kitahara, S.~Mishima, D.~Ueda and K.~Yamamoto,
  ``Gluino-mediated electroweak penguin with flavor-violating trilinear couplings'',
  arXiv:1712.04959 [hep-ph].

\bibitem{Lellouch:2000pv}
  L.~Lellouch and M.~Luscher,
  ``Weak transition matrix elements from finite volume correlation functions'',
  Commun.\ Math.\ Phys.\  {\bf 219} (2001) 31
  [hep-lat/0003023].

\bibitem{Luscher:1986pf}
  M.~Luscher,
  ``Volume Dependence of the Energy Spectrum in Massive Quantum Field Theories. 2. Scattering States'',
  Commun.\ Math.\ Phys.\  {\bf 105} (1986) 153.

\bibitem{Luscher:1990ux}
  M.~Luscher,
  ``Two particle states on a torus and their relation to the scattering matrix'',
  Nucl.\ Phys.\ B {\bf 354} (1991) 531.

\bibitem{Feng:2017voh}
  X.~Feng,
  ``Recent progress in applying lattice QCD to kaon physics'',
  arXiv:1711.05648 [hep-lat].

\bibitem{Cirigliano:2003nn}
  V.~Cirigliano, A.~Pich, G.~Ecker and H.~Neufeld,
  ``Isospin violation in $\varepsilon'$'',
  {\it Phys. Rev. Lett.}  {\bf 91} (2003) 162001
  [hep-ph/0307030].

\bibitem{Cirigliano:2003gt}
  V.~Cirigliano, G.~Ecker, H.~Neufeld and A.~Pich,
  ``Isospin breaking in $K\to\pi\pi$ decays'',
  {\it Eur. Phys. J.} C {\bf 33} (2004) 369
  [hep-ph/0310351];

\bibitem{Cirigliano:2009rr}
  V.~Cirigliano, G.~Ecker and A.~Pich,
  ``Reanalysis of pion pion phase shifts from $K\to\pi\pi$ decays'',
  {\it Phys. Lett.} B {\bf 679} (2009) 445
  [arXiv:0907.1451 [hep-ph]].








\bibitem{Aoki:2016frl}
  S.~Aoki {\it et al.},
  ``Review of lattice results concerning low-energy particle physics'',
  Eur.\ Phys.\ J.\ C {\bf 77} (2017) no.2,  112
  [arXiv:1607.00299 [hep-lat]].

\bibitem{Ecker:1988te}
  G.~Ecker, J.~Gasser, A.~Pich and E.~de Rafael,
  ``The Role of Resonances in Chiral Perturbation Theory'',
  Nucl.\ Phys.\ B {\bf 321} (1989) 311.

\bibitem{Ecker:1989yg}
  G.~Ecker, J.~Gasser, H.~Leutwyler, A.~Pich and E.~de Rafael,
  ``Chiral Lagrangians for Massive Spin 1 Fields'',
  Phys.\ Lett.\ B {\bf 223} (1989) 425.

\bibitem{Pich:2002xy}
  A.~Pich,
  ``Colorless mesons in a polychromatic world'',
Proc. Int. Workshop on {\it Phenomenology of Large $N_C$ QCD} (Tempe, Arizona, 2002), ed. R. F. Lebed, Proc. Institute for Nuclear Theory -- Vol. 12 (World Scientific, Singapore, 2002), p.~239  [hep-ph/0205030].

\bibitem{Cirigliano:2006hb}
  V.~Cirigliano, G.~Ecker, M.~Eidemuller, R.~Kaiser, A.~Pich and J.~Portol\'es,
  ``Towards a consistent estimate of the chiral low-energy constants'',
  Nucl.\ Phys.\ B {\bf 753} (2006) 139
  [hep-ph/0603205].

\bibitem{Kaiser:2007zz}
  R.~Kaiser,
  ``$\eta'$ contributions to the chiral low-energy constants'',
  Nucl.\ Phys.\ Proc.\ Suppl.\  {\bf 174} (2007) 97.
  
\bibitem{Cirigliano:2004ue}
  V.~Cirigliano, G.~Ecker, M.~Eidemuller, A.~Pich and J.~Portol\'es,
  ``$\langle VAP\rangle$  Green function in the resonance region'',
  Phys.\ Lett.\ B {\bf 596} (2004) 96
  [hep-ph/0404004].

\bibitem{Cirigliano:2005xn}
  V.~Cirigliano, G.~Ecker, M.~Eidemuller, R.~Kaiser, A.~Pich and J.~Portol\'es,
  ``The  Green function and SU(3) breaking in $K_{l3}$ decays'',
  JHEP {\bf 0504} (2005) 006
  [hep-ph/0503108].

\bibitem{RuizFemenia:2003hm}
  P.~D.~Ruiz-Femenia, A.~Pich and J.~Portol\'es,
  ``Odd intrinsic parity processes within the resonance effective theory of QCD'',
  JHEP {\bf 0307} (2003) 003
  [hep-ph/0306157].

\bibitem{Jamin:2004re}
  M.~Jamin, J.~A.~Oller and A.~Pich,
  ``Order $p^{6}$ chiral couplings from the scalar $K \pi$ form-factor'',
  JHEP {\bf 0402} (2004) 047
  [hep-ph/0401080].
  
\bibitem{Rosell:2004mn}
  I.~Rosell, J.~J.~Sanz-Cillero and A.~Pich,
  ``Quantum loops in the resonance chiral theory: The Vector form-factor'',
  JHEP {\bf 0408} (2004) 042
  [hep-ph/0407240].

\bibitem{Rosell:2006dt}
  I.~Rosell, J.~J.~Sanz-Cillero and A.~Pich,
  ``Towards a determination of the chiral couplings at NLO in $1/N_C$: $L^r_8(\mu)$'',
  JHEP {\bf 0701} (2007) 039
  [hep-ph/0610290].

\bibitem{Pich:2008jm}
  A.~Pich, I.~Rosell and J.~J.~Sanz-Cillero,
  ``Form-factors and current correlators: Chiral couplings $L_{10}^r(\mu)$ and $C_{87}^r(\mu)$ at NLO in $1/N_C$'',
  JHEP {\bf 0807} (2008) 014
  [arXiv:0803.1567 [hep-ph]].
 
\bibitem{GonzalezAlonso:2008rf}
  M.~Gonzalez-Alonso, A.~Pich and J.~Prades,
  ``Determination of the Chiral Couplings $L_{10}$ and $C_{87}$ from Semileptonic Tau Decays'',
  Phys.\ Rev.\ D {\bf 78} (2008) 116012
  [arXiv:0810.0760 [hep-ph]].
  
\bibitem{Pich:2010sm}
  A.~Pich, I.~Rosell and J.~J.~Sanz-Cillero,
  ``The vector form factor at the next-to-leading order in $1/N_C$: chiral couplings $L_9(\mu)$ and $C_{88}(\mu) - C_{90}(\mu)$'',
  JHEP {\bf 1102} (2011) 109
  [arXiv:1011.5771 [hep-ph]].

\bibitem{Bijnens:2014lea}
  J.~Bijnens and G.~Ecker,
  ``Mesonic low-energy constants'',
  Ann.\ Rev.\ Nucl.\ Part.\ Sci.\  {\bf 64} (2014) 149
  [arXiv:1405.6488 [hep-ph]].

\bibitem{Rodriguez-Sanchez:2016jvw}
  M.~González-Alonso, A.~Pich and A.~Rodríguez-Sánchez,
  ``Updated determination of chiral couplings and vacuum condensates from hadronic $\tau$ decay data'',
  Phys.\ Rev.\ D {\bf 94} (2016) no.1,  014017
  [arXiv:1602.06112 [hep-ph]].

\bibitem{Ananthanarayan:2017qmx}
  B.~Ananthanarayan, J.~Bijnens, S.~Friot and S.~Ghosh,
  ``An analytic representation of $F_K/F_{\pi}$'',
  arXiv:1711.11328 [hep-ph].

\bibitem{Weinberg:1978kz}
  S.~Weinberg,
  ``Phenomenological Lagrangians'',
  Physica A {\bf 96} (1979) 327.

\bibitem{Gasser:1983yg}
  J.~Gasser and H.~Leutwyler,
  ``Chiral Perturbation Theory to One Loop'',
  Annals Phys.\  {\bf 158} (1984) 142.

\bibitem{Gasser:1984gg}
  J.~Gasser and H.~Leutwyler,
  ``Chiral Perturbation Theory: Expansions in the Mass of the Strange Quark'',
  Nucl.\ Phys.\ B {\bf 250} (1985) 465.

\bibitem{Ecker:1994gg}
  G.~Ecker,
  ``Chiral perturbation theory'',
  Prog.\ Part.\ Nucl.\ Phys.\  {\bf 35} (1995) 1
  [hep-ph/9501357].

\bibitem{Pich:1995bw}
  A.~Pich,
  ``Chiral perturbation theory'',
  Rept.\ Prog.\ Phys.\  {\bf 58} (1995) 563
  [hep-ph/9502366].

\bibitem{Bijnens:1999sh}
  J.~Bijnens, G.~Colangelo and G.~Ecker,
  ``The Mesonic chiral Lagrangian of order $p^6$'',
  JHEP {\bf 9902} (1999) 020
  [hep-ph/9902437].

\bibitem{Bijnens:1999hw}
  J.~Bijnens, G.~Colangelo and G.~Ecker,
  ``Renormalization of chiral perturbation theory to order $p^6$'',
  Annals Phys.\  {\bf 280} (2000) 100
  [hep-ph/9907333].




\bibitem{Antonelli:2010yf} M.~Antonelli {\it et al.},
  ``An Evaluation of $|V_{us}|$ and precise tests of the Standard Model from world data on leptonic and semileptonic kaon decays'',
  {\it Eur. Phys. J.} C {\bf 69} (2010) 399.
  [arXiv:1005.2323 [hep-ph]].

\bibitem{Ecker:1999kr}
  G.~Ecker, G.~Muller, H.~Neufeld and A.~Pich,
  ``$\pi^0$-$\eta$ mixing and CP violation'',
  Phys.\ Lett.\ B {\bf 477} (2000) 88
  [hep-ph/9912264].

\bibitem{Altarelli:1974exa}
  G.~Altarelli and L.~Maiani,
  ``Octet Enhancement of Nonleptonic Weak Interactions in Asymptotically Free Gauge Theories'',
  Phys.\ Lett.\  {\bf 52B} (1974) 351.

\bibitem{Gaillard:1974nj}
  M.~K.~Gaillard and B.~W.~Lee,
  ``$\Delta I = 1/2$ Rule for Nonleptonic Decays in Asymptotically Free Field Theories'',
  Phys.\ Rev.\ Lett.\  {\bf 33} (1974) 108.

\bibitem{Vainshtein:1975sv}
  A.~I.~Vainshtein, V.~I.~Zakharov and M.~A.~Shifman,
  ``A Possible mechanism for the Delta T = 1/2 rule in nonleptonic decays of strange particles'',
  JETP Lett.\  {\bf 22} (1975) 55
   [Pisma Zh.\ Eksp.\ Teor.\ Fiz.\  {\bf 22} (1975) 123].

\bibitem{Shifman:1975tn}
  M.~A.~Shifman, A.~I.~Vainshtein and V.~I.~Zakharov,
  ``Light Quarks and the Origin of the $\Delta I = 1/2$ Rule in the Nonleptonic Decays of Strange Particles'',
  Nucl.\ Phys.\ B {\bf 120} (1977) 316.

\bibitem{Gilman:1979bc}
  F.~J.~Gilman and M.~B.~Wise,
  ``Effective Hamiltonian for $\Delta S = 1$ Weak Nonleptonic Decays in the Six Quark Model'',
  Phys.\ Rev.\ D {\bf 20} (1979) 2392.
  
  

\bibitem{Bijnens:1983ye}
  J.~Bijnens and M.~B.~Wise,
  ``Electromagnetic Contribution to $\varepsilon'/\varepsilon$'',
  Phys.\ Lett.\  {\bf 137B} (1984) 245.

\bibitem{Buras:1987wc}
  A.~J.~Buras and J.~M.~Gerard,
  ``Isospin Breaking Contributions to $\varepsilon'/\varepsilon$'',
  Phys.\ Lett.\ B {\bf 192} (1987) 156.

\bibitem{Sharpe:1987cx}
  S.~R.~Sharpe,
  ``On the Contribution of Electromagnetic Penguins to $\varepsilon^\prime$'',
  Phys.\ Lett.\ B {\bf 194} (1987) 551.
  
\bibitem{Lusignoli:1988fz}
  M.~Lusignoli,
  ``Electromagnetic Corrections to the Effective Hamiltonian for Strangeness Changing Decays and $\varepsilon^\prime / \varepsilon$'',
  Nucl.\ Phys.\ B {\bf 325} (1989) 33.

\bibitem{Flynn:1989iu}
  J.~M.~Flynn and L.~Randall,
  ``The Electromagnetic Penguin Contribution to $\epsilon' / \epsilon$ for Large Top Quark Mass'',
  Phys.\ Lett.\ B {\bf 224} (1989) 221
   [Erratum: Phys.\ Lett.\ B {\bf 235} (1990) 412].

\bibitem{Buchalla:1995vs}
  G.~Buchalla, A.~J.~Buras and M.~E.~Lautenbacher,
  ``Weak decays beyond leading logarithms'',
  {\it Rev. Mod. Phys.}  {\bf 68} (1996) 1125
  [hep-ph/9512380].
  
\bibitem{Buras:1991jm}
  A.~J.~Buras, M.~Jamin, M.~E.~Lautenbacher and P.~H.~Weisz,
  ``Effective Hamiltonians for $\Delta S = 1$ and $\Delta B = 1$ nonleptonic decays beyond the leading logarithmic approximation'',
  Nucl.\ Phys.\ B {\bf 370} (1992) 69
   [Addendum: Nucl.\ Phys.\ B {\bf 375} (1992) 501].
  
\bibitem{Buras:1992tc} 
  A.~J.~Buras, M.~Jamin, M.~E.~Lautenbacher and P.~H.~Weisz,
  ``Two loop anomalous dimension matrix for $\Delta S = 1$ weak nonleptonic decays. 1. $O(\alpha_s^2)$'',
  {\it Nucl. Phys.} B {\bf 400} (1993) 37
  [hep-ph/9211304].

\bibitem{Buras:1992zv}
  A.~J.~Buras, M.~Jamin and M.~E.~Lautenbacher,
  ``Two loop anomalous dimension matrix for $\Delta S = 1$ weak nonleptonic decays. 2. $O(\alpha\alpha_s)$'',
  Nucl.\ Phys.\ B {\bf 400} (1993) 75
  [hep-ph/9211321].

\bibitem{Ciuchini:1993vr}
  M.~Ciuchini, E.~Franco, G.~Martinelli and L.~Reina,
  ``The $\Delta S = 1$ effective Hamiltonian including next-to-leading order QCD and QED corrections'',
  {\it Nucl. Phys.} B {\bf 415} (1994) 403
  [hep-ph/9304257].


\bibitem{Buras:1999st}
  A.~J.~Buras, P.~Gambino and U.~A.~Haisch,
  ``Electroweak penguin contributions to nonleptonic $\Delta F = 1$ decays at NNLO'',
  Nucl.\ Phys.\ B {\bf 570} (2000) 117
  [hep-ph/9911250].

\bibitem{Gorbahn:2004my}
  M.~Gorbahn and U.~Haisch,
  ``Effective Hamiltonian for non-leptonic $|\Delta F| = 1$ decays at NNLO in QCD'',
  Nucl.\ Phys.\ B {\bf 713} (2005) 291
  [hep-ph/0411071].

\bibitem{Cerda-Sevilla:2016yzo}
  M.~Cerdà-Sevilla, M.~Gorbahn, S.~Jäger and A.~Kokulu,
  ``Towards NNLO accuracy for $\epsilon'/\epsilon$'',
  J.\ Phys.\ Conf.\ Ser.\  {\bf 800} (2017) no.1,  012008
  [arXiv:1611.08276 [hep-ph]].

\bibitem{tHooft:1972tcz}
  G.~'t Hooft and M.~J.~G.~Veltman,
  ``Regularization and Renormalization of Gauge Fields'',
  Nucl.\ Phys.\ B {\bf 44} (1972) 189.

\bibitem{Breitenlohner:1977hr}
  P.~Breitenlohner and D.~Maison,
  ``Dimensional Renormalization and the Action Principle'',
  Commun.\ Math.\ Phys.\  {\bf 52} (1977) 11.

\bibitem{Breitenlohner:1975hg}
  P.~Breitenlohner and D.~Maison,
  ``Dimensionally Renormalized Green's Functions for Theories with Massless Particles. 1.'',
  Commun.\ Math.\ Phys.\  {\bf 52} (1977) 39.
    
\bibitem{Breitenlohner:1976te}
  P.~Breitenlohner and D.~Maison,
  ``Dimensionally Renormalized Green's Functions for Theories with Massless Particles. 2.'',
  Commun.\ Math.\ Phys.\  {\bf 52} (1977) 55.

\bibitem{Pich:2016yfh}
  A.~Pich,
  ``Precision physics with QCD'',
  EPJ Web Conf.\  {\bf 137} (2017) 01016
  [arXiv:1612.05010 [hep-ph]].

\bibitem{Wolfenstein:1983yz}
  L.~Wolfenstein,
  ``Parametrization of the Kobayashi-Maskawa Matrix'',
  Phys.\ Rev.\ Lett.\  {\bf 51} (1983) 1945.

\bibitem{tHooft:1973alw} 
  G.~'t Hooft,
  ``A Planar Diagram Theory for Strong Interactions'',
  Nucl.\ Phys.\ B {\bf 72}, 461 (1974).

\bibitem{Witten:1979kh}
  E.~Witten,
  ``Baryons in the 1/n Expansion'',
  Nucl.\ Phys.\ B {\bf 160} (1979) 57.

\bibitem{Buras:1985yx}
  A.~J.~Buras and J.~M.~Gerard,
  ``$1/N$ Expansion for Kaons'',
  Nucl.\ Phys.\ B {\bf 264} (1986) 371.

\bibitem{Bardeen:1986uz}
  W.~A.~Bardeen, A.~J.~Buras and J.~M.~Gerard,
  ``The $K \to\pi \pi$ Decays in the Large n Limit: Quark Evolution'',
  Nucl.\ Phys.\ B {\bf 293} (1987) 787.






\bibitem{Wess:1971yu}
  J.~Wess and B.~Zumino,
  ``Consequences of anomalous Ward identities'',
  Phys.\ Lett.\  {\bf 37B} (1971) 95.
  
\bibitem{Witten:1983tw}
  E.~Witten,
  ``Global Aspects of Current Algebra'',
  Nucl.\ Phys.\ B {\bf 223} (1983) 422.
  
\bibitem{Urech:1994hd}
  R.~Urech,
  ``Virtual photons in chiral perturbation theory'',
  Nucl.\ Phys.\ B {\bf 433} (1995) 234
  [hep-ph/9405341].

    
\bibitem{Kambor:1989tz}
  J.~Kambor, J.~H.~Missimer and D.~Wyler,
  ``The Chiral Loop Expansion of the Nonleptonic Weak Interactions of Mesons'',
  Nucl.\ Phys.\ B {\bf 346} (1990) 17.

\bibitem{Ecker:1992de}
  G.~Ecker, J.~Kambor and D.~Wyler,
  ``Resonances in the weak chiral Lagrangian'',
  Nucl.\ Phys.\ B {\bf 394} (1993) 101.

\bibitem{Ecker:2000zr}
  G.~Ecker, G.~Isidori, G.~Muller, H.~Neufeld and A.~Pich,
  ``Electromagnetism in nonleptonic weak interactions'',
  Nucl.\ Phys.\ B {\bf 591} (2000) 419
  [hep-ph/0006172].

\bibitem{Knecht:1999ag}
  M.~Knecht, H.~Neufeld, H.~Rupertsberger and P.~Talavera,
  ``Chiral perturbation theory with virtual photons and leptons'',
  Eur.\ Phys.\ J.\ C {\bf 12} (2000) 469
  [hep-ph/9909284].
  
\bibitem{Cronin:1967jq}
  J.~A.~Cronin,
  ``Phenomenological model of strong and weak interactions in chiral $\mathrm{U}(3)\times \mathrm{U}(3)$'',
  Phys.\ Rev.\  {\bf 161} (1967) 1483.

\bibitem{Grinstein:1985ut}
  B.~Grinstein, S.~J.~Rey and M.~B.~Wise,
  ``{CP} Violation in Charged Kaon Decay'',
  Phys.\ Rev.\ D {\bf 33} (1986) 1495.

\bibitem{Cirigliano:2003yq}
  V.~Cirigliano, G.~Ecker, H.~Neufeld and A.~Pich,
  ``Meson resonances, large $N_c$ and chiral symmetry'',
  JHEP {\bf 0306} (2003) 012
  [hep-ph/0305311].

\bibitem{Dowdall:2013rya}
  R.~J.~Dowdall, C.~T.~H.~Davies, G.~P.~Lepage and C.~McNeile,
  ``$V_{us}$ from $\pi$ and $K$ decay constants in full lattice QCD with physical u, d, s and c quarks'',
  Phys.\ Rev.\ D {\bf 88} (2013) 074504
  [arXiv:1303.1670 [hep-lat]].
  
\bibitem{Moussallam:1997xx}
  B.~Moussallam,
  ``A Sum rule approach to the violation of Dashen's theorem'',
  Nucl.\ Phys.\ B {\bf 504} (1997) 381
  [hep-ph/9701400].
  
\bibitem{Bijnens:1996kk}
  J.~Bijnens and J.~Prades,
  ``Electromagnetic corrections for pions and kaons: Masses and polarizabilities'',
  Nucl.\ Phys.\ B {\bf 490} (1997) 239
  [hep-ph/9610360].

\bibitem{Kambor:1991ah}
  J.~Kambor, J.~H.~Missimer and D.~Wyler,
  ``$K \to 2 \pi$ and $K \to 3 \pi$ decays in next-to-leading order chiral perturbation theory'',
  Phys.\ Lett.\ B {\bf 261} (1991) 496.

\bibitem{Bijnens:1998mb}
  J.~Bijnens, E.~Pallante and J.~Prades,
  ``Obtaining $K \to\pi \pi$ from off-shell $K \to\pi$ amplitudes'',
  Nucl.\ Phys.\ B {\bf 521} (1998) 305
  [hep-ph/9801326].
  
\bibitem{Pallante:1998gk}
  E.~Pallante,
  ``The Generating functional for hadronic weak interactions and its quenched approximation'',
  JHEP {\bf 9901} (1999) 012
  [hep-lat/9808018].

\bibitem{Weinberg:1966kf}
  S.~Weinberg,
  ``Pion scattering lengths'',
  Phys.\ Rev.\ Lett.\  {\bf 17} (1966) 616.

\bibitem{Cirigliano:1999pv}
  V.~Cirigliano and E.~Golowich,
  ``Analysis of $O(p^2)$ corrections to $\langle\pi \pi | Q_{7,8} | K \rangle$'',
  Phys.\ Lett.\ B {\bf 475} (2000) 351
  [hep-ph/9912513].

\bibitem{Cirigliano:2001hs}
  V.~Cirigliano and E.~Golowich,
  ``Comment on `Analysis of $O(p^2)$ corrections to $\langle\pi \pi | Q_{7,8} | K \rangle$''',
  Phys.\ Rev.\ D {\bf 65} (2002) 054014
  [hep-ph/0109265].

\bibitem{Bijnens:1989mr}
  J.~Bijnens,
  ``$K_{l4}$ Decays and the Low-energy Expansion'',
  Nucl.\ Phys.\ B {\bf 337} (1990) 635.

\bibitem{Riggenbach:1990zp}
  C.~Riggenbach, J.~Gasser, J.~F.~Donoghue and B.~R.~Holstein,
  ``Chiral symmetry and the large $N_c$ limit in $K_{l4}$ decays'',
  Phys.\ Rev.\ D {\bf 43} (1991) 127.
  
\bibitem{Bijnens:1994ie}
  J.~Bijnens, G.~Colangelo and J.~Gasser,
  ``$K_{l4}$ decays beyond one loop'',
  Nucl.\ Phys.\ B {\bf 427} (1994) 427
  [hep-ph/9403390].

\bibitem{Amoros:1999qq}
  G.~Amoros, J.~Bijnens and P.~Talavera,
  ``Low-energy constants from $K_{\ell 4}$ form-factors'',
  Phys.\ Lett.\ B {\bf 480} (2000) 71
  [hep-ph/9912398].

\bibitem{Amoros:2000mc}
  G.~Amoros, J.~Bijnens and P.~Talavera,
  ``$K_{\ell 4}$ form-factors and $\pi-\pi$ scattering'',
  Nucl.\ Phys.\ B {\bf 585} (2000) 293
   Erratum: [Nucl.\ Phys.\ B {\bf 598} (2001) 665]
  [hep-ph/0003258].

\bibitem{Bijnens:1995yn}
  J.~Bijnens, G.~Colangelo, G.~Ecker, J.~Gasser and M.~E.~Sainio,
  ``Elastic $\pi \pi$ scattering to two loops'',
  Phys.\ Lett.\ B {\bf 374} (1996) 210
  [hep-ph/9511397].

\bibitem{Bijnens:1997vq}
  J.~Bijnens, G.~Colangelo, G.~Ecker, J.~Gasser and M.~E.~Sainio,
  ``Pion-pion scattering at low energy'',
  Nucl.\ Phys.\ B {\bf 508} (1997) 263
   Erratum: [Nucl.\ Phys.\ B {\bf 517} (1998) 639]
  [hep-ph/9707291].


\bibitem{Ananthanarayan:2000ht}
  B.~Ananthanarayan, G.~Colangelo, J.~Gasser and H.~Leutwyler,
  ``Roy equation analysis of $\pi \pi$ scattering'',
  Phys.\ Rept.\  {\bf 353} (2001) 207
  [hep-ph/0005297].

\bibitem{Colangelo:2000jc}
  G.~Colangelo, J.~Gasser and H.~Leutwyler,
  ``The $\pi \pi$ S wave scattering lengths'',
  Phys.\ Lett.\ B {\bf 488} (2000) 261
  [hep-ph/0007112].

\bibitem{Colangelo:2001df}
  G.~Colangelo, J.~Gasser and H.~Leutwyler,
  ``$\pi \pi$ scattering'',
  Nucl.\ Phys.\ B {\bf 603} (2001) 125
  [hep-ph/0103088].

\bibitem{Caprini:2003ta}
  I.~Caprini, G.~Colangelo, J.~Gasser and H.~Leutwyler,
  ``On the precision of the theoretical predictions for $\pi \pi$ scattering'',
  Phys.\ Rev.\ D {\bf 68} (2003) 074006
  [hep-ph/0306122].
  
\bibitem{Caprini:2011ky}
  I.~Caprini, G.~Colangelo and H.~Leutwyler,
  ``Regge analysis of the $\pi \pi$ scattering amplitude'',
  Eur.\ Phys.\ J.\ C {\bf 72} (2012) 1860
  [arXiv:1111.7160 [hep-ph]].

\bibitem{DescotesGenon:2001tn}
  S.~Descotes-Genon, N.~H.~Fuchs, L.~Girlanda and J.~Stern,
  ``Analysis and interpretation of new low-energy $\pi \pi$ scattering data'',
  Eur.\ Phys.\ J.\ C {\bf 24} (2002) 469
  [hep-ph/0112088].
  
\bibitem{GarciaMartin:2011cn}
  R.~Garcia-Martin, R.~Kaminski, J.~R.~Pelaez, J.~Ruiz de Elvira and F.~J.~Yndurain,
  ``The Pion-pion scattering amplitude. IV: Improved analysis with once subtracted Roy-like equations up to 1100 MeV'',
  Phys.\ Rev.\ D {\bf 83} (2011) 074004
  [arXiv:1102.2183 [hep-ph]].


\bibitem{Colangelo:2015kha}
  G.~Colangelo, E.~Passemar and P.~Stoffer,
  ``A dispersive treatment of $K_{\ell 4}$ decays'',
  Eur.\ Phys.\ J.\ C {\bf 75} (2015) 172
  [arXiv:1501.05627 [hep-ph]].

\bibitem{Christ:2017pze}
  N.~Christ and X.~Feng,
  ``Including electromagnetism in $K\to\pi\pi$ decay calculations'',
  arXiv:1711.09339 [hep-lat].

\bibitem{Lehner:2015jga}
  C.~Lehner, E.~Lunghi and A.~Soni,
  ``Emerging lattice approach to the K-Unitarity Triangle'',
  Phys.\ Lett.\ B {\bf 759} (2016) 82
  [arXiv:1508.01801 [hep-ph]].

\bibitem{Baikov:2014qja}
  P.~A.~Baikov, K.~G.~Chetyrkin and J.~H.~Kühn,
  ``Quark Mass and Field Anomalous Dimensions to ${\cal O}(\alpha_s^5)$'',
  JHEP {\bf 1410} (2014) 076
  [arXiv:1402.6611 [hep-ph]].

\bibitem{VincenzoEtAl} V.~Cirigliano {\it et. al.}, work in progress.

\bibitem{Donoghue:1993xd}
  J.~F.~Donoghue and E.~Golowich,
  ``Anatomy of a weak matrix element'',
  Phys.\ Lett.\ B {\bf 315} (1993) 406
  [hep-ph/9307263].

\bibitem{Knecht:1998nn}
  M.~Knecht, S.~Peris and E.~de Rafael,
  ``Matrix elements of electroweak penguin operators in the $1/N_c$ expansion'',
  Phys.\ Lett.\ B {\bf 457} (1999) 227
  [hep-ph/9812471].
    
\bibitem{Donoghue:1999ku}
  J.~F.~Donoghue and E.~Golowich,
  ``Dispersive calculation of $B^{(3/2)}_7$ and $B^{(3/2)}_8$ in the chiral limit'',
  Phys.\ Lett.\ B {\bf 478} (2000) 172
  [hep-ph/9911309].
  
\bibitem{Knecht:2001bc}
  M.~Knecht, S.~Peris and E.~de Rafael,
  ``A critical reassessment of $Q_7$ and $Q_8$ matrix elements'',
  Phys.\ Lett.\ B {\bf 508} (2001) 117
  [hep-ph/0102017].
  
\bibitem{Narison:2000ys}
  S.~Narison,
  ``New QCD estimate of the kaon penguin matrix elements and $\epsilon' / \epsilon$'',
  Nucl.\ Phys.\ B {\bf 593} (2001) 3
  [hep-ph/0004247].
    
\bibitem{Cirigliano:2001qw}
  V.~Cirigliano, J.~F.~Donoghue, E.~Golowich and K.~Maltman,
  ``Determination of $\langle(\pi \pi)_{I=2}| Q_{7,8} |K^0\rangle$ in the chiral limit'',
  Phys.\ Lett.\ B {\bf 522} (2001) 245
  [hep-ph/0109113].

\bibitem{Bijnens:2001ps}
  J.~Bijnens, E.~Gamiz and J.~Prades,
  ``Matching the electroweak penguins $Q_7$, $Q_8$ and spectral correlators'',
  JHEP {\bf 0110} (2001) 009
  [hep-ph/0108240].

\bibitem{Cirigliano:2002jy}
  V.~Cirigliano, J.~F.~Donoghue, E.~Golowich and K.~Maltman,
  ``Improved determination of the electroweak penguin contribution to $\varepsilon' / \varepsilon$ in the chiral limit'',
  Phys.\ Lett.\ B {\bf 555} (2003) 71
  [hep-ph/0211420].



\bibitem{Antonio} A.~Rodríguez-Sánchez and A.~Pich, work in progress.


\bibitem{Boucaud:2004aa}
  P.~Boucaud, V.~Gimenez, C.~J.~D.~Lin, V.~Lubicz, G.~Martinelli, M.~Papinutto and C.~T.~Sachrajda,
  ``An Exploratory lattice study of $\Delta I = 3/2$ $K \to\pi \pi$ decays at next-to-leading order in the chiral expansion'',
  Nucl.\ Phys.\ B {\bf 721} (2005) 175
  [hep-lat/0412029].

\bibitem{Buchler:2001nm}
  M.~Buchler, G.~Colangelo, J.~Kambor and F.~Orellana,
  ``Dispersion relations and soft pion theorems for $K \to\pi\pi$'',
  Phys.\ Lett.\ B {\bf 521} (2001) 22
  [hep-ph/0102287].

\bibitem{Buchler:2005xn}
  M.~Buchler,
  ``The Chiral logs of the $K \to\pi\pi$ amplitude'',
  Phys.\ Lett.\ B {\bf 633} (2006) 497
  [hep-ph/0511087].




\bibitem{Davier:2017zfy}
  M.~Davier, A.~Hoecker, B.~Malaescu and Z.~Zhang,
  ``Reevaluation of the hadronic vacuum polarisation contributions to the Standard Model predictions of the muon $g-2$ and $\alpha(m_Z)$ using newest hadronic cross-section data'',
  arXiv:1706.09436 [hep-ph].
  
\bibitem{Fuster:2017rev}
  J.~Fuster, A.~Irles, D.~Melini, P.~Uwer and M.~Vos,
  ``Extracting the top-quark running mass using $t\bar t +$ 1-jet events produced at the Large Hadron Collider'',
  arXiv:1704.00540 [hep-ph].
  
\bibitem{Aad:2015waa}
  G.~Aad {\it et al.} [ATLAS Collaboration],
  ``Determination of the top-quark pole mass using $ t\overline{t} $ + 1-jet events collected with the ATLAS experiment in 7 TeV pp collisions'',
  JHEP {\bf 1510} (2015) 121
  [arXiv:1507.01769 [hep-ex]].




%

    
\end{thebibliography}
\end{document}